# Technology interactions reshape the economics of China's coal power decarbonization

Yun-Long Zhang[1,2,3,4,*], Jia-Ning Kang[1,2,3,5,*], Xiaoming Kan[6], Lan-Cui Liu[7,*], Zhimin Huang [8], Song Peng[1,2,3,5], Biying Yu[1,2,3,5,*], Yi-Ming Wei[1,2,3,5,*]

**Abstract:** Decarbonizing existing coal-fired power plants can contribute to near-term climate mitigation, but identifying cost-effective retrofit strategies is complicated by interactions among mitigation technologies. Here we develop an interaction-aware optimization framework that jointly evaluates energy conservation, biomass co-firing, and carbon capture across 1,885 coal-fired power plants in China while accounting for plant heterogeneity and shared biomass and $CO_2$ storage resources. We find that technology interactions alter both mitigation costs and the emission reductions attributable to individual measures, thereby changing cost-optimal technology portfolios and marginal abatement cost curve at the fleet level. Approximately 1.2 Gt $CO_2$ $yr^{-1}$ can be mitigated at negative marginal cost, while reaching carbon neutrality requires a marginal abatement cost of US$56 t $CO_2^{-1}$. Progressively deeper mitigation shifts the cost-optimal portfolio from energy conservation toward biomass co-firing and ultimately carbon capture, with biomass combined with carbon capture enabling net-negative emissions. Explicitly accounting for interactions among mitigation technologies therefore provides a more consistent basis for evaluating coal-power decarbonization and coordinating retrofit investment, infrastructure development, and climate policy.

[1] Center for Energy and Environmental Policy Research, Beijing Institute of Technology, Beijing 100081, China. [2] Beijing Lab for System Engineering of Carbon Neutrality, Beijing Municipal Education Commission, Beijing 100081, China. [3] Basic Science Center for Energy and Climate Change, Beijing 100081, China. [4] Division of Physical Resource Theory, Department of Environmental and Energy Sciences, Chalmers University of Technology, 412 96, Göteborg, Sweden. [5] School of Management, Beijing Institute of Technology, Beijing 100081, China. [6] Department of Computer and Systems Sciences, Stockholm University, Stockholm, Sweden. [7] School of National Safety and Emergency Management, Beijing Normal University, Beijing 100875, China. [8] Robert B. Willumstad School of Business, Adelphi University, Garden City, NY 11530, USA.

* Corresponding authors: wei@bit.edu.cn (Y.-M. Wei), zhangyl_9@163.com (Y.-L. Zhang), kangjianing@bit.edu.cn (J.-N. Kang), liulancui@163.com (L.-C. Liu), yubiying_bj@bit.edu.cn (B. Yu)

Coal-fired power generation remains one of the largest sources of energy-related $CO_2$ emissions, and decarbonizing the existing coal fleet is therefore central to near- and medium-term climate mitigation. China plays a pivotal role in this transition, hosting approximately 51% of global coal power capacity, with nearly 90% of its generating units entering operation after 2000 [1]. Continued operation of this relatively young fleet could lock in nearly 140 Gt $CO_2$ emissions over its remaining lifetime [2], whereas achieving the Paris climate goals would require substantial early retirement of many existing coal-fired power plants [3].

However, despite the rapid expansion of renewable electricity, rapidly phasing out coal capacity in China remains heavily constrained by both technical [4] and socioeconomic considerations [5,6]. Extreme weather events and the Russia–Ukraine conflict have heightened concerns about energy supply security during the transition from fossil fuels to renewables [7], and similar challenges have emerged in advanced economies, where several countries temporarily extended coal-fired generation during recent energy crises despite long-term phase-out commitments [8]. For China, an aggressive and rapid retirement of relatively young units risks massive stranded assets, economic and job losses [9], as well as potential grid instability [10] as variable wind and solar generation expands. These practical constraints suggest that retrofitting the existing coal fleet will remain an important component of China's transition toward carbon neutrality. However, multiple retrofit options are available, each differing in mitigation potential, cost, and engineering characteristics. When multiple technologies coexist within the same plant, interactions among them can further alter fuel requirements, resource use, mitigation potential, and economic performance. Determining cost-effective technology portfolios while accounting for these interdependencies under progressively more stringent emission reduction targets therefore remains a central challenge for coal power decarbonization.

A range of measures has been proposed to address carbon lock-in in coal power [11], spanning operational adjustments such as reducing running hours (RRH) to technical retrofits like energy conservation (EC) [12], biomass co-firing (BC) [13–15], and carbon capture and storage (encompassing $CO_2$ capture, transport, and geological storage, hereafter referred to as CC) [16–19]. Early studies primarily examined gradually phasing out strategy [20] or assessed individual retrofit technologies, whereas more recent studies have increasingly evaluated multiple mitigation options within integrated modelling frameworks and derived plant- or fleet-level abatement costs. For example, joint evaluations have considered heat-rate improvements, biomass co-firing, and carbon capture for the U.S. coal fleet [21], while recent studies have investigated integrated coal-transition options in China, including biomass co-firing combined with carbon capture and storage (BECCS) [16,22–24]. As the range of technologies considered expands, interactions among retrofit measures become increasingly relevant: when multiple measures coexist within the same plant, one technology can alter fuel consumption, resource requirements, operating conditions, or the remaining emissions attributable to another technology, thereby affecting both mitigation costs and achievable emission reductions [25]. Accounting for such interdependencies is particularly important in bottom-up assessments used to construct system-level marginal abatement cost (MAC) curves, an issue long recognized in the MAC literature [26,27] and recently addressed through generalized interaction-consistent frameworks [28]. Yet, in coal-power decarbonization studies, these interaction effects are often embedded within integrated modelling results rather than explicitly isolated and quantified. Consequently, relatively little is known about how individual technology interactions propagate from plant-level techno-economic performance to fleet-wide MAC curves and cost-optimal technology portfolios.

In this study, we develop an interaction-aware optimization framework for decarbonizing China's coal-fired power fleet (research framework can be found in Supplementary Fig. S1). We first establish consistent plant-level benchmarks for individual mitigation measures and quantify pairwise interactions among energy conservation, biomass co-firing, and carbon capture. These interactions are then incorporated into a unified fleet-wide optimization framework that jointly determines cost-effective technology portfolios while accounting for plant heterogeneity, biomass resource competition, and $CO_2$ transport and storage constraints across 1,885 coal-fired power plants. Using this framework, we construct interaction-aware fleet-wide MAC curves and examine how cost-optimal technology portfolios evolve under progressively more ambitious emission reduction targets. By linking plant-level technology interactions with system-level optimization, this study provides a more realistic assessment of cost-effective coal power decarbonization and offers practical insights for coordinating retrofit investment and policy support toward carbon neutrality.

**Plant-level unit abatement costs of individual mitigation measures**

We first compared the plant-level unit abatement costs (UACs) of individual mitigation measures across China's coal-fired power fleet, using generation curtailment through reducing running hours (RRH) as a common benchmark (Fig. 1). Energy conservation (EC) achieves a lower UAC than RRH across 100% of existing power plants, followed by direct biomass co-firing (DB; ~92%), gasified biomass co-firing (GB; ~27%), and carbon capture (CC; 0%). Average UACs peak for CC (\$58 t $CO_2^{-1}$) and GB (up to \$26 t $CO_2^{-1}$). In contrast, DB reaches costs as low as − \$1.4 t $CO_2^{-1}$, and EC achieves values as low as − \$26 t $CO_2^{-1}$, both delivering negative average UACs. These results establish the standalone cost ranking of the mitigation options before accounting for interactions among technologies.

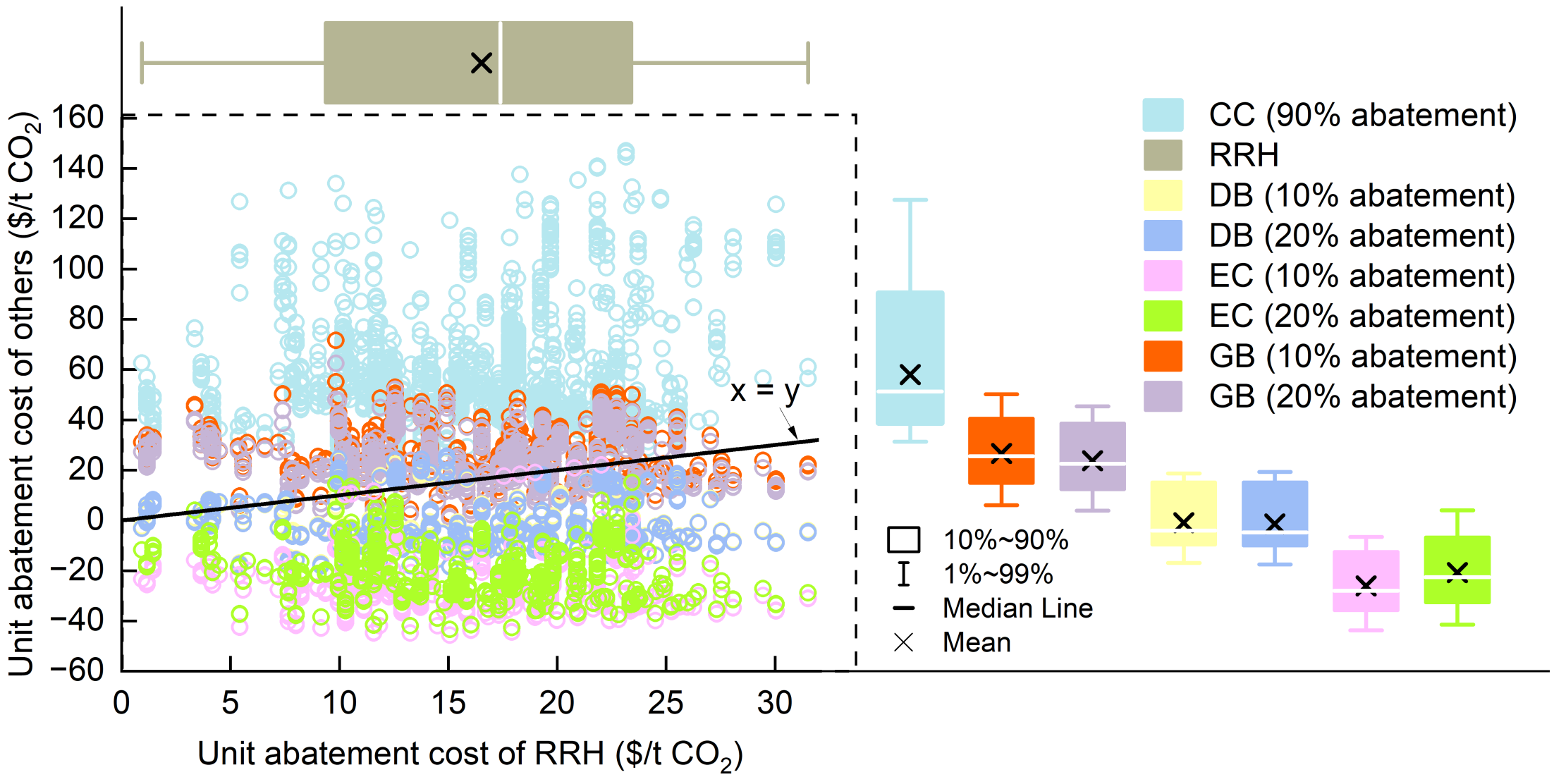

**Fig. 1** Plant-level unit abatement costs of individual mitigation measures. Unit abatement cost (UAC) distributions for energy conservation (EC), direct biomass co-firing (DB), gasified biomass co-firing (GB), and carbon capture (CC), benchmarked against reducing running hours (RRH). EC, DB, and GB are evaluated at 10% and 20% $CO_2$ reduction rates, and CC at a 90% capture rate. RRH has a constant UAC across curtailment levels.

**Impacts of interaction among $CO_2$ reduction technologies**

We next quantified pairwise interactions among EC, BC, and CC by comparing standalone and paired technology configurations (Fig. 2). Combining EC with BC or CC reduces the incremental levelized cost of electricity by 1–17% on average, reflecting improved overall plant efficiency and, for CC, a partial offset of the capture-related energy penalty (Fig. 2a,c), while simultaneously increasing their UACs because the attributable $CO_2$ reduction of the paired technology decreases (Fig. 2b,d). EC also lowers the unit cost of biomass-generated electricity: under 10% EC abatement, the reduction reaches \$2.4 $MWh^{-1}$ for DB and \$3.4 $MWh^{-1}$ for GB (Fig. 2f), increasing to \$5.2 $MWh^{-1}$ and \$6.8 $MWh^{-1}$, respectively, under 20% EC abatement. For BC, the average UAC increase ranges from \$0.5 to 2.6 t $CO_2^{-1}$ under 10% EC abatement and from \$1.1 to 6.0 t $CO_2^{-1}$ under 20% EC abatement, depending on the co-firing pathway and blending ratio; the corresponding increases for CC are \$1.1 and \$2.8 t $CO_2^{-1}$, respectively. In contrast, the BC–CC interaction produces the opposite effect: biomass co-firing reduces the UAC of CC by approximately \$0.5–1.0 t $CO_2^{-1}$ (Fig. 2e). The variation in these interaction effects across individual plants reflects heterogeneity in plant-specific technical and spatial characteristics, including baseline generation efficiency and plant scale, as well as differences in local coal prices, biomass availability and transport requirements for BC, and proximity to geological $CO_2$ storage sites for CC.

Together, these results show that technology combinations can alter both mitigation costs and attributable emission reductions, providing the basis for incorporating their interactions into the unified fleet-wide optimization.

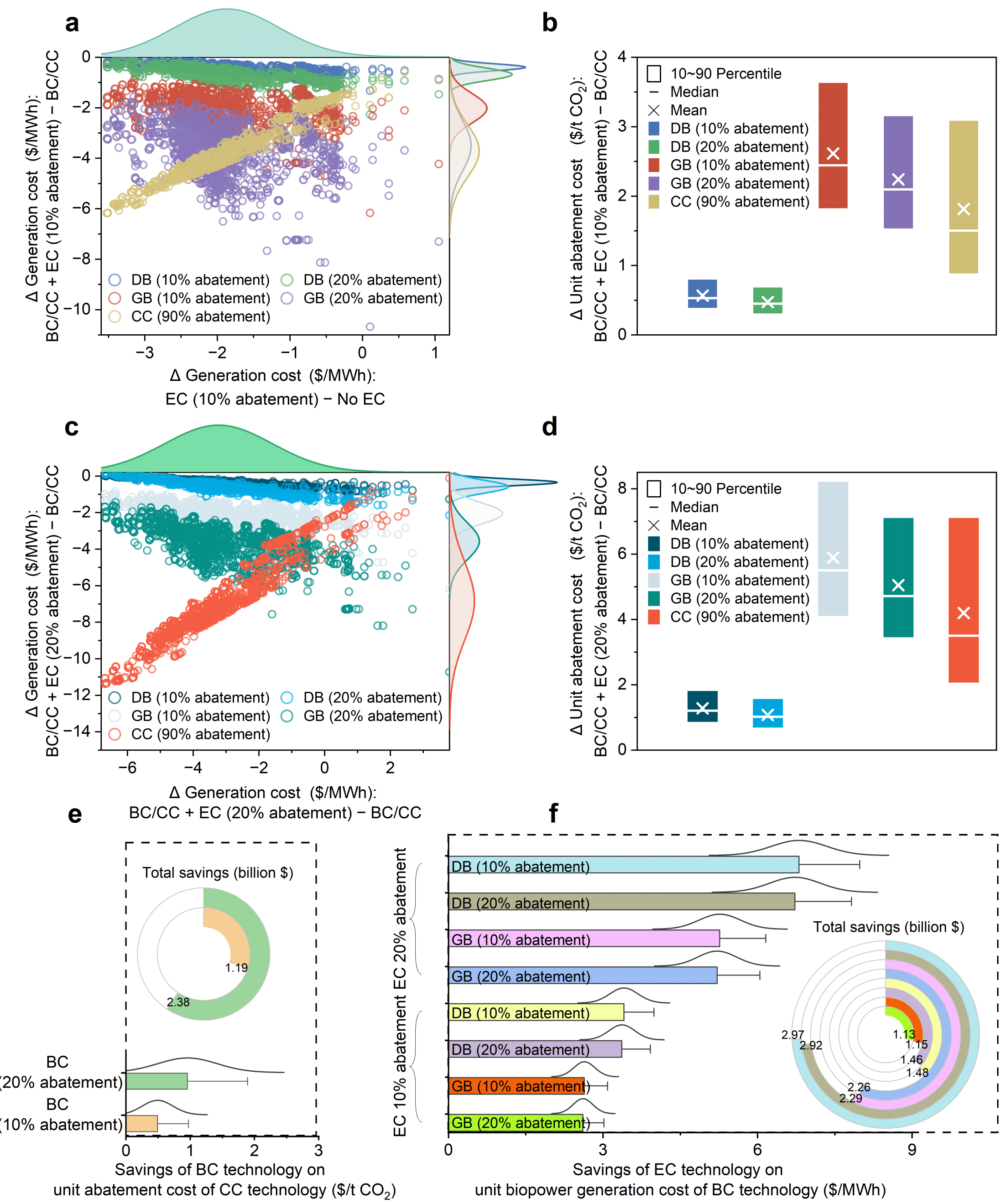

**Fig. 2** Interaction effects among abatement technologies on the economics of $CO_2$ reduction. Panels (a) and (c) compare the change in unit generation cost induced by EC alone (x-axis) with the corresponding change when EC is combined with BC or CC (y-axis), for EC packages achieving 10% and 20% $CO_2$ abatement, respectively. Negative values indicate cost reductions associated with EC. Panels (b) and (d) show the corresponding interaction-induced changes in unit abatement cost (UAC) of BC and CC after combining them with the 10% and 20% EC packages, respectively; positive values indicate that EC increases the UAC of the combined technology. Panel (e) quantifies

the BC–CC interaction as the reduction in the UAC of CC when carbon capture is combined with biomass co-firing relative to CC without biomass co-firing. Panel (f) quantifies the EC–BC interaction as the reduction in the unit cost of biomass-generated electricity for direct biomass co-firing (DB) and gasified biomass co-firing (GB) when combined with EC relative to biomass co-firing without EC. The unit savings in (e) and (f) capture the economic benefits arising from these technology interactions and are further used to assess their implications for technology-specific subsidy requirements. Total savings are calculated by multiplying the corresponding unit savings by the optimized amount of $CO_2$ abatement or biomass-based electricity generation in each scenario (see Supplementary Note 6 for details).

**Cost-optimal technology portfolios under progressively stringent mitigation targets**

Using the unified optimization framework, we derived the fleet-wide marginal abatement cost (MAC) curve and corresponding cost-optimal technology portfolios under progressively increasing $CO_2$ mitigation targets (Fig. 3). Three representative mitigation levels are identified along the MAC curve (Fig. 3a). At the zero-marginal-cost node (P12), approximately 1.2 Gt $CO_2$ can be mitigated while reducing total system costs, defining the range of profitable retrofit opportunities. The MAC subsequently rises to \$56 t $CO_2^{-1}$ as the fleet reaches carbon neutrality at P36 (approximately 3.6 Gt $CO_2$ of abatement) and to \$125 t $CO_2^{-1}$ at the maximum-abatement point P47. At P47, total $CO_2$ abatement reaches approximately 4.7 Gt, with EC reducing the original coal demand by 18.2% and biomass supplying approximately one-third of total fuel input (Supplementary Fig. S20). At this mitigation level, the large-scale combination of biomass co-firing and carbon capture enables fleet-wide net-negative emissions. Correspondingly, average fleet carbon intensity declines continuously from 805 g $CO_2$ $kWh^{-1}$ to −257 g $CO_2$ $kWh^{-1}$ at P47 (Fig. 3a). Benchmarking the MAC curve against carbon-market prices further shows that approximately 1.3, 4.0, and 4.4 Gt $CO_2$ of abatement can be achieved at marginal costs below the 2025 average carbon prices in China, the European Union, and the United Kingdom, respectively. In addition, approximately 1.75 Gt $CO_2$ of abatement can be achieved through technology retrofits at marginal costs below the highest plant-level cost of reducing running hours (RRH; approximately \$31 t $CO_2^{-1}$), indicating that retrofit-based mitigation remains more economical than generation curtailment over this range.

The contributions of individual technologies change substantially across the MAC curve (Fig. 3b). Energy conservation dominates at lower mitigation levels, contributing more than 68% of total abatement before cumulative reductions reach approximately 1 Gt $CO_2$. Biomass co-firing expands as targets increase, with gasified co-firing becoming more prominent at higher mitigation levels. Carbon capture expands rapidly under deep mitigation targets and ultimately accounts for nearly 60% of total abatement at the maximum-abatement node.

The optimized portfolios at the three representative nodes further illustrate this evolution (Fig. 3c). P12 is dominated by negative-cost EC measures with limited DB deployment, yielding annual system cost savings of approximately $20.5 billion through reduced coal consumption. At the carbon-neutrality threshold (P36), GB and CC expand substantially, with annualized system costs reaching approximately $75.4 billion. At P47, widespread CC deployment increases annualized system costs to approximately $163.7 billion. Overall, the optimized portfolios exhibit a clear shift in technology composition as mitigation targets tighten, from EC-dominated portfolios at lower mitigation levels toward increasing BC deployment and ultimately large-scale CC (Supplementary Fig. S14). Alternative technology-ordering analyses further show that this endogenous portfolio evolution is more cost-efficient than imposing alternative technology sequences (Supplementary Fig. S18).

Spatial deployment patterns also vary systematically across mitigation levels (Fig. 3d). At the profitable retrofit level, widespread EC and localized DB deployment dominate. At higher mitigation levels, GB expands throughout central and northern China, while CCS becomes increasingly prevalent across northern and western China as the fleet approaches carbon neutrality. At the maximum-abatement level, CCS is deployed broadly across southern and eastern China, and integrated multi-technology configurations prevail, enabling net-negative emissions across the coal-fired power fleet.

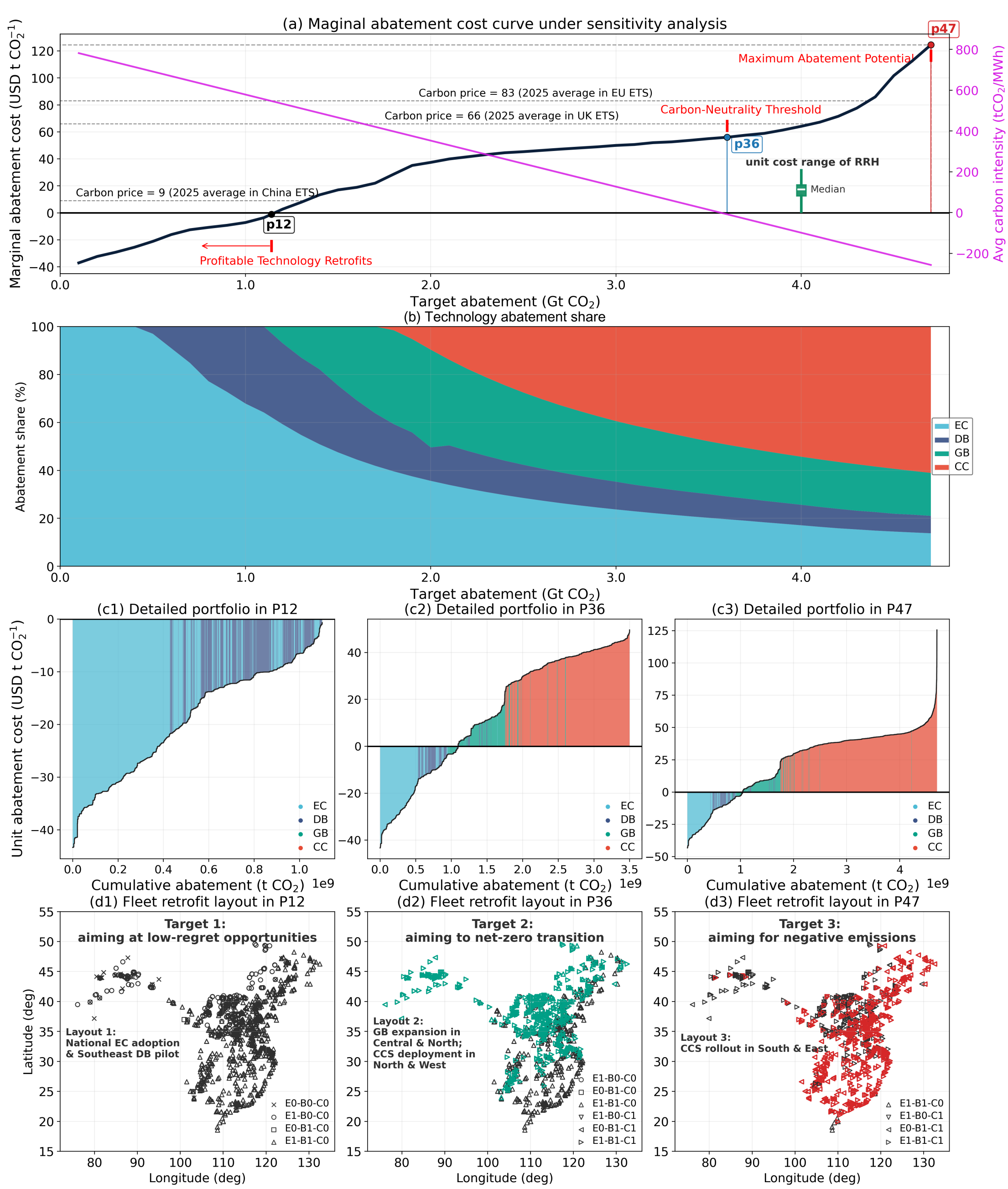

**Fig. 3 Fleet-wide marginal abatement cost (MAC) curve, technology deployment, and spatial retrofit patterns.** Fleet-wide marginal abatement cost curve and optimized technology portfolios. a, Fleet-wide marginal abatement cost (MAC) curve and average carbon intensity under progressively increasing mitigation targets. Three representative nodes are highlighted: P12, the zero-marginal-cost profitable retrofit stage; P36, the carbon-neutrality threshold; and P47, the maximum-abatement node. The boxplot shows the distribution of unit abatement costs for reducing running hours (RRH). b, Contributions of energy conservation (EC), direct biomass co-firing (DB), gasified biomass co-firing (GB), and carbon capture (CC) to total $CO_2$ abatement. c, Cost-optimal

technology portfolios at P12, P36, and P47. d, Corresponding spatial distributions of plant-level retrofit configurations. Marker codes denote technology combinations (E=EC, C=CC, B=BC; e.g., E1-C0-B1 denotes the co-adoption of EC and BC without CC), while highlighted markers indicate changes in plant-level retrofit configurations between representative targets.

**Robustness of the optimized strategy under alternative assumptions and uncertainties**

To evaluate the robustness of the optimized technology portfolios and assess the implications of alternative modelling approaches, we first compared the baseline joint optimization framework with two configurations adopted in previous studies: a conventional standalone technology-ranking approach [29,30] and a joint optimization framework without considering energy conservation [22,23]. We then examined the influence of key economic, engineering, and resource uncertainties (Fig. 4).

The baseline joint optimization yields higher marginal abatement costs and a lower maximum abatement potential than the conventional standalone technology-ranking approach across most of the mitigation range (Fig. 4a). The standalone approach underestimates the corresponding abatement cost by up to 27% at the zero-marginal-cost node P12, while overestimating the maximum mitigation potential by 8%. These differences indicate that independently ranking technologies provides a systematically more optimistic assessment when technology interactions and shared resource constraints are not fully represented. In contrast, excluding EC from the joint optimization increases marginal abatement costs by up to \$25 t $CO_2^{-1}$ and reduces the maximum mitigation potential by approximately 2%, demonstrating the contribution of EC to both the economics and mitigation potential of the optimized portfolios.

The MAC curve remains broadly robust to variations in economic parameters (Fig. 4b). Coal-price changes produce the largest deviations, reaching up to ± \$12 t $CO_2^{-1}$, particularly at low and intermediate mitigation levels, whereas variations in CC capital costs primarily affect the high-abatement portion of the curve. Changes in the discount rate have comparatively bigger effects at around 1.6 Gt $CO_2$ when gasified biomass co-firing start to deploy (Supplementary Fig. S21). Despite these variations, the overall MAC profile and major shifts in technology contributions remain stable

Resource availability and engineering performance primarily affect the extent of the mitigation frontier (Fig. 4c). Reduced biomass availability substantially constrains the maximum achievable abatement; when only 25% of the baseline biomass resource is available, the maximum mitigation potential is approximately sufficient to reach fleet-wide carbon neutrality, leaving little potential for net-negative emissions. Variations in EC effectiveness produce comparatively modest changes. Increasing the CC capture rate from 90% to 95% or 99.7% extends the maximum mitigation potential with limited changes to the overall MAC profile. Across these scenarios, the principal evolution of the optimized portfolios remains consistent, supporting the robustness of the main system-level findings.

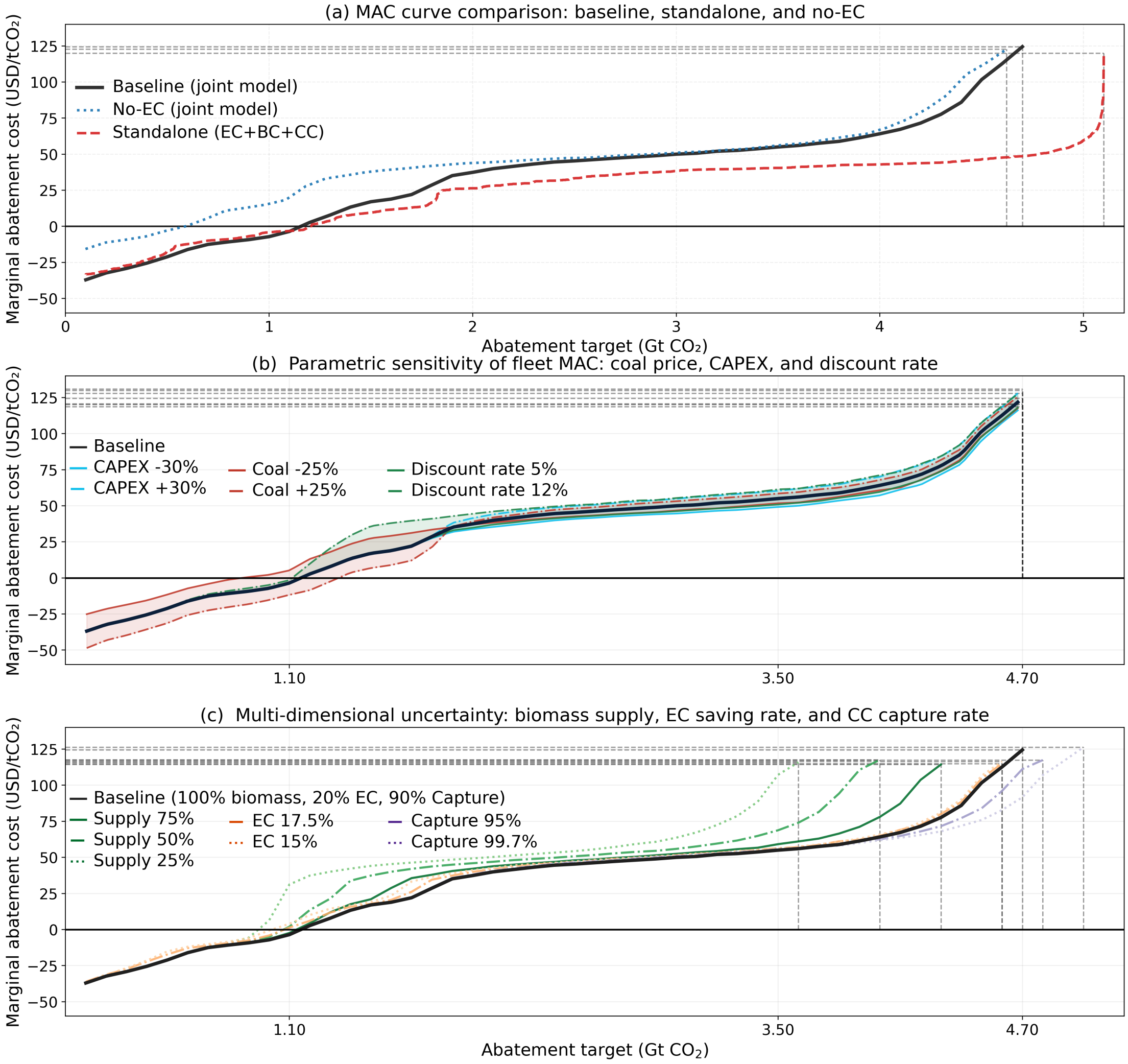

**Fig. 4 Robustness of the optimized MAC curve under alternative modeling assumptions and uncertainties.** Panel (a) compares the marginal abatement cost (MAC) curves obtained from the baseline joint optimization, a conventional standalone technology-ranking approach, and a joint optimization scenario excluding energy conservation. Panel (b) illustrates the sensitivity of the optimized MAC curve to uncertainties in coal prices (±25%), carbon capture capital expenditure (±30%), and discount rates (5% and 12%). Panel (c) evaluates the influence of biomass availability (100%, 75%, 50%, and 25% of the baseline supply), energy conservation performance (15%, 17.5%, and 20% energy savings), and carbon capture rates (90%, 95%, and 99.7%).

**Discussion**

Our findings have two main implications for coal-power decarbonization: retrofit technologies should be coordinated rather than evaluated independently, and their deployment should adapt as mitigation requirements become more stringent. Although the broad cost hierarchy of energy conservation (EC), biomass co-firing (BC), and carbon capture (CC) remains relatively consistent across the fleet, their economic performance changes when technologies are combined because each measure alters the operating conditions, resource requirements, and remaining mitigation opportunities of the others. At the system level, these interactions are further coupled with plant heterogeneity and competition for spatially constrained biomass and $CO_2$ storage resources. Cost-effective retrofit strategies therefore depend on both technology interactions and the depth of mitigation required.

The interaction analysis illustrates why these effects matter for both technology assessment and the construction of marginal abatement cost (MAC) curves. EC improves plant efficiency and reduces fuel requirements, thereby lowering the generation-cost penalty associated with biomass co-firing. At the same time, lower fuel consumption reduces the $CO_2$ emissions remaining for other mitigation measures, which can increase their calculated unit abatement costs (UACs) even when absolute operating costs decline. Biomass co-firing creates a different interaction with carbon capture: the biogenic $CO_2$ generated from biomass combustion enters the capture system and, when captured and geologically stored, increases the net mitigation attributed to CC, thereby reducing its effective abatement cost. Importantly, changes in UAC do not necessarily correspond to equivalent changes in physical resource costs. The engineering–accounting decomposition shows that the interaction effect for biomass co-firing is driven predominantly by changes in engineering costs, whereas changes in the UAC of carbon capture are more strongly influenced by changes in the amount of $CO_2$ attributed to CC—the denominator used to calculate its UAC (Fig. S19). This distinction is important because MAC estimates based solely on standalone technologies can obscure whether apparent cost changes reflect genuine engineering synergies or changes in the attribution of mitigation.

Accounting for these interactions yields a distinct evolution of cost-optimal technology portfolios as mitigation requirements tighten. At relatively low targets, EC and selected biomass co-firing retrofits provide substantial low-cost or profitable mitigation, with approximately 1.2 Gt $CO_2$ $yr^{-1}$ of abatement achievable before the system-level marginal cost becomes positive. As deeper reductions are required, the limited mitigation potential of these low-cost options necessitates increasing deployment of biomass co-firing and ultimately carbon capture, with the combination of biomass co-firing and carbon capture enabling net-negative emissions at the highest mitigation levels. This progression highlights an important distinction between cost-effectiveness and mitigation potential: technologies with the lowest standalone UACs are attractive early options but cannot alone deliver deep decarbonization, whereas higher-cost technologies become indispensable as the remaining emissions decline.

Investment characteristics add another dimension to these technology choices. Among plants for which both EC and direct biomass co-firing have negative UACs, direct co-firing generally requires less upfront capital but has a longer payback period, whereas EC typically requires greater initial investment but recovers that investment more rapidly (Fig. S24). Thus, even among nominally profitable mitigation options, financing availability can affect practical technology choices. Where capital is constrained, lower-investment biomass retrofits may be more accessible; where sufficient financing is available, EC may be more attractive because of its shorter payback period. At the full-fleet scale, fleet-wide levelized cost of electricity (LCOE) analysis indicates that low-cost EC and BC retrofits cut average power generation costs in the early energy transition phase. In contrast, large-scale carbon capture deployment significantly raises generation costs under deep decarbonization targets; for certain power plants, these elevated costs can even surpass documented generation costs of offshore wind and solar-plus-storage facilities (Fig. S13). Overall, the results reinforce the distinction between relatively low-cost retrofit opportunities and the substantially greater financing requirements associated with near-zero and net-negative emissions.

The interaction effects also have direct implications for policy design. Deep-decarbonization technologies remain difficult to deploy on purely commercial grounds, and different policy instruments affect different portions of the MAC curve. Biomass electricity incentives primarily improve the economics of biomass-based mitigation, whereas support for captured and geologically stored $CO_2$ becomes increasingly relevant as CC expands under stringent mitigation targets (Fig. S16). Importantly, technology interactions can reduce the fiscal support required to make these options economically viable. EC-induced reductions in biopower generation costs could lower the required biomass-related subsidy by more than \$1.1 billion $yr^{-1}$, while large-scale biomass co-firing could reduce required support for $CO_2$ capture and storage by up to \$2.38 billion $yr^{-1}$ (Fig. 2). These savings are spatially concentrated, with Inner Mongolia, Shandong, Xinjiang, Jiangsu, and Shanxi showing particularly large potential reductions in subsidy requirements (Fig. S23). Coordinating technology support according to local retrofit opportunities and interaction benefits could therefore achieve the same mitigation objectives with lower fiscal expenditure than policies that treat technologies independently.

Carbon pricing provides a complementary but, at current levels, insufficient incentive for deep decarbonization. The carbon-price analysis indicates that relatively low carbon prices primarily activate mitigation options that are already close to economic competitiveness, whereas widespread CC deployment requires substantially stronger incentives (Fig. S22). This suggests that a uniform carbon price alone may not address the heterogeneous investment barriers facing technologies at different stages of the transition. A combination of carbon pricing, technology-specific incentives, and financing mechanisms may therefore be more effective, particularly where high upfront investment rather than lifetime profitability constrains adoption.

Implementation feasibility depends equally on technological maturity, physical infrastructure, and resource availability. EC technologies are relatively mature and can often be incorporated into scheduled plant upgrades with limited additional infrastructure requirements [12,31]. Biomass co-firing requires reliable feedstock supply and supporting storage, preprocessing, and transportation systems [32], while high biomass shares additionally depend on technically and economically viable gasification. Large-scale CC similarly requires low-energy-penalty capture technologies, coordinated $CO_2$ transport networks, and access to geological storage. As mitigation deepens, increasing biomass and $CO_2$ transport requirements make regional infrastructure coordination increasingly important (Fig. S15) [33]. Biomass availability is particularly consequential, whereas plausible variations in EC performance and $CO_2$ capture rates have more moderate effects on the optimized portfolios. Future scenarios further show that technology learning and accelerated coal retirement alter mitigation costs and deployment scales without fundamentally changing the broader evolution of cost-optimal technology portfolios (Fig. S17).

Retrofit deployment should nevertheless be considered alongside progressive coal phase-out. Long-lived retrofit investments may create additional carbon lock-in if they prolong the operation of units that would otherwise be retired. Our results should therefore not be interpreted as supporting retrofit of the entire existing coal fleet. Rather, the RRH comparison and retirement scenarios indicate that the economic relevance of retrofit declines for some plants as mitigation deepens and the operating fleet contracts. Retrofit is thus best viewed as a complementary strategy for reducing emissions from the residual operating coal fleet, rather than as a substitute for coal retirement and renewable-energy expansion.

Several limitations should be considered when interpreting these results. First, although the framework represents plant heterogeneity, cross-technology interactions, biomass competition, and shared $CO_2$ storage resources, site-specific engineering constraints and additional integration costs associated with multi-technology retrofits cannot be consistently parameterized across the national fleet and are therefore not explicitly represented. Second, biomass availability remains uncertain because competing uses, sustainability constraints, and future supply-chain development may reduce the quantity economically available to the power sector. Biomass is also treated as carbon-neutral at the point of combustion, without explicitly accounting for life-cycle greenhouse-gas emissions from collection, processing, transportation, and other supply-chain activities. Incorporating these emissions would reduce the net mitigation attributed to biomass co-firing and BECCS and could affect their estimated abatement potential and cost-effectiveness. Third, the CCS pathway considered here includes $CO_2$ capture, transport, and geological storage but does not incorporate potential $CO_2$ utilization pathways, such as enhanced oil recovery (EOR). Potential revenues from $CO_2$ utilization are therefore excluded, which may lead to conservative estimates of CCS economics where economically viable utilization opportunities exist. Finally, the framework evaluates progressively stringent mitigation targets rather than an explicit intertemporal investment pathway; construction lead times, endogenous technology learning, infrastructure expansion, and investment lock-in are therefore not dynamically optimized. These limitations primarily affect the estimated costs and feasible scale of deployment and provide directions for future extensions of the framework.

Overall, our findings show that the economics of coal-power retrofits cannot be fully understood by ranking mitigation technologies independently. Cross-technology interactions alter both engineering costs and the attribution of emission reductions, while shared resources and heterogeneous plant conditions translate these effects into different fleet-level technology portfolios as mitigation requirements tighten. Explicitly representing these relationships provides a more informative basis for constructing system-level MAC curves and for coordinating retrofit investment, infrastructure development, and policy support. Although developed for China's coal-fired power fleet, the interaction-aware approach may also be applicable to other multi-technology decarbonization problems in which mitigation options compete for shared resources or modify one another's costs and abatement potential.

**Methods**

The methodological framework combines standalone technology assessment, pairwise interaction analysis, fleet-wide joint optimization, and scenario analysis to evaluate cost-effective decarbonization strategies for China's existing coal-fired power fleet (Supplementary Fig. S1). Standalone models first quantify the plant-level techno-economic performance of energy conservation (EC), biomass co-firing (BC), carbon capture (CC), and reducing running hours (RRH) under standardized mitigation targets. Pairwise analyses then quantify how combining retrofit technologies alters generation costs, attributable $CO_2$ reductions, and unit abatement costs (UACs). These technology models and interaction mechanisms are subsequently integrated into a fleet-wide mixed-integer linear programming (MILP) framework that optimizes technology portfolios under progressively increasing emission-reduction targets while accounting for biomass competition, $CO_2$ transport and storage constraints, and plant heterogeneity. Sensitivity, policy, and future-transition scenarios are further evaluated to test the robustness of the optimized portfolios.

The analysis covers 2,907 generating units across 1,885 coal-fired power plants with capacities exceeding 30 MW. Detailed data sources, technology parameters, scenario assumptions, and supplementary model formulations are provided in the Supplementary Notes 1-9.

**Energy conservation technology portfolio optimization**

The standalone EC model identifies the least-cost portfolio of energy-saving measures required for each generating unit to achieve a prescribed $CO_2$ reduction target. Twenty representative EC technologies covering boilers, steam turbines, auxiliary equipment, waste-heat recovery, and operational management are considered, with technology-specific investment costs, lifetimes, energy-saving potentials, and applicability constraints (Supplementary Table S3).

Multiple EC technologies can be implemented simultaneously[12,31,34]. For each plant $i$, the model minimizes the annualized retrofit cost:

$$\min F_i^{EC} \tag{1}$$

$$F_i^{EC} = \sum_{k=1}^{N_k} \left( C_{i,k}^{inv_EC} \cdot X_i \cdot CRF_{i,k}^{EC} \cdot O_{i,k} \right) \tag{2}$$

subject to the prescribed plant-level $CO_2$ reduction target:

$$ER_i^{EC} = \sum_{k=1}^{N_k} \left( ER_{i,k} \cdot X_i \cdot T_i \cdot O_{i,k} \right) \geq CRR \times E_{i,0} \tag{3}$$

$$E_{i,0} = \frac{3.6 \times X_i \cdot T_i \cdot e_{coal}}{\eta_i} \tag{4}$$

with the binary technology-selection variable:

$$O_{i,k} \in \{0,1\} \tag{5}$$

where $O_{i,k}$ is the decision variable of the model, determining whether power plant $i$ implements technology $k$ retrofit. $C_{i,k}^{inv_EC}$ and $CRF_{i,k}^{EC}$ denote the plant-specific investment cost and capital recovery factor (CRF); $ER_{i,k}$ is the unit emission-reduction potential; $CRR$ is the $CO_2$ reduction rate of the power plant; $E_{i,0}$ is the current annual carbon emissions of power plant $i$; $X_i$, $T_i$ and $\eta_i$ denote plant capacity, annual operating hours, and baseline generation efficiency; 3.6 is the unit conversion factor of power generation MWh to heat GJ; $e_{coal}$ is the carbon emission factor of coal combustion. Plant-specific investment costs and capital recovery factors account for equipment scale, technology lifetime, and the remaining lifetime of each coal unit, as detailed in the Supplementary Note 2.1.

After optimization, the plant-level unit abatement cost of EC is calculated as:

$$UAC_i^{EC} = \frac{\overline{F_i^{EC}}}{\sum_{k=1}^{N_k}\left(ER_{i,k} \cdot X_i \cdot T_i \cdot \overline{O_{i,k}}\right)} \tag{6}$$

where $\overline{O_{i,k}}$ and $\overline{F_i^{EC}}$ denote the optimized technology-selection decisions and corresponding annualized EC cost, respectively.

**Biomass co-firing technology minimum cost planning**

The standalone biomass co-firing (BC) model determines the least-cost biomass supply strategy for each generating unit under a prescribed co-firing level. Agricultural and forestry residues are spatially allocated to 10 × 10 km biomass supply hubs, with available resources and transportation distances linked to individual power plants. Detailed biomass resource estimation and spatialization procedures are provided in the Supplementary Note 3. In the standalone assessment, each plant is optimized independently without inter-plant competition for biomass resources.

For plant *i*, the model minimizes the annualized BC cost:

$$min\ F_i^{BC} \tag{7}$$

$$F_i^{BC} = F_i^{inv_BC} + F_i^{om_BC} + F_i^{fuel_BC} - F_i^{fuel_saving_BC} \tag{8}$$

where

$$F_i^{inv_BC} = C_i^{inv_BC} \cdot X_i \cdot \theta_i \cdot CRF_i^{BC} \tag{9}$$

$$F_i^{om_BC} = F_i^{inv_BC} \cdot a + b \cdot X_i \cdot T_i \cdot \theta_i \tag{10}$$

$$F_i^{fuel_BC} = \sum_{j=1}^{N_j}\left(C_j + C_{ij} \cdot D_{ij} \cdot \sigma\right) \cdot Z_{ji} \tag{11}$$

$$F_i^{fuel_saving_BC} = C_i^{coal} \cdot \frac{3.6 \cdot X_i \cdot T_i}{\eta_i} \cdot \theta_i \tag{12}$$

Biomass procurement is constrained by the energy demand corresponding to the prescribed co-firing ratio and the resource availability of each supply site:

$$\sum_{j=1}^{N_j} Z_{ji} \geq CRR \times P_i = \theta_i \cdot \frac{3.6 \cdot X_i \cdot T_i}{\eta_i} \quad (13)$$

$$Z_{ji} \leq P_j^{within10km} \quad (14)$$

Here, $\theta_i$ is the biomass co-firing ratio (biomass energy share) of power plant *i*, with fuel energy replacement ratio considered as the emission reduction rate of the power plant, $Z_{ji}$ is the biomass supplied from hub *j* to plant *i*, and $P_j^{within10km}$ denotes the available biomass resource associated with hub $j$. $C_i^{inv_BC}$ and $CRF_i^{BC}$ denote the plant-specific investment cost and capital recovery factor; *a* and *b* are fixed and variable O&M parameters; $F_i^{fuel_BC}$ is the cost of biomass fuel, which is the sum of the cost of biomass collection, storage, processing and transportation, among them, $C_j$ includes the unit cost of biomass collection, storage and processing, while $C_{ij}$ is the biomass unit transportation cost, and $\sigma$ is the distance curvature factor. $F_i^{fuel_saving_BC}$ is the coal cost savings from fuel substitution, and $C_i^{coal}$ is the unit cost of coal combustion for power plant *i* [35]. Plant-specific capital costs and other techno-economic parameters are detailed in the Supplementary Note 2.1.

The optimized plant-level UAC of BC is calculated as

$$UAC_i^{BC} = \frac{\overline{F_i^{BC}}}{E_{i,0} \cdot CRR} \quad (15)$$

where $\overline{F_i^{BC}}$ is the optimized annual mitigation cost and CRR is the corresponding $CO_2$ reduction rate. Direct biomass co-firing (DB) and gasified biomass co-firing (GB) are evaluated using the same framework but with technology-specific investment costs derived from Chinese demonstration projects, yielding separate $UAC_i^{DB}$ and $UAC_i^{GB}$.

**Carbon capture and storage technology source-sink matching optimization**

The standalone carbon capture and storage (CCS; hereafter CC) model determines the least-cost $CO_2$ source–sink matching solution for each power plant under a prescribed capture rate, considering the locations, storage capacities, and transport distances of candidate geological storage sites. Each plant is evaluated independently in the standalone assessment, without inter-plant competition for storage capacity.

For each power plant i, the optimization objective is to minimize the total annualized CCS cost:

$$\min F_i^{CC} \tag{16}$$

which comprises capture-system investment, operation and maintenance (O&M), and $CO_2$ transport and storage costs:

$$F_i^{CC} = F_i^{inv_CC} + F_i^{om_CC} + F_i^{TS} \tag{17}$$

The annualized capture-system investment and O&M costs are calculated as:

$$F_i^{inv_CC} = C_i^{inv_CC} \cdot CRF_i^{CC} \tag{18}$$

Similar to biomass co-firing, the fixed O&M cost of carbon capture technology is calculated as a percentage (*c*) of the investment cost. The variable O&M cost includes the cost of MEA solvent $C_i^{MEA}$ , corrosion inhibitor $C_i^{inhibiter}$ other reagents $C_i^{other}$, water $C_i^{water}$, waste treatment $C_i^{waste}$, steam $C_i^{steam}$ and electricity consumption $C_i^{electricity}$[36]. The specific calculations are as Equation (19):

$$F_i^{om_CC} = c \cdot F_i^{inv_CC} + \begin{pmatrix} C_i^{MEA} + C_i^{inhibiter} + C_i^{other} + C_i^{water} \\ +C_i^{waste} + C_i^{steam} + C_i^{electricity} \end{pmatrix} \cdot E_i^{CC} \tag{19}$$

The $CO_2$ transport and geological storage cost is calculated as:

$$F_i^{TS} = \sum_{u=1}^{N_u} CO2_{iu}(c_{iu}^{C} \cdot D_{iu} + c^{store}) \tag{20}$$

where $CO2_{iu}$ is the annual $CO_2$ flow from plant *i* to storage site *u*. $D_{iu}$ is the corresponding transport distance, $c_{iu}^{C}$ is the unit transport cost, and $c^{store}$ is the unit geological storage cost. Pipeline transport is assumed in the baseline analysis. Detailed capture-cost parameters, capital-cost scaling, transport assumptions, and storage data are provided in the Supplementary Note 2.1.

Neglecting $CO_2$ losses during transport and storage, captured and stored $CO_2$ is treated as the corresponding emission reduction. The annual capture requirement is:

$$E_i^{CC} = CRR \times E_{i,0} \tag{21}$$

where $E_i^{CC}$ is the baseline annual $CO_2$ emission of power plant *i*, and CRR is the prescribed $CO_2$ capture rate. A capture rate of 90% is adopted in the standalone and baseline assessment.

The standalone source–sink matching optimization is subject to the following $CO_2$ allocation and storage-capacity constraints:

$$\sum_{u=1}^{N_u} CO2_{iu} = E_i^{CC} \tag{22}$$

$$CO2_{iu} \leq Q_u \tag{23}$$

where $Q_u$ is the annualized storage capacity of site *u*. These constraints ensure that the prescribed capture volume is fully allocated without exceeding the capacity of any individual storage site.

The optimized plant-level UAC of CCS is then calculated as:

$$UAC_i^{CC} = \frac{\overline{F_i^{CC}}}{E_i^{CC}} \tag{24}$$

Where $\overline{F_i^{CC}}$ is the minimum annualized CCS cost obtained from the standalone optimization.

**Reducing running hours abatement loss evaluation**

Reducing running hours (RRH) lowers $CO_2$ emissions by reducing coal-fired electricity generation rather than through additional retrofit investment. Its economic cost is therefore measured as the foregone operating profit associated with reduced generation. For plant i, annual generation revenue and cost at operating level h are calculated as:

$$PG_{i,h} = X_i \times T_{i,h} \times Tariff_i \tag{25}$$

$$CG_{i,h} = C_i^{inv_coal} \cdot CRF_i^{RRH} + X_i \times T_{i,h} \times \left(OM_i^{coal} + FUEL_i^{coal}\right) \tag{26}$$

where $PG_{i,h}$ is the revenue achievable by power plant *i* at an annual generation hour of *h*; $CG_{i,h}$ is the cost required for power plant *i* to generate *h* hours of electricity. $Tariff_i$, $OM_i^{coal}$ and $FUEL_i^{coal}$ denote the feed-in tariff, O&M cost, and fuel cost per unit of electricity generation, respectively; $C_i^{inv_coal}$ is the initial investment cost of the coal power plant *i*, and $CRF_i^{RRH}$ is the capital recovery factor for coal power generation.

The plant-level UAC of RRH is calculated from the change in operating profit between the baseline generation level (*h0*) and the reduced-generation scenario (*h1*) conditions:

$$UAC_i^{RRH} = \frac{\left(PG_{i,h0} - CG_{i,h0}\right) - \left(PG_{i,h1} - CG_{i,h1}\right)}{X_i \times \left(T_{i,h0} - T_{i,h1}\right) \times \frac{3.6 \cdot e_{coal}}{\eta_i}} = \frac{Tariff_i - \left(OM_i^{coal} + FUEL_i^{coal}\right)}{\frac{3.6 \cdot e_{coal}}{\eta_i}} \tag{27}$$

Where $T_{i,h0}$ is the baseline annual generation hours, and $T_{i,h1}$ represents the reduced operating hours corresponding to the prescribed $CO_2$ reduction rate. Because the fixed capital-cost term cancels when calculating the change in operating profit, $UAC_i^{RRH}$ provides a plant-specific benchmark that is independent of the magnitude of the reduction in operating hours.

**Pairwise technology interaction analysis**

To quantify how retrofit technologies affect one another when combined within the same power plant, pairwise interaction analyses were conducted for energy conservation (EC), biomass co-firing (BC), and carbon capture (CC), considering three combinations: EC–BC, EC–CC, and BC–CC. For each pair, the combined configuration was compared with the corresponding standalone case under consistent plant characteristics and technology assumptions. Changes in fuel consumption, resource requirements, costs, $CO_2$ emissions, and attributable abatement were propagated through the technology-specific models described above.

The three pairs capture distinct interaction mechanisms. EC alters plant fuel consumption and efficiency, thereby affecting biomass requirements and the amount of $CO_2$ available for capture; biomass co-firing changes fuel composition and the amount of biogenic $CO_2$ entering the capture system. These effects are quantified through changes in unit generation costs, unit abatement costs (UACs), and attributable $CO_2$ reductions.

For a generic technology pair *m* and *n*, the interaction-induced change in the unit cost indicator of technology *n* is defined as:

$$\Delta UC_{n,i} = UC_{n,i}^{(m+n)} - UC_{n,i}^{(n)} \tag{28}$$

where $UC_{n,i}^{(n)}$ represents the unit cost of technology *n* for plant *i* when evaluated independently, and $UC_{n,i}^{(m+n)}$ represents its corresponding unit cost when technologies *m* and *n* coexist within the same plant. Depending on the technology considered, UC represents either the unit electricity-generation cost or the unit abatement cost. A negative value of $\Delta UC_{n,i}$ indicates that the paired configuration improves the economic performance of technology *n*, whereas a positive value indicates an increase in its unit cost.

**Unified fleet-wide joint optimization framework for marginal abatement cost**

（1）**Unified fleet-wide optimization model**

Building on the standalone technology models and pairwise interaction analysis, we develop a fleet-wide mixed-integer linear programming (MILP) framework to identify least-cost retrofit portfolios under progressively increasing $CO_2$-abatement targets. The model jointly determines technology adoption, biomass allocation, $CO_2$ source–sink matching, and plant-level mitigation contributions while accounting for cross-technology interactions, heterogeneous plant characteristics, and shared biomass and geological storage resources.

The objective is to minimize the total annualized mitigation cost across all $N_i$ coal-fired power plants:

$$\min Z = \sum_{i=1}^{N_i} \left( F_i^{EC} + F_i^{BC} + F_i^{CC} \right) \tag{29}$$

Where, $F_i^{EC}$, $F_i^{BC}$, and $F_i^{CC}$ represent the annualized costs associated with EC, BC, and CC, respectively.

Technology-specific investment, O&M, fuel, and transport and storage costs follow the formulations of the standalone models described above, with modifications introduced below to represent joint deployment and resource competition.

**Energy conservation**

For EC retrofits, the binary decision variable $O_i^{EC}$ determines whether EC sub-technology *k* is implemented at plant *i*. The annualized EC investment cost is:

$$F_i^{EC} = \sum_{k=1}^{N_k} \left( C_{i,k}^{inv_EC} \cdot X_i \cdot CRF_{i,k}^{EC} \cdot O_{i,k} \right) \equiv C_i^{EC} O_i^{EC} \tag{30}$$

**Biomass co-firing**

The unified model distinguishes direct biomass co-firing (DB) from gasified biomass co-firing (GB). Globally accepted commercial direct biomass co-firing is limited at a 20% co-firing ratio [37], while gasified co-firing enables higher biomass shares and can achieve a 100% co-firing [38,39]. This engineering boundary is represented through the binary variable $h_i$, which activates the additional investment requirement associated with GB.

The annualized BC investment cost is expressed as:

$$F_i^{inv_BC} = \left[C_i^{inv_{DB}} \cdot \theta_i^{eff} \cdot X_i + \Delta C_i^{inv_GB_DB} \cdot h_i \cdot X_i\right] \cdot CRF_i^{BC} \tag{31}$$

$$\Delta C_i^{inv_GB_DB} = C_i^{inv_GB} - C_i^{inv_{DB}} \tag{32}$$

where $C_i^{inv_DB}$ and $C_i^{inv_GB}$ denote the unit investment expenditures for direct and gasified co-firing technologies, respectively.

When EC and BC coexist, efficiency improvements reduce the absolute biomass requirement associated with a nominal biomass blending ratio $\theta_i$. The effective biomass requirement is therefore represented as:

$$\theta_i^{eff} = \theta_i - \eta_{ec} \times brxEC_i \tag{33}$$

Where $brxEC_i = \theta_i O_i^{EC}$ is an auxiliary variable representing the interaction between EC deployment and biomass blending. The corresponding linearization constraints are provided in Supplementary Note 2.2.

BC O&M, biomass procurement, transportation, and avoided coal costs follow the standalone BC formulation but are evaluated using the endogenous biomass share $\theta_i^{eff}$, and biomass resources are now allocated simultaneously among competing power plants. Biomass is treated as carbon-neutral at the point of combustion, and its life-cycle greenhouse-gas emissions are outside the system boundary and are therefore not included in the $CO_2$-abatement calculation.

**Carbon capture and storage**

CC deployment at plant *i* is represented by the binary variable $O_i^{CC}$. The annualized capture-system investment cost $F_i^{inv_CC}$ is:

$$F_i^{inv_CC} = C_i^{inv_CC} \cdot CRF_i^{CC} \cdot O_i^{CC} \tag{34}$$

Fixed and variable O&M costs and $CO_2$ transport and geological storage costs follow the formulations introduced in the standalone CCS model, but are evaluated using the actual capture volume determined endogenously by the joint technology configuration.

When EC, BC, and CC coexist, the $CO_2$ volume entering the capture system depends on both the change in energy requirements associated with EC and the fuel composition determined by biomass co-firing. The actual capture volume is therefore formulated as:

$$\begin{aligned} E_i{}^{CC} &= 90\% \times E_{i,0} \times \left[\gamma_c O_i{}^{CC} + (e_{bio} - e_{coal}) \times brxCC_i\right] \\ &-90\% \times \eta_{ec} \times E_{i,0} \times \left[\gamma_c y_i{}^{EC,CC} + (e_{bio} - e_{coal}) \times brxECCC_i\right] \end{aligned} \tag{35}$$

The nonlinear products are represented using the following auxiliary variables:

$$brxCC_i = \theta_i O_i^{CC} \tag{36}$$

$$y_i^{EC,CC} = O_i^{EC} O_i^{CC} \tag{37}$$

$$brxECCC_i = \theta_i y_i^{EC,CC} \tag{38}$$

The complete linearization constraints and variable bounds for $h_i$, $brxCC_i$, $y_i^{EC,CC}$, $brxECCC_i$ are provided in Supplementary Note 2.2.

The CCS pathway considered in the optimization is limited to $CO_2$ capture, pipeline transport, and geological storage; $CO_2$ utilization pathways and associated revenues, such as those from enhanced oil recovery (EOR), are not included.

**Shared resource and fleet-wide abatement constraints**

Unlike the standalone technology assessments, the unified model explicitly accounts for competition for finite biomass resources and geological $CO_2$ storage capacity across the fleet. For each biomass supply location j, total biomass allocated to all power plants cannot exceed the locally available resource:

$$\sum_{i=1}^{N_i} Z_{ji} \le P_j^{within10km} \quad (39)$$

The biomass supplied to plant *i* must satisfy its endogenous biomass requirement:

$$\sum_{j=1}^{N_j} Z_{ji} = BC_i = P_i \theta_i^{eff} \quad (40)$$

Similarly, the $CO_2$ captured at each plant must be allocated to geological storage sites:

$$\sum_{u=1}^{N_u} CO2_{iu} = E_i^{CC} \quad (41)$$

while the cumulative $CO_2$ allocated from all plants to storage site *u* cannot exceed its available annualized capacity:

$$\sum_{i=1}^{N_i} CO2_{iu} \le Q_u \quad (42)$$

Finally, the optimized technology portfolio must satisfy the prescribed fleet-wide $CO_2$-abatement target $A_p$:

$$\sum_{i=1}^{N_i} \left( ER_i^{EC} O_i^{EC} + E_i^{CC} + E_{i,0} \cdot \theta_i^{eff} \right) \ge A_p \quad (43)$$

The model is solved repeatedly over a series of progressively increasing fleet-wide abatement targets. At each target $p$, the optimization determines the minimum annualized system cost $Z_p$ and the corresponding achieved $CO_2$ abatement $A_p$. The fleet-level marginal abatement cost between two consecutive cost-optimal solutions is calculated as

$$MAC_p = \frac{Z_p - Z_{p-1}}{A_p - A_{p-1}} \quad (44)$$

The resulting MAC curve therefore represents the marginal slope of the optimized system-cost frontier under progressively increasing mitigation requirements, with cross-technology interactions and shared resource constraints incorporated directly into each solution.

（2）**Representation of cross-technology interactions**

Cross-technology interactions are represented through their effects on plant energy requirements, biomass demand, and $CO_2$ capture volumes. The EC–BC interaction is captured through $\theta_i^{eff}$ in Eq. (33): efficiency improvements reduce the thermal-energy requirement and hence the absolute biomass demand associated with a given blending ratio, affecting both plant-level biomass costs and fleet-wide competition for biomass resources. The EC–CC interaction is represented through $y_i^{EC,CC}$ in Eq. (35) and Eq. (37), which adjusts the $CO_2$ available for capture when fuel consumption is reduced by EC. The BC–CC interaction is represented through $brxCC_i$ in Eq. (35) and Eq. (36), which links biomass blending to the biogenic $CO_2$ entering the capture system and consequently to capture, transport, and storage requirements.

When EC, BC, and CC coexist, EC also modifies the biomass-related contribution to the $CO_2$ capture stream. This higher-order interaction is represented by $brxECCC_i$ defined in Eq. (35) and Eq. (38), which couples biomass blending with joint EC and CC deployment and ensures that the biomass-related $CO_2$ contribution is consistently adjusted for the EC-induced reduction in total fuel demand. Together, these interaction terms propagate plant-level changes in fuel use and $CO_2$ flows into technology costs, shared-resource allocation, and fleet-wide mitigation outcomes.

**Scenarios setting**

To enable consistent comparison of the techno-economic performance of different mitigation measures across China's coal-fired power fleet, standardized mitigation scenarios were first defined for the standalone technology assessment. Reducing running hours (RRH), energy conservation (EC), direct biomass co-firing (DB), and gasified biomass co-firing (GB) were each evaluated at representative $CO_2$ reduction rates of 10% and 20%, whereas carbon capture (CC) was evaluated using a baseline capture rate of 90%, consistent with current large-scale post-combustion capture systems. These standardized scenarios provide the basis for the plant-level comparison of individual mitigation technologies and the subsequent pairwise interaction analysis.

For the unified fleet-wide optimization, a baseline scenario was established using a 90% $CO_2$ capture rate, 20% EC effectiveness, full biomass availability, baseline economic parameters, and no policy subsidies. Additional sensitivity analyses and alternative scenarios were designed to evaluate the robustness of the optimized technology portfolios and MAC curves under uncertainties in economic parameters, engineering performance, biomass resource availability, and technology interactions. Benchmark and model-validation scenarios were further constructed to examine the influence of alternative modelling assumptions, while future time-slice scenarios incorporate projected technology learning and coal-plant retirement. Policy scenarios evaluate the responses of optimized mitigation portfolios to carbon pricing and technology-specific subsidy mechanisms. A complete overview of the scenario definitions, objectives, assumptions, and corresponding analyses is provided in Supplementary Table S1.

Unless otherwise specified, the results presented in the main text refer to the baseline scenario, while the remaining scenarios are used to assess model robustness, examine key sources of uncertainty, and explore alternative policy and future transition conditions.

**Data availability**

The plant-level and technology-specific datasets used in this study were compiled from publicly available databases, government statistics, engineering demonstration projects, and published literature. The principal datasets include the characteristics of 2,907 coal-fired generating units across 1,885 power plants, techno-economic parameters for 20 energy-conservation retrofit technologies, spatially explicit agricultural and forestry biomass resources, geological $CO_2$ storage sites and capacities, provincial operating hours and electricity tariffs, coal prices, and technology-specific investment and O&M costs. All data sources and corresponding values used in the analysis are provided in Supplementary Tables S2–S8. The input data required to reproduce the principal analyses are also provided with the code repository described in the Code Availability section.

**Code availability**

The optimization models and data-analysis scripts developed for this study are publicly available through GitHub https://github.com/yun-long-zhang/coal-power-mac-optimization-model/tree/main and are archived via Zenodo at: https://zenodo.org/records/21526396 [40]. The repository contains the computational implementation used for the standalone technology assessments, unified fleet-wide optimization, scenario and sensitivity analyses, and processing of the principal model outputs presented in this study. The computational framework was implemented in Python 3.11, and the fleet-wide joint optimization problem was formulated as a mixed-integer linear programming (MILP) model and solved using Gurobi Optimizer 13.0.

**Acknowledgements**

This work was supported by National Natural Science Foundation of China (grant nos. 72488101, 72293605, 72474023, 52570118, 72073014, 72104025 and 72225010). Y.-L.Z. also acknowledges funding from the Swedish Energy Agency, project numbers P2023-00888, through the CETPartnership (https://cetpartnership.eu/) project RESILIENT, supported by the European Union's Horizon Europe Research And Innovation Programme under grant agreement no. 101069750. The authors also thank Markus Millinger for useful comments. The responsibility for the contents lies with the authors.

**Author contributions**

Y.-M.W., L.-C.L., B.Y., J.-N.K., and Y.-L.Z. conceived the study. Y.-L.Z. contributed to the data collection and processing. Y.-L.Z., L.-C.L., and J.-N.K. designed and performed the models' runs. Y.-L.Z., L.-C.L., J.-N.K., S.P. and B.Y. implemented the data presentation and visualization. Y.-M.W., L.-C.L., J.-N.K., Y.-L.Z., Z.H., X.K. and B.Y. contributed to the interpretation of the results. Y.-M.W., Y.-L.Z., Z.H., and J.-N.K. prepared the first draft. Y.-M.W., Y.-L.Z., L.-C.L., J.-N.K. and B.Y. contributed to the Supplementary Information. Y.-M.W., Z.H., J.-N.K., L.-C.L., X.K. and B.Y. worked on the review and editing. All authors approved and contributed to writing the paper.

**Competing interests**

The authors declare no competing interests.

**Declaration of generative AI and AI-assisted technologies in the writing process**

During the preparation of this work, the authors used ChatGPT to improve readability. After using this tool/service, the authors reviewed and edited the content as needed and take full responsibility for the content of the published article.

**Supplementary Information**

Supplementary Information including:

1. Overall research framework, subject, and scenarios (Figs. S1-S2 and Table S1)
2. Supplementary model formulation and parameter definitions (Table S2 and Eqs. S1-S23)
3. Additional models and data for Biomass Co-firing technology (Figs. S3-S8 and Eqs. S24-S27)
4. Additional models and data for Energy Conservation technology (Figs. S9-S10, Table S3 and Eqs. S28-S32)
5. Additional models and data for CCS technology (Figs. S11-S12)
6. Calculation of EC – BC and BC – CC Interaction Cost Savings (Eqs. S33-S36)
7. Decomposition of engineering and accounting contributions to interaction-induced changes in unit abatement cost (Eqs. S37-S41)
8. Derivation of the optimal abatement response under carbon pricing (Eqs. S42-S46)
9. Other Data used in the model (Tables S4-S7)
10. Supplementary results (Figs. S13-S24)
11. Reference (1-38)

**Lead contact**

Correspondence and requests for materials should be addressed to Yi-Ming Wei, Yun-Long Zhang, Lan-Cui Liu, Biying Yu or Jia-Ning Kang.

**Supplementary Information for "Technology interactions reshape the economics of China's coal power decarbonization"**

Yun-Long Zhang[1,2,3,4,*], Jia-Ning Kang[1,2,3,5,*], Xiaoming Kan[6], Lan-Cui Liu[7,*], Zhimin Huang [8], Song Peng[1,2,3,5], Biying Yu[1,2,3,5,*], Yi-Ming Wei[1,2,3,5,*]

Supplementary Information including Notes below:

[1] Center for Energy and Environmental Policy Research, Beijing Institute of Technology, Beijing 100081, China. [2] Beijing Lab for System Engineering of Carbon Neutrality, Beijing Municipal Education Commission, Beijing 100081, China. [3] Basic Science Center for Energy and Climate Change, Beijing 100081, China. [4] Division of Physical Resource Theory, Department of Environmental and Energy Science, Chalmers University of Technology, 412 96, Göteborg, Sweden. [5] School of Management, Beijing Institute of Technology, Beijing 100081, China. [6] Department of Computer and Systems Sciences, Stockholm University, Stockholm, Sweden. [7] School of National Safety and Emergency Management, Beijing Normal University, Beijing 100875, China. [8] Robert B. Willumstad School of Business, Adelphi University, Garden City, NY 11530, USA.
* Corresponding authors: wei@bit.edu.cn (Y.-M. Wei), zhangyl_9@163.com (Y.-L. Zhang), kangjianing@bit.edu.cn (J.-N. Kang), liulancui@163.com (L.-C. Liu), yubiying_bj@bit.edu.cn (B. Yu)

Supplementary Information

# 1. Overall research framework, subject, and scenarios

## 1.1 Overall research framework

The overall analytical framework of this study is illustrated in Fig. S1. The workflow consists of three major stages: model & data construction, technology evaluation & optimization, and results & robustness discussion.

First, technology-specific optimization models and datasets were developed for the three major retrofit measures considered in this study, namely energy conservation (EC), biomass co-firing (BC), and carbon capture (CC). Each model incorporates the engineering characteristics and economic constraints of the corresponding technology while preserving plant-level heterogeneity, including capacity, operating efficiency, commissioning year, fuel consumption, biomass resource accessibility, and $CO_2$ storage availability.

Second, the technology evaluation and optimization stage consists of three complementary analyses. **The first analysis** evaluates individual mitigation technologies independently by optimizing each power plant separately under standardized $CO_2$ reduction targets. This provides a consistent comparison of the unit abatement costs of different retrofit technologies across the national coal-fired power fleet, also with the curtailment loss assessment of reducing running hours (RRH) option. **The second analysis** explicitly quantifies pairwise interactions among retrofit technologies by evaluating how upstream measures alter the generation costs, unit abatement costs, and attributable emission reductions of downstream technologies. This intermediate step isolates the underlying synergistic and trade-off mechanisms before system-level optimization and provides the basis for interpreting the interaction effects embedded in the unified optimization results. **The third analysis i**ntegrates all mitigation technologies into a unified mixed-integer optimization framework, which simultaneously optimizes technology portfolios across all power plants under progressively increasing fleet-wide emission reduction targets. Unlike the previous analyses, this integrated framework simultaneously accounts for technology interactions, competition for biomass resources, shared $CO_2$ transport and storage infrastructure, and spatial resource allocation, thereby identifying the cost-optimal technology portfolios at the system level.

Finally, the optimized technology portfolios are used to construct fleet-wide marginal abatement cost (MAC) curves and to evaluate the evolution of technology deployment, spatial allocation, fuel transition, economic performance, and carbon intensity under progressively more stringent decarbonization targets. Additional sensitivity analyses, policy scenarios, and future transition scenarios are further conducted to assess the robustness of the proposed interaction-aware optimization framework and to explore its implications for future coal power decarbonization.

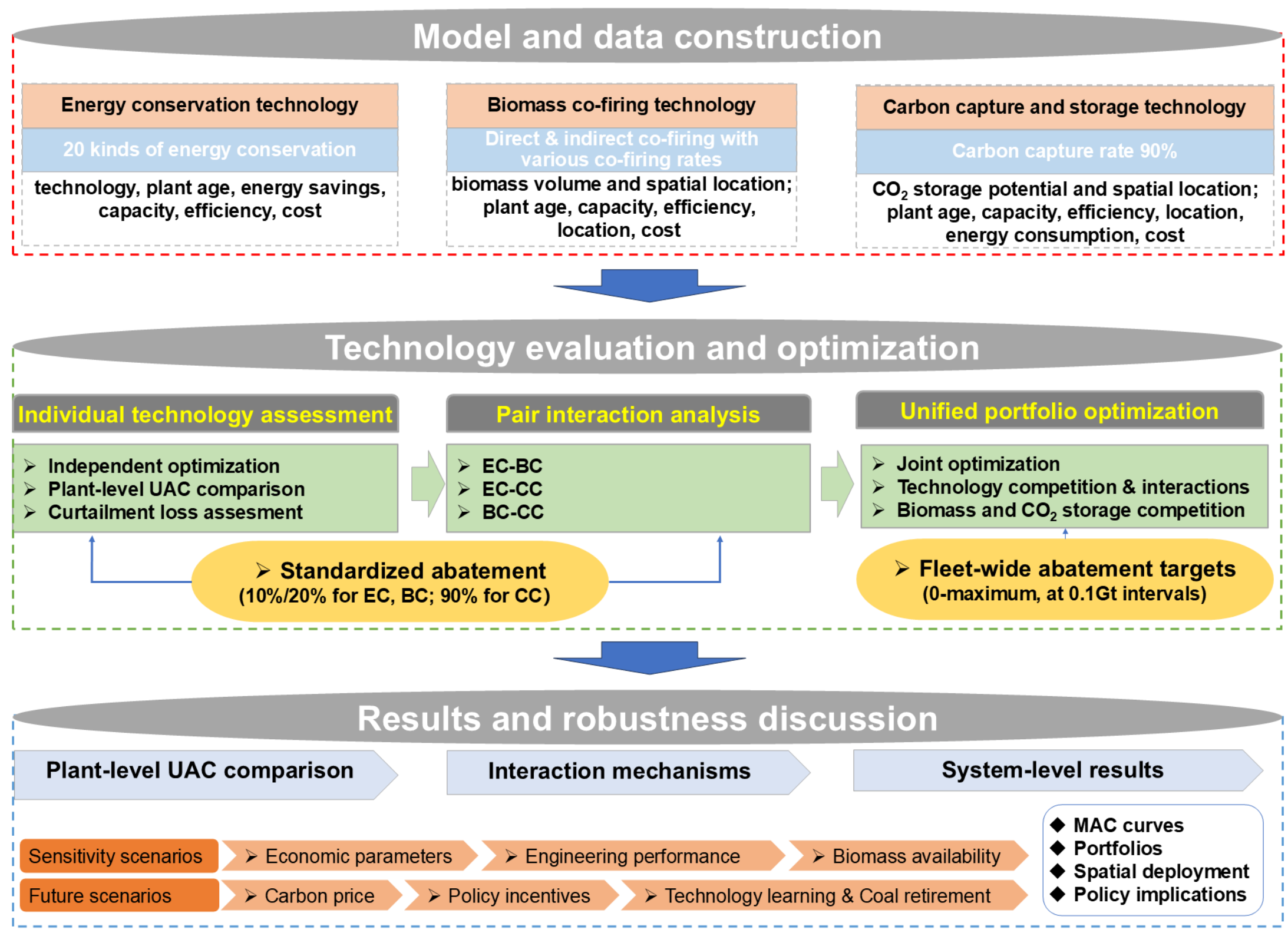

Fig. S1 research framework

## 1.2 Coal-fired power plant dataset

This study encompasses 1,885 coal-fired power plants containing 2,907 generating units each with a minimum installed capacity of 30 MW. Collectively, these facilities amount to a total installed capacity of 1,019 GW, accounting for roughly 93% of China’s operational coal-fired power generation capacity.

As shown in **Fig. S2a**, the fleet is relatively young, with an average operational age of approximately 14 years by 2024. More than 90% of the installed capacity entered operation after 2000, reflecting the relatively recent expansion of China's coal-fired power sector. Large generating units dominate the fleet, with 752 units having capacities exceeding 600 MW and 1,940 units exceeding 300 MW, highlighting the prevalence of modern large-capacity units.

The spatial distribution of the coal-fired power plants is shown in **Fig. S2b**. Most generating capacity is concentrated in central and eastern China, corresponding to regions with high electricity demand and dense industrial activity. In western China, coal-fired power plants are primarily located in Xinjiang Province, where abundant coal resources support large-scale electricity generation and long-distance power transmission.

Annual electricity generation for each plant was estimated using historical average operating hours (Supplementary Table S4), from which annual $CO_2$ emissions were calculated based on plant-specific fuel consumption and emission factors. The resulting fleet emits approximately 3.57 Gt $CO_2$ $yr^{-1}$, corresponding to an average carbon intensity of 805 kg $CO_2$ $MWh^{-1}$, which serves as the baseline emission inventory throughout the optimization analyses.

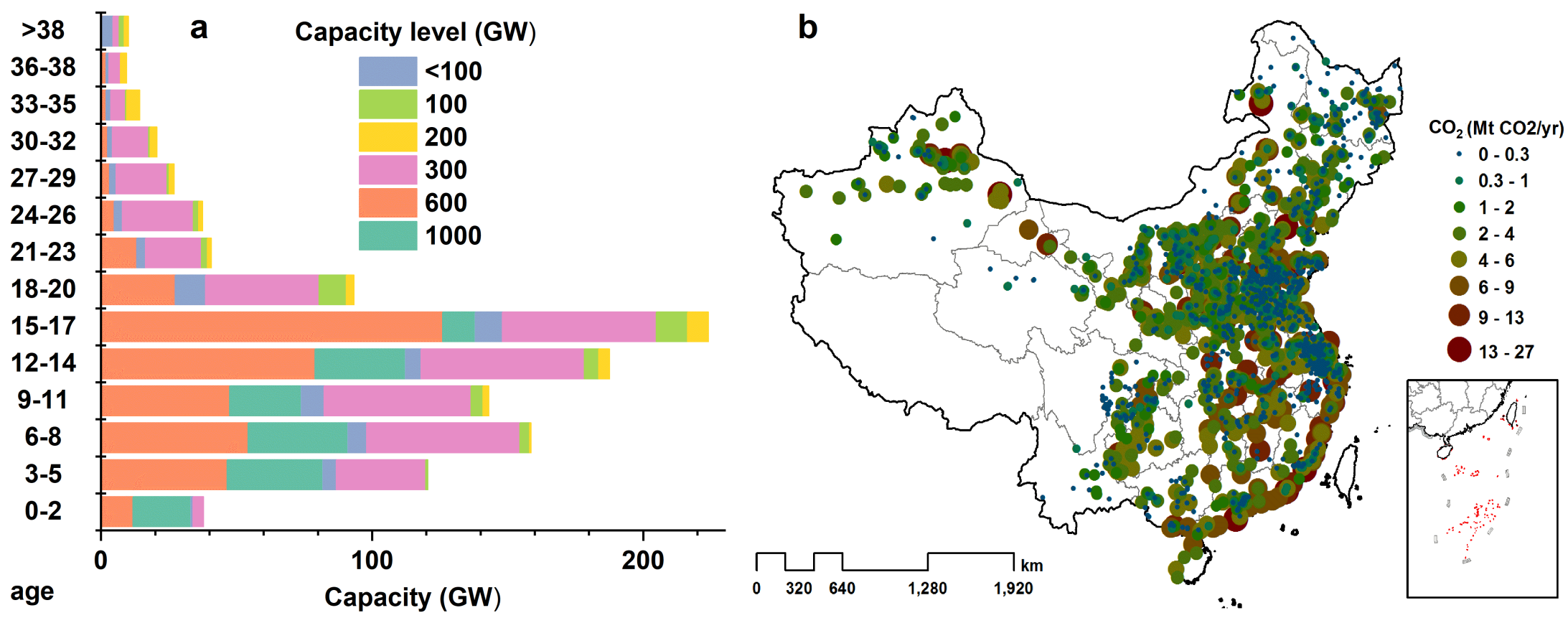

Fig. S2 Characteristics of the coal-fired power plant dataset. **(a)** Distribution of installed generating capacity by commissioning period and unit size for the 2,907 generating units included in this study. **(b)** Spatial distribution of the 1,885 coal-fired power plants. Marker size is proportional to annual $CO_2$ emissions estimated from historical electricity generation, while the inset indicates the geographic location of the study area within China.

### 1.3 Scenario design

To systematically evaluate the robustness and policy implications of the proposed interaction-aware optimization framework, a series of baseline, sensitivity, future-transition, and policy scenarios were designed. These scenarios serve different purposes, ranging from benchmarking individual mitigation technologies to assessing uncertainties in engineering assumptions, resource availability, and future policy conditions. **Unless otherwise stated, all results presented in the main text correspond to the baseline scenario**, while the remaining scenarios are used to evaluate the robustness of the optimization framework or to explore alternative future conditions. **Table S1** summarizes the objectives, key assumptions, and corresponding results presented in the main text and Supplementary Information.

Table S1 Summary of baseline, benchmark, sensitivity, and policy scenarios considered in this study

| Scenario category | Scenario | Purpose | Main assumptions | Results presented |
|---|---|---|---|---|
| Baseline | Baseline joint optimization | Construct the interaction-aware fleet-wide optimization results and baseline MAC curve | 20% EC effectiveness; 90% CC capture rate; 100% biomass availability; baseline economic parameters; no policy subsidies | Figs. 3–4 |
| Technology evaluation | Individual technology assessment | Compare the plant-level cost-effectiveness of individual mitigation options | EC, BC, CC and RRH optimized independently under standardized CO2 reduction targets | Fig. 1 |
| | Pairwise interaction analysis | Quantify interaction effects between retrofit technologies | Combined EC-BC, EC-CC and BC-CC implementation while isolating interaction effects | Fig. 2; Fig. S23 |
| Benchmark comparison | Standalone technology ranking | Compare the proposed interaction-aware framework with the conventional MAC construction approach | Technologies ranked independently according to unit abatement costs without interactions | Fig. 4a |
| | No-EC portfolio | Evaluate the contribution of energy conservation to the optimized portfolios | Unified optimization excluding EC technologies | Fig. 4a |
| | Alternative deployment | Verify whether the recommended technology ordering | Alternative technology-order constraints | Fig. S18 |
| | sequences | emerges endogenously | imposed during optimization | |
| Economic sensitivity | Coal price | Evaluate sensitivity to fuel-price uncertainty | ±25% coal price | Fig. 4b; Fig. S21 |
| | Carbon capture CAPEX | Evaluate uncertainty in CC investment costs | ±30% capital expenditure | Fig. 4b; Fig. S21 |
| | Discount rate | Evaluate sensitivity to financing conditions | 5%, 8% (baseline), and 12% | Fig. 4b; Fig. S21 |
| Engineering sensitivity | EC effectiveness | Evaluate uncertainty in energy-saving performance | 15%, 17.5%, and 20% | Fig. 4c |
| | Carbon capture rate | Evaluate alternative capture-performance scenarios | 90%, 95%, and 99.7% | Fig. 4c |
| | EC overlap assumptions | Evaluate non-additive effects among bundled EC technologies | Alternative overlap coefficients among EC technology groups | Fig. S9, Fig. S10 |
| Resource sensitivity | Biomass availability | Assess biomass resource competition and supply uncertainty | 25%, 50%, 75%, and 100% biomass availability | Fig. 4c |
| Future transition scenarios | Time-slice analysis | Explore future evolution of optimized technology portfolios | Technology learning combined with alternative coal-retirement assumptions | Fig. S17 |
| Policy scenarios | Biomass electricity subsidy | Evaluate the influence of biomass generation incentives | Alternative biomass electricity subsidy levels | Fig. S16 |
| | CO2 storage subsidy | Evaluate the influence of CO2 storage incentives | Alternative geological storage subsidy levels | Fig. S16 |
| | Carbon price response | Identify economically optimal mitigation levels under alternative carbon prices | Uniform carbon price applied to the optimized fleet | Fig. S22 |
| Economic interpretation | Fleet LCOE comparison | Compare optimized coal-retrofit costs with renewable electricity benchmarks | Fleet-wide LCOE under progressively increasing mitigation targets | Fig. S13 |
| | EC versus DB investment comparison | Compare investment characteristics of the two negative-cost retrofit options | Comparison of UAC, CAPEX, and payback period between EC and DB | Fig. S24 |

## 2. Supplementary model formulation and parameter definitions

This section provides the mathematical details of the unified mixed-integer linear programming (MILP) framework presented in the main text. It documents the capital cost scaling and annualization methods, exact linearization procedures used to transform nonlinear terms into an equivalent MILP formulation, together with complete definitions of all decision variables and model parameters. These supplementary descriptions are intended to facilitate model transparency and reproducibility without interrupting the flow of the main manuscript.

### 2.1 Capital cost scaling and annualization

To distinguish the heterogeneous investment costs of different specifications of equipment, the scale effect was considered. A scaling factor (SF) was used to in the equation (S1) [1]:

$$C_{req}^{inv_tech} = C_{ref}^{inv_tech} \times \left(\frac{Q_{req}}{Q_{ref}}\right)^{SF} \qquad (S1)$$

where $C_{req}^{inv_tech}$ is unit investment cost of required equipment, with a known capacity of $Q_{req}$; $C_{ref}^{inv_tech}$ is known unit investment cost of reference equipment, with a known capacity of $Q_{ref}$; $SF$ is scale factor for the technology.

To recover the investment cost of technological retrofit, a certain number of years of operation must be guaranteed. However, the remaining life length of coal-fired power plants varies, so this paper uses the heterogeneous capital recovery factor to calculate the annual capital costs for each power generation unit, as following equation (S2-S4).

$$CRF_i = \frac{r_{tech}}{[1-(1+r_{tech})^{-ST}]} \qquad (S2)$$

$$ST_{tech} = \min\left(RL_{coal_plant}, SL_{tech}\right) \qquad (S3)$$

$$RL_{coal_plant} = 40 - \text{Current age of power plant} \qquad (S4)$$

In the equation (S2), $r_{tech}$ refers to the return of investment (ROI) of the retrofit technology project, and ST refers to the investment cost sharing time. The investment sharing time for each unit is determined by considering the remaining life (RL) of the coal unit and the service life (SL) of the retrofit technology (**Eq. S3**). The remaining life of the unit is calculated based on the average total life of 40 years using 2020 as the base year (**Eq. S4**).

### 2.2 Exact linearization of nonlinear terms

The unified optimization model contains several nonlinear product terms arising from technology interactions and piecewise investment formulations. To enable efficient solution using a mixed-integer linear programming (MILP) solver, all nonlinear expressions were reformulated into mathematically equivalent linear constraints using exact binary-product reformulations and McCormick envelopes. Because all continuous variables involved have finite bounds, these transformations introduce no approximation error and preserve the global optimum of the original formulation.

The model contains five types of nonlinear products that are linearized exactly as follows.

(1) Binary × binary product: $y_i^{EC,CC} = O_i^{EC} O_i^{CC}$

$$y_i^{EC,CC} \le O_i^{EC}, \forall i \tag{S5}$$

$$y_i^{EC,CC} \le O_i^{CC}, \forall i \tag{S6}$$

$$y_i^{EC,CC} \ge O_i^{EC} + O_i^{CC} - 1, \forall i \tag{S7}$$

$$y_i^{EC,CC} \ge 0, \forall i \tag{S8}$$

These four constraints exactly enforce $y_i^{EC,CC} = O_i^{EC} O_i^{CC}$. When both binaries are 1, the first two give $y \le 1$ and the third forces $y \ge 1$, so $y = 1$. In the other three combinations, at least one of the first two constraints gives $y \le 0$ while the third gives $y \ge -1$ or $y \ge 0$, so $y = 0$. This reformulation exactly represents the logical product of two binary variables.

(2) Continuous × binary product (McCormick envelope): $brxEC_i = \theta_i O_i^{EC}$

$$brxEC_i \leq \theta_i^{max} O_i^{EC}, \forall i \tag{S9}$$

$$brxEC_i \leq \theta_i, \forall i \tag{S10}$$

$$brxEC_i \geq \theta_i - \theta_i^{max}\left(1 - O_i^{EC}\right), \forall i \tag{S11}$$

$$brxEC_i \geq 0, \forall i \tag{S12}$$

When $O_i^{EC} = 1$, these constraints simplify to $brxEC_i \leq \theta_i^{max}$, $brxEC_i \leq \theta_i$, $brxEC_i \geq \theta_i$, $brxEC_i \geq 0 \rightarrow brxEC_i = \theta_i$. When $O_i^{EC} = 0$, they give $brxEC_i \leq 0$, $brxEC_i \leq \theta_i$, $brxEC_i \geq \theta_i - \theta_i^{max} \leq 0$, $brxEC_i \geq 0 \rightarrow brxEC_i = 0$. This is the standard McCormick envelope for a bounded continuous variable multiplied by a binary variable.

(3) Same formulation is applied to: $brxCC_i = \theta_i O_i^{CC}$

$$brxCC_i \leq \theta_i^{max} O_i^{CC}, \forall i \tag{S13}$$

$$brxCC_i \leq \theta_i, \forall i \tag{S14}$$

$$brxCC_i \geq \theta_i - \theta_i^{max}\left(1 - O_i^{CC}\right), \forall i \tag{S15}$$

$$brxCC_i \geq 0, \forall i \tag{S16}$$

(4) Same as (2): $brxECCC_i = \theta_i y_i^{EC,CC}$

$$brxECCC_i \leq \theta_i^{max} y_i^{EC,CC}, \forall i \tag{S17}$$

$$brxECCC_i \geq \theta_i - \theta_i^{max}\left(1 - y_i^{EC,CC}\right), \forall i \tag{S18}$$

$$brxECCC_i \geq 0, \forall i \tag{S19}$$

(5) The two-tier formulation separates low-ratio direct co-firing from high-ratio gasified co-firing, allowing different investment cost functions to be represented within a linear optimization framework:

$$\theta_i \leq \tau_i + (\theta_i^{max} - \tau_i) y_i^H, \forall i \tag{S20}$$

$$h_i \leq \theta_i, \forall i \tag{S21}$$

$$h_i \leq \theta_i^{max} y_i^H, \forall i \tag{S22}$$

$$h_i \geq \theta_i - \theta_i^{max}(1 - y_i^H), \forall i \tag{S23}$$

All reformulations presented above are exact rather than approximate. Since every continuous variable has a known finite upper bound and all binary variables are explicitly represented, the resulting MILP is mathematically equivalent to the original nonlinear formulation. Consequently, the global optimum obtained by Gurobi is identical to that of the original optimization problem.

### 2.3 Decision variables and parameter definitions

To facilitate interpretation of the unified optimization framework, Table S2 summarizes all symbols used throughout the mathematical formulation, including decision variables, parameters, sets, and indices. These definitions apply consistently to the objective function, technology-specific constraints, and system-level optimization model presented in the following sections. Technology-specific techno-economic parameters are introduced separately in the corresponding supplementary tables and data sources.

Table S2 Model Parameters

| symbol | definition | Unit | Value | Description |
|---|---|---|---|---|
| Sets and indices | | | | |
| $i \in I$ | | | 1–1885 | Coal-fired power plant index |
| $j \in J$ | | | 1–20 | Energy conservation (EC) technology index |
| $k \in K$ | | | Plant-specific | Biomass supply node index |
| $u \in U$ | | | Plant-specific | $CO_2$ geological storage site index |
| $s \in S$ | | | Scenario-dependent | Scenario index (baseline, sensitivity, subsidy) |
| $(j,i) \in A_B$ | | | | Biomass transportation network |
| $(i,u) \in A_C$ | | | | $CO_2$ transportation network |
| Current parameters of the power plant | | | | |
| $X_i$ | Rated power of power plant i | MW | heterogeneous | Global Energy Monitor data[2] |
| $T_i$ | Annual operating hours | h | heterogeneous | Allocation according to the average power generation hours by province according to the power level difference, see Table S4 |
| $\eta_i$ | Power generation efficiency before retrofitting | % | heterogeneous | Calculated from Heat rate data, see Table S7 |
| $e_{coal}$ | Carbon emission factors of coal combustion | ton/GJ | heterogeneous | Global Energy Monitor data[2] |
| $AF_i$ | Annuity factor for annualization | | Calculated | |
| HR | Heat rate of coal power plants | Btu/kWh | heterogeneous | Global Energy Monitor data[2] |
| $E_{i,0}$ | Power plant i annual Carbon Emissions | t/yr | Plant-specific | Calculation based on power generation and carbon emission factors |
| $P_i$ | Power plant i annual energy demand | GJ/yr | Plant-specific | Calculation based on power generation and heat rate |
| Energy conservation retrofit technology | | | | |
| $F_i^{EC}$ | total investment cost for EC technology portfolio | | | Optimized |
| $C_{i,k}^{inv_EC}$ | unit investment cost for EC technology k | | heterogeneous | Reference data See Table S3 |

| $CRF_{i,k}^{EC}$ | capital recovery factor (CRF) for investment of EC technology k | | heterogeneous | Calculated by formula (S2) |
|---|---|---|---|---|
| $r_{EC}$ | Return on Investment of EC technology | % | 8 | Literature[3] |
| $CRR$ | Power Plant Emission Reduction Rate | % | 10/20/90 | Scenario Setting |
| $ER_{i,k}$ | Emission reduction of unit power generation of EC technology k | t/MWh | | Pilot projects, See Table S3 |
| $ER_{i}^{EC}$ | Annual emission reduction of plant i | t $CO_2$ | Calculated | |
| $SF^{EC}$ | scale factor of EC technology | | 0.82 | literature [1] |
| $SL^{EC}$ | service life of EC technology | | heterogeneous | literature [3] |
| Biomass co-firing generation technology | | | | |
| $F_{i}^{BC}$ | total system cost for BC technology | | | Optimized |
| $C_{i}^{inv_BC}$ | unit investment cost for BC technology | RMB/kW | | Optimized |
| $C_{i}^{inv_DB}$ | unit investment cost for DB technology | RMB/kW | 1153 | Based on Liaoning Diaobingshan Coal Mine Power Plant pilot projects, see Note 3.2 |

| $C_i^{inv_GB}$ | unit investment cost for GB technology | RMB/kW | 5781 | Based on 52 pilot projects of agroforestry biomass co-firing, see Note 3.2 |
|---|---|---|---|---|
| $CRF_i^{BC}$ | capital recovery factor (CRF) for investment of BC technology | | heterogeneous | Calculated by formula (S2) |
| $r_{BC}$ | Return on Investment of BC technology | % | 8 | Literature[4] |
| a | Fixed O&M cost factor | % | 2 | Literature[5] |
| b | Variable O&M costs | $/MWh | 1.9 | Literature[6] |
| $C_j$ | Collection storage processing costs | RMB/t | 294.3 | Literature[7] |
| $C_{ij}$ | Collection transport costs | $/t/km | 0.0118 | Literature[8] |
| $D_{ij}$ | Transportation distance | km | heterogeneous | Get in GIS from latitude and longitude coordinates |
| $\sigma$ | Curvature factor | | 1.5 | Literature[7] |
| $\boldsymbol{\theta_i}$ | Biomass co-firing ratio | % | 0-100 | Literature [9,10] |
| $e_{bio}$ | Biomass carbon emission factor | ton/GJ | 0.112 | IPCC [11] |
| LHV-bio | Low calorific value of biomass | GJ/ton | 17.3 | For unit conversions, from Literature [12] |
| LHV-coal | Coal-fired low calorific value | GJ/ton | 20.93 | For unit conversions, from Literature [13] |
| $\eta_i'$ | Power generation efficiency with EC technology | % | heterogeneous | Calculated by energy savings |
| $\boldsymbol{P_j^{within10km}}$ | Biomass in a 10km grid | GJ | heterogeneous | See supplementary materials Fig. S6 |
| $\gamma$ | US Dollar Exchange Rate | RMB/$ | 6.8985 | for unit conversions[14] |
| $SF^{BC}$ | scale factor of EC technology | | 0.795 | literature [1] |
| $SL^{BC}$ | service life of EC technology | yr | 20 | literature [3] |
| Carbon Capture and Storage Technology | | | | |
| $F_i^{CC}$ | total system cost for CC technology | | | Optimized |
| $C_i^{inv_CC}$ | unit investment cost for CC technology | RMB/kW | heterogeneous | Calculated by formula (S1), Reference data See [15] |
| $CRF_i^{CC}$ | capital recovery factor (CRF) for investment of CC technology | | heterogeneous | Calculated by formula (S2) |
| $r_{CC}$ | Discount rate of CC technology | % | 8 | Baseline, Literature[4] |
| c | Fixed O&M cost factor | % | 4 | Literature[15] |
| $C_i^{MEA}$ | MEA solvent unit cost | RMB/t $CO_2$ | 19.5 | 13000RMB/t MEA [16] ×1.5kg MEA/t $CO_2$ [15] |
| $C_i^{inhibiter}$ | Corrosion inhibitor cost | RMB/t $CO_2$ | 3.9 | 20% of $C_{i,CC}^{MEA}$ [15] |
| $C_i^{other}$ | Other reagent costs | RMB/t $CO_2$ | 1.405 | Includes cost of caustic and activated carbon[15] |

| $C_i^{water}$ | Unit cost of water consumption | RMB/t $CO_2$ | 5.125 | According to the unit water consumption [15] and the average price of water in China, we get |
|---|---|---|---|---|
| $C_i^{waste}$ | Unit cost of waste treatment | RMB/t $CO_2$ | 1.45 | According to the cost of waste treatment and the amount of waste generated, we get [15] |
| $C_i^{steam}$ | Steam cost | RMB/t $CO_2$ | heterogeneous | Based on the heat consumption of 3GJ/t $CO_2$ [17,18] Calculated by converting into an equivalent electricity cost assuming a thermal-to-electric conversion efficiency of 25% [15] |
| $C_i^{electricity}$ | Unit cost of electricity consumption | RMB/t $CO_2$ | heterogeneous | Calculated by multiplying the power consumption of 150kWh/t $CO_2$ [17,18] multiplied by each power plant's own generation cost, which varies due to the different generation costs of each power plant |
| $Q_u$ | Available $CO_2$ storage capacity at point u | t $CO_2$ $yr^{-1}$ | heterogeneous | Open source dataset [19] |
| $c_{iu}^{C}$ | $CO_2$ transportation cost | USD t $CO_2^{-1}$ $km^{-1}$ | 0.1 | Literature [20] |
| $c^{store}$ | Geological $CO_2$ storage cost | USD t $CO_2^{-1}$ | 5 | Literature [20] |
| $\eta_{cap}$ | Carbon capture rate | % | 90 | Default setting |
| $SF^{CC}$ | scale factor of CC technology | | 0.67 | literature [21] |
| $SL^{CC}$ | service life of CC technology | yr | 20 | literature [3] |
| Compressed generation hours | | | | |
| $Tariff_i$ | Feed-in tariffs for power plants | RMB/kWh | heterogeneous | Benchmark feed-in tariff for coal power, from China Development and Reform Commission |
| $OM_i^{coal}$ | Coal power O&M costs | RMB/kWh | heterogeneous | Table S5, Literature[22] |
| $FUEL_i^{coal}$ | Fuel costs for power plant generation | RMB/kWh | heterogeneous | Calculated from coal to-gate price, coal consumption and coal calorific value |
| $C^{coal}$ | Electricity coal price | RMB/ton | heterogeneous | Table S6, Average 2015-2019, 5000 kcal/kg power coal[13] |
| LHV-coal | Standard coal low calorific value | MJ/tca | 29308 | For coal price calculation [23] |

## 3. Additional models and data for Biomass Co-firing technology

### 3.1 High-precision spatial distribution of agricultural and forestry biomass energy sources in China

The effective utilization of biomass energy sources has put forward higher requirements to grasp more detailed spatial distribution of biomass energy sources. Accordingly, this study creates a 1km high-precision spatial distribution dataset of agricultural and forestry biomass energy sources in China based on the existing research and combining multiple sources of spatial data with the ArcGIS GIS platform.

Agricultural and forestry resources are the two main constituent categories of biomass energy sources, which are spatialized separately in our study due to the differences in their collection methods and transportation costs.

#### 3.1.1 Assessment of the distribution of agricultural residue resources in China

Based on the global high-precision crop spatial data products [24] , the latest research results of the Chinese Academy of Agricultural Sciences (CAAS) in conjunction with several international research institutions, this study combines land use data and elevation data to spatialize agricultural biomass energy resources. Specifically, the global high-precision crop spatial data product provides the results of 10km×10km yield spatial distribution data of 42 crops in 2010, among which 27 crops, such as wheat, maize and rice, are available in the Chinese region. According to the existing literature [25,26], the theoretical amount of agricultural biomass energy is calculated by multiplying the crop yield by the residual ratio ($RPR_i$), which is the ratio of residual base such as straw, stems and leaves to crop yield. The usable potential of agricultural biomass energy ($UP_{AR}$) can be obtained by considering the crop collection method, straw utilization method and calorific value. The calculation formula is as follows:

$$UP_{AR} = \sum_{i=1}^{n} P_i \times RPR_i \times C_i \times U_{AR} \times LHV_{AR} \qquad (S24)$$

Where, $P_i$ is the yield of the ith crop (t); $C_i$ is the collectability coefficient of the ith crop; $U_{AR}$ is the availability coefficient of agricultural residues; $LHV_{AR}$ is the low heat value of agricultural residues, 17.3 GJ/ton [25]

From the above equations, the available potential of crop biomass energy in each 10km×10km grid in the Chinese region is calculated, and its total is 12.17 EJ. Considering that the size of slope may affect the collection difficulty of biomass energy, the spatial distribution data of 1km altitude in China released by the Resource and Environmental Science and Data Center of the Chinese Academy of Sciences (data from https:// www.resdc.cn/data.aspx?DATAID=99) to calculate the slope of each grid. The principles of collection ratio are: no collection for slope greater than 25 degrees, 30% collection for slope greater than or equal to 20 degrees and less than 25 degrees, 50% collection for slope greater than or equal to 15 degrees and less than 20 degrees, 80% collection for slope greater than or equal to 10 degrees and less than 15 degrees, and full collection for slope less than 10 degrees. In addition, combined with the 2010 China 1km land use type data released by this data center (data from: https://www.resdc.cn/data.aspx?DATAID=99), the amount of crop biomass energy available in 10km×10km precision was downscaled to the 1km×1km precision agricultural land use type. The final results of 1km high-precision spatial distribution of agricultural biomass energy in China were obtained, as shown in Fig. S3.

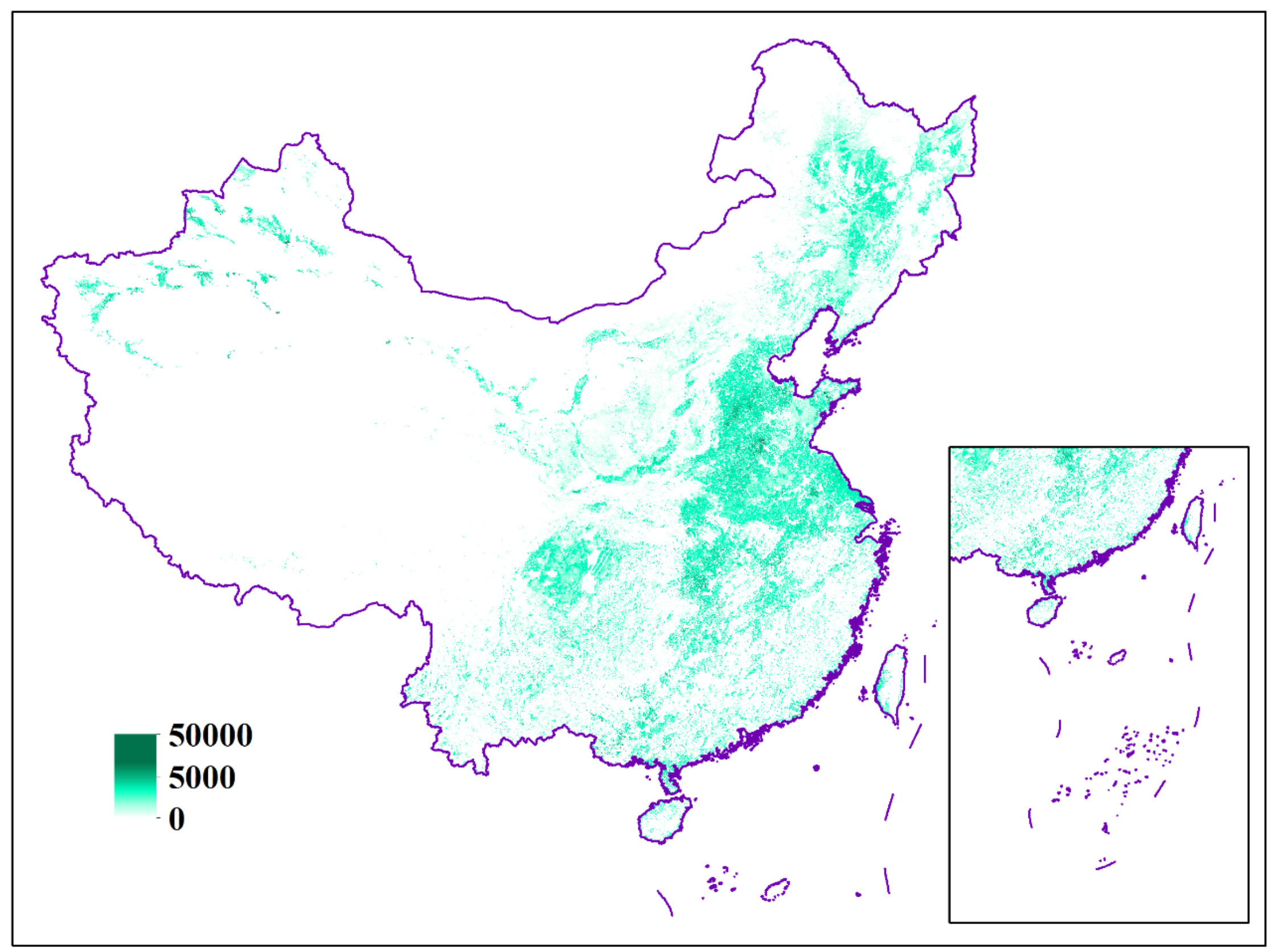

Fig. S3 1km high-precision spatial distribution of agricultural biomass energy sources

### 3.1.2 Assessment of the distribution of forestry residual resources in China

The accounting of China's forestry biomass energy harvestable potential ($CP_{FR}$) refers to existing research results [27] and calculates the amount of resources on four land use types: forested land, shrubland, open forest land and other forest land, respectively, with the calculation formula shown in (S25).

$$CP_{FR} = \sum_{i=1}^{n} A_i \times Y_i \times K_i \times LHV_{FR} \qquad (S25)$$

Where, $A_i$ is the area of species i, ha; $Y_i$ is the firewood production rate of species i (kg/ha); $K_i$ is the collectability factor of species i; $LHV_{FR}$ is the low-level heat generation of forestry residues, 18.1 GJ/ton [25].

The amount of forestry biomass energy was calculated to be 4.89, 0.63, 0.06, and 0.25 EJ for forested land, shrubland, open forest land, and other forested land, respectively, for a total of 5.83 EJ.

Then the spatialization of forestry biomass energy was carried out using the vegetation net primary productivity index (NPP) as weights with reference to existing studies [28]. Based mainly on China 1km land use type data (data from Resource and Environmental Science and Data Center: https://www.resdc.cn/data.aspx?DATAID=99) and China 2015 NPP raster data (data from NASA: https://ladsweb.modaps.eosdis.nasa.gov/), the NPP values on the four types of forestry corresponding land use types were extracted and used as weights to spatialize them according to Equation (S26), respectively:

$$P_{i,j} = \frac{S_i \times npp_{i,j}}{NPP_i} \qquad (S26)$$

where $P_{i,j}$ is the forestry biomass energy potential expressed in grid (i, j); $S_i$ is the biomass energy potential of species i (t); $npp_{i,j}$ is the net primary productivity index expressed in grid (i, j) (g/m$^2$); $NPP_i$ is the sum of net primary productivity of species i (g/m$^2$).

Similarly, considering that forestry collection is also affected by slope, the forestry residues with a slope of 36° or more are not considered for collection with reference to the provisions of the “Beijing Technical Regulations for the Cultivation of Ecological Public Welfare Forests in Mountainous Areas". Finally, the spatial results of the four forestry biomass energy sources were summarized to obtain the 1 km high-precision spatial distribution results of forestry biomass energy sources in China, as shown in Fig. S4.

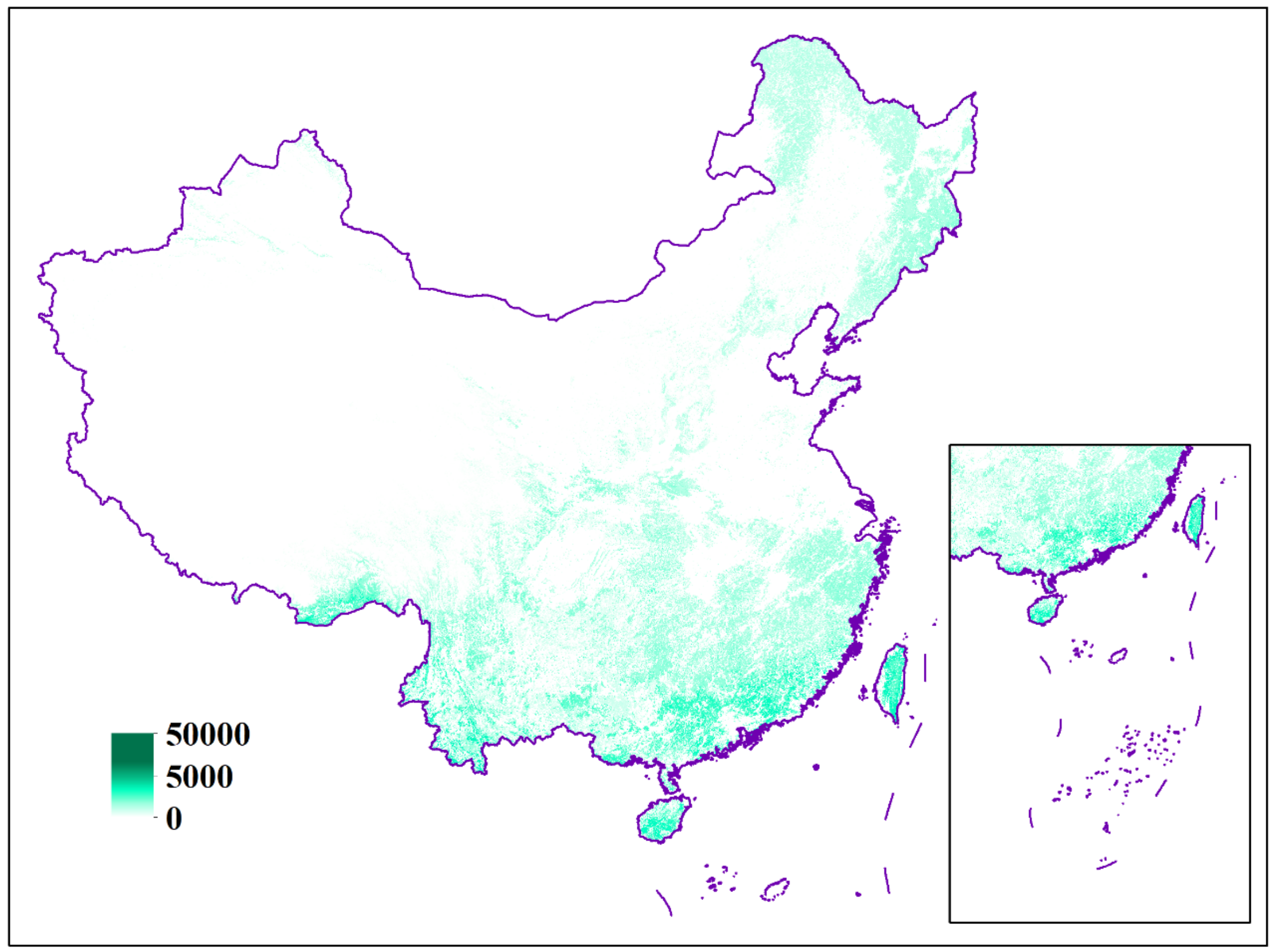

Fig. S4 1km high-precision spatial distribution of forestry biomass energy sources (GJ)

### 3.1.3 The potential of agricultural and forestry residue resources in China

The assessment of the total amount of residual resources in agroforestry category by adopting the above method is in the middle level of the assessment amount of multiple literature, as shown in Fig. S5.

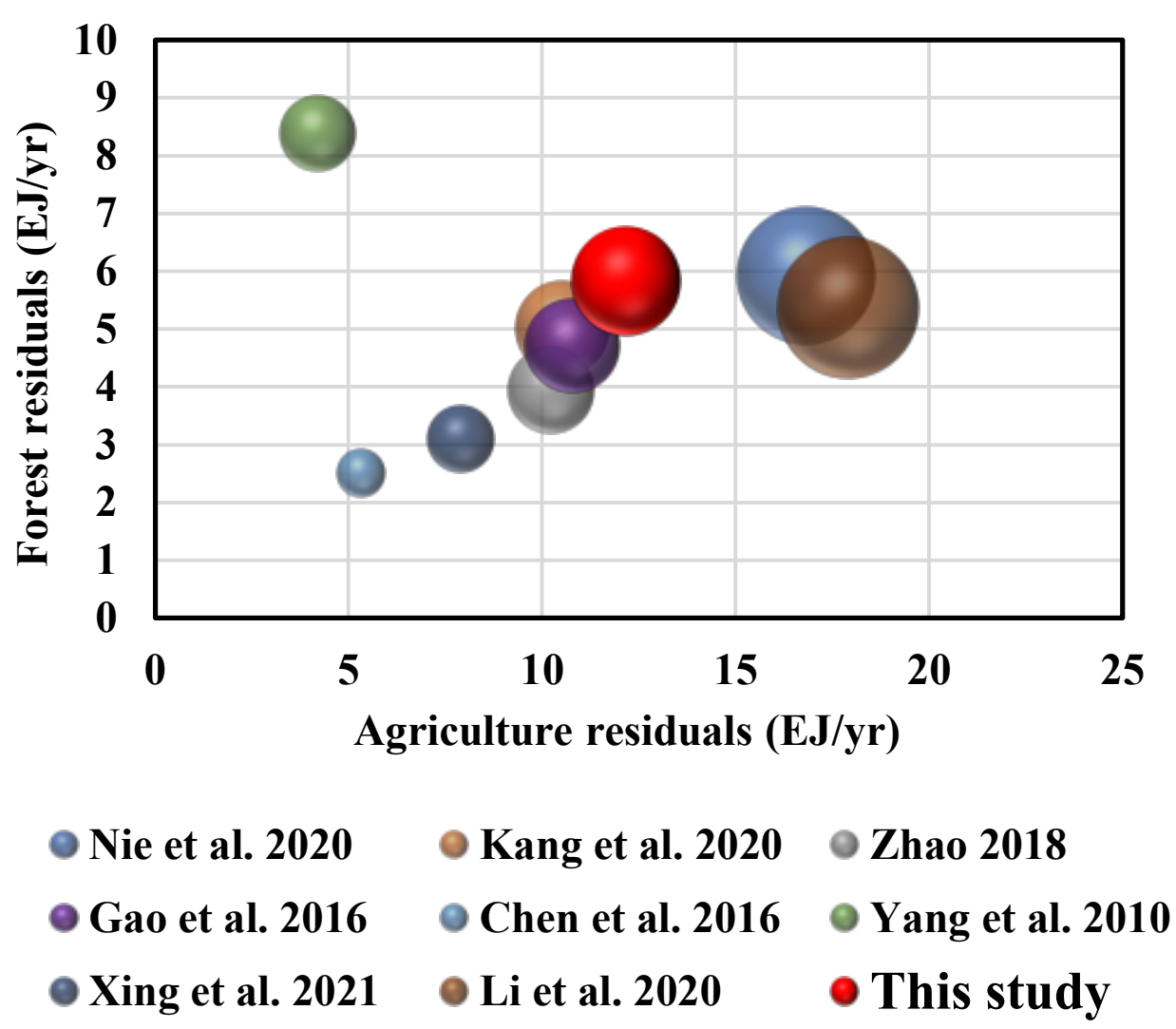

Fig. S5 Assessment of biomass residue potential in China (Bubble size represents the total amount of biomass resources. Data from the literature [25–27,29–33])

The results of 1km high-precision spatial distribution of agricultural and forestry biomass energy sources were summed in GIS to finally obtain the 1km high-precision spatial distribution data set of agricultural and forestry biomass energy sources in China, as shown in Fig. S6.

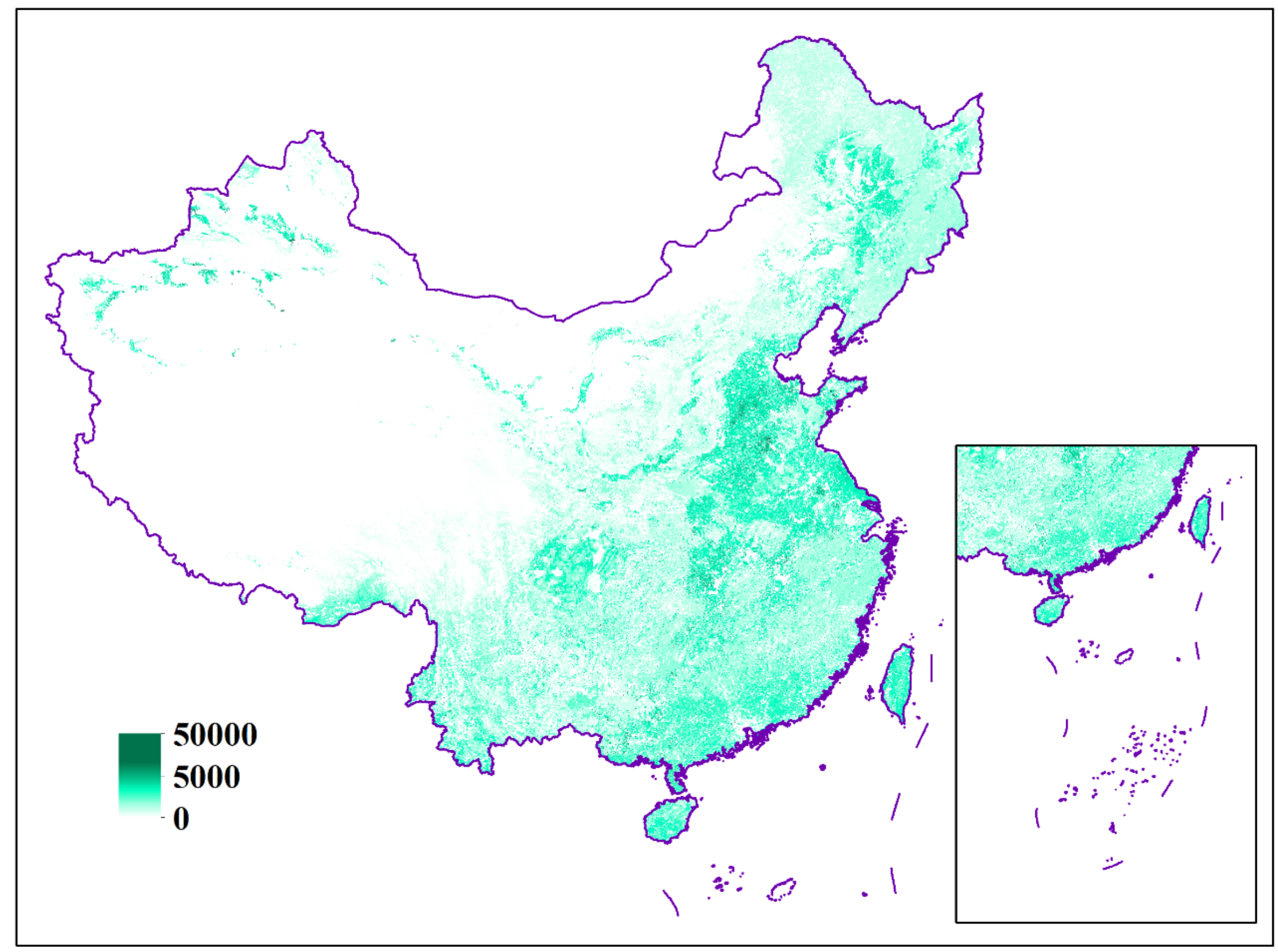

Fig. S6 1km high-precision spatial distribution of agricultural and forestry

biomass energy sources (GJ)

Finally, we scale up the 1km raster data of total agroforestry biomass to 10km again, and take the 10km raster as each biomass collection range [7], the center of which is each biomass storage site, where pre-processing such as drying and processing is carried out before transporting biomass fuel to surrounding power plants, forming a total of 18840 possible and appropriate sites, as shown in Fig. S7.

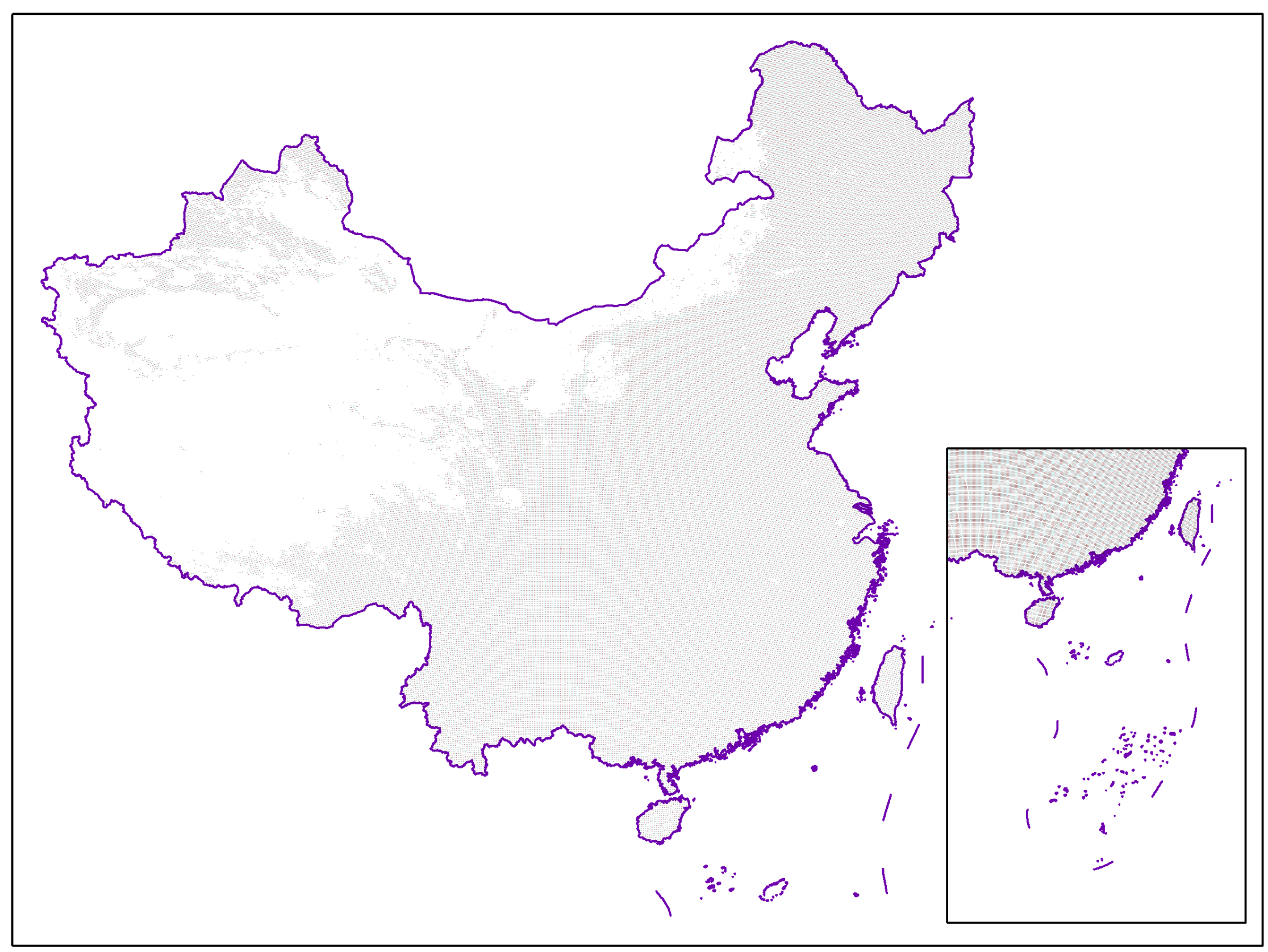

Fig. S7 18840 biomass collection sites

### 3.2 Biomass co-firing projects retrofit investment cost in China

The investment cost function for biomass co-firing was derived from engineering data collected from 58 commercial biomass co-firing demonstration projects (81 units) across China [34]. Unlike generic engineering assumptions, these projects represent actual retrofit investments implemented in operating coal-fired power plants, thereby providing an empirical basis for parameterizing the optimization model.

For each project, the reported total investment budget was converted into a unit investment cost (RMB/kW) by dividing the total investment by the installed generating capacity. For direct co-firing, we took cost data of Liaoning Diaobingshan Coal Mine Power Plant (30MW), with 10% co-firing rate and unit investment cost of 115.267 RMB/kW. As for gasified biomass co-firing, 51 projects data available. A linear regression was then performed against the biomass co-firing ratio to derive the investment-cost function used in this study (Fig. S8).

The resulting relationship shows a strong positive linear correlation:

$$y = 5781.3x \text{ with } R^2 = 0.9056 \quad \text{(S27)}$$

where *x* denotes the biomass co-firing ratio and *y* represents the corresponding unit retrofit investment cost (RMB $kW^{-1}$).

To further evaluate the realism of the derived relationship, we compared the fitted coefficient with published international datasets. The resulting unit investment cost of approximately **843 USD2019/kW** is close to the value reported for U.S. biomass co-firing projects [35] (**1,152 USD2019/kW**, converted using the Chemical Engineering Plant Cost Index), with the lower Chinese cost mainly reflecting lower domestic equipment and construction costs.

Accordingly, the biomass retrofit cost function used throughout the optimization framework is calibrated directly from commercial engineering projects and benchmarked against international evidence, providing an empirically grounded representation of biomass retrofit investments.

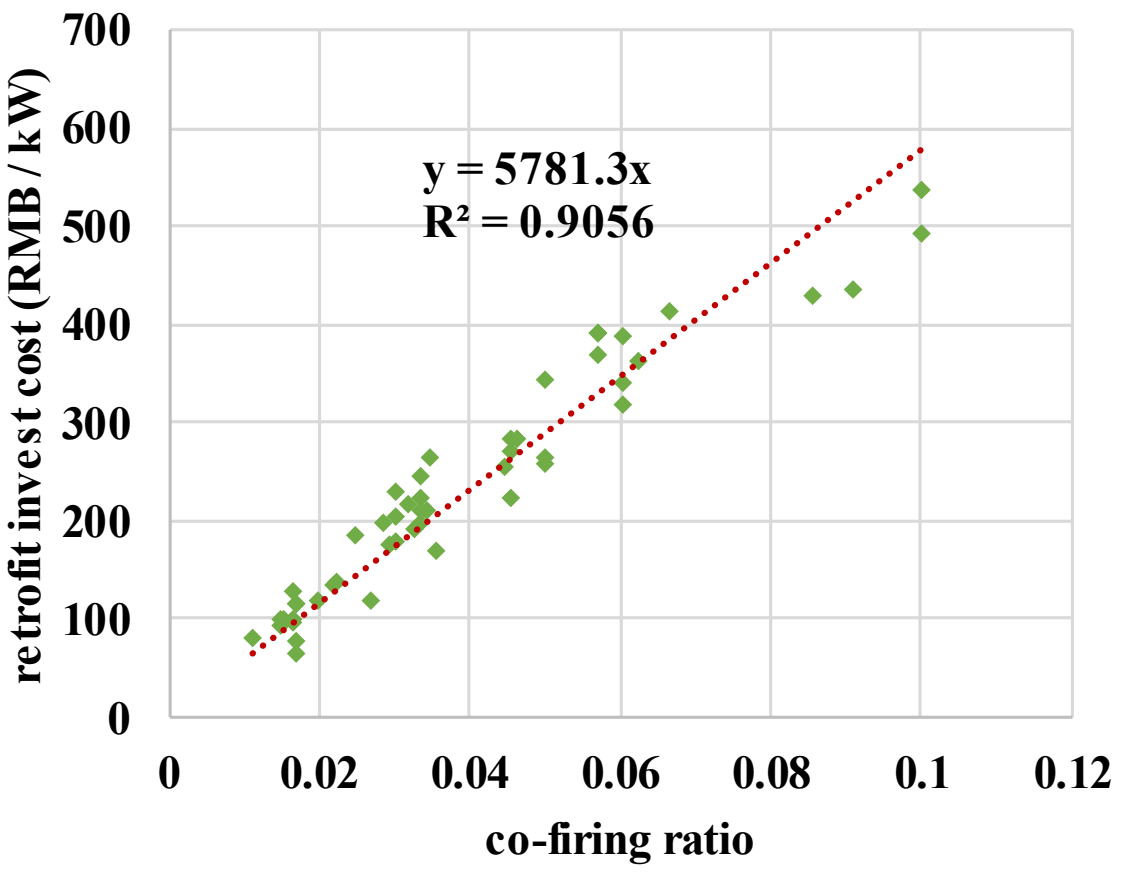

Fig. S8 GB technology retrofit investment cost fitting line

## 4. Additional models and data for Energy Conservation technology

### 4.1 optimization for EC technology considering Overlap

In conventional bottom-up engineering frameworks, marginal greenhouse gas abatement potentials are frequently assumed to be perfectly separable, leading to a simple linear aggregation of impacts [36–38]. However, in practical thermal power engineering, deploying multiple retrofits within a single physical facility inevitably triggers thermodynamic or electrical overlapping effects that diminish the marginal returns of clustered interventions. To model these non-separable interactions endogenously, we classify the 20 discrete EC sub-technologies into six dedicated (Table S3), process-specific categories based on their underlying engineering mechanisms and thermodynamic boundaries.

The first category, boiler combustion and milling, contains sub-technologies {2, 5, 6, 8, 11, 19}, which govern combustion organization, fuel milling efficiency, boiler startup cycles, and furnace-side thermal optimization. The second category, steam turbine body and flow path, covers sub-technologies {1, 13, 18} focused on optimizing internal cylinder sealing and aerodynamic flow passage efficiencies. The third category, cooling and condensing systems, includes sub-technologies {3, 9, 10, 15} governing condenser vacuum maintenance, backpressure limits, and cooling tower heat rejection. The fourth category, air preheater sealing, comprises sub-technologies {4, 7} that address rotary air preheater air leakage and contact seal modifications. The fifth category, flue-gas heat recovery, links sub-technologies {12, 16} to execute advanced flue gas cooling and deep waste heat recovery at the backend of the boiler gas pathway. The final category, electrical auxiliaries and control, contains sub-technologies {14, 17, 20} targeting house power consumption, electrostatic precipitator efficiency, and frequency conversion for liquid-coupled feed pumps. The system-wide interaction matrix is governed by a strict intra-category penalty rule: the overlap penalty applies exclusively when two co-selected sub-technologies belong to the same process category, whereas cross-category pairs incur no penalty and their mitigation potentials remain fully additive.

We establish a Mixed-Integer Linear Programming (MILP) framework to determine the plant-level optimal portfolio of the 20 discrete EC sub-technologies. The optimization space is mapped across two critical dimensions: the Target Mitigation Intensity and the Intra-Category Overlap Coefficient. This allows us to quantify the economic viability of the EC pathway and evaluate how interaction assumptions reshape tech-selection topologies.

**The Target Ratio Dimension:** we define three incremental mitigation targets achieving 15%, 17.5%, and 20% reduction of the fleet's baseline emissions.

**The Overlap Coefficient Dimension**: we introduce an intra-category overlap coefficient, $\delta_{ov}$, evaluated through parametric sweeping at discrete intervals (0%, 25%, and 50%). For any pair of co-selected sub-technologies (k, l) belonging to the same category in plant i, a penalty term $\delta_{ov}$ is subtracted from the gross abatement. Cross-category pairs incur zero penalty. The default value is $\delta_{ov} = 0.0$, meaning sub-technologies are treated as fully additive even within the same category unless the user explicitly sets a positive coefficient.

The overall energy-saving technology combination optimization model still follows formulas (1) to (5) in the main text. However, with overlap penalty, abatement constraint becomes:

$$\sum_{k=1}^{N_k}\left(ER_{i,k}\cdot X_i\cdot T_i\cdot O_{i,k}\right)-\sum_{(k,l)\in O_i}\delta_{ov}\cdot min(a_{ik},a_{il})O_{i,(k,l)}\geq CRR\times E_{i,0}\qquad(S28)$$

where $(k,l)\in O_i$is the set of same-category technology pairs at plant i.

Moreover, in order to solve it, it is necessary to have the overlap binary-product linearization (McCormick):

$$O_{i,(k,l)}\leq O_{i,k}\qquad(S29)$$

$$O_{i,(k,l)}\leq O_{i,l}\qquad(S30)$$

$$O_{i,(k,l)}\geq O_{i,k}+O_{i,l}-1\qquad(S31)$$

With new constraint, we can get optimal decision for every plant. And the total abatement expression with overlap penalty will be:

$$A=\sum_{i=1}^{N_i}\left(\sum_{k=1}^{N_k}\left(ER_{i,k}\cdot X_i\cdot T_i\cdot O_{i,k}\right)-\sum_{(k,l)\in O_i}\delta_{ov}\,min(a_{ik},a_{il})\,O_{i,(k,l)}\right)\qquad(S32)$$

The new decision variables including:

$O_{i,k}\in\{0,1\}$: binary selection of sub-tech k at plant i

$O_{i,(k,l)} \in \{0,1\}$: binary product $O_{i,k} \cdot O_{i,l}$ (same category only)

## 4.2 Database for EC technology projects

This study analyzes 20 energy conservation (EC) retrofit technologies documented in "Action Plan for Upgrading and Retrofitting of Coal Power Energy Conservation and Emission Reduction (2014-2020)" (http://zfxxgk.nea.gov.cn/auto84/201409/t20140919_1840.html) and "National Key Energy Conservation and Low-carbon Technology Promotion Catalogue" (2015-2017) [3]. The techno-economic database for the 20 energy conservation (EC) technologies was compiled entirely from completed retrofit projects implemented in China's coal-fired power sector. All investment costs, technical lifetimes, and energy-saving performances were obtained from documented industrial retrofit projects catalogue or engineering reports (http://www.cqjnw.org/main/index.html).

As summarized in Table S3, each EC technology is linked to one or more representative commercial applications, including projects implemented at Waigaoqiao Power Station, Shanwei Honghaiwan Power Station, Kaifeng Power Station, Zouxian Power Station, and several other large coal-fired generating units.

For each technology, the database includes the reported retrofit investment, expected service life, verified annual energy savings (expressed as tonnes of coal equivalent, tce), and plant operating conditions required for subsequent annualization and unit-cost calculations.

The complete database of the 20 EC technologies is provided in Table S3.

Table S3 Energy conservation technical parameters

| Tech | Energy-saving and low-carbon technologies | Technical life (yr) | Project size (MW) | Project Investment ($10^6$ RMB) | Energy saving ($10^3$ tca/yr) | Emission reduction t/yr | Unit investment cost RMB/MW | Annual power generation hours h/yr | Unit emission reduction (kg/MWh) | Retrofitting cases |
|---|---|---|---|---|---|---|---|---|---|---|
| 1 | Integrated technology of improved performance of steam turbine in power plant | 30 | 600 | 7.8 | 12.2 | 32279 | 13000 | 5500 | 9.78 | Shanwei Honghaiwan power station |
| 2 | Stable combustion and oil-saving technology of oxygen-enriched ignition | 30 | 600 | 4.7 | 11.6 | 30624 | 7833 | 5500 | 9.28 | Chengdu Jintang power station |
| 3 | Efficient centrifugal spray device in cooling tower | 10 | 300 | 0.4 | 1.8 | 4792 | 1380 | 5500 | 2.90 | Guixi power station |
| 4 | Contacting sealing technology of rotary air preheater | 15 | 600 | 3.8 | 7.3 | 19166 | 6333 | 5500 | 5.81 | Guizhou Fa'er power station |
| 5 | High parameter and large capacity technology for brown coal powder boilers | 30 | 600 | 100.0 | 108.0 | 285119 | 166667 | 5500 | 86.40 | Yimin power station |

| | | | | | | | | | |
|---|---|---|---|---|---|---|---|---|---|
| 6 | Steam heating startup technology of utility boiler from neighboring unit | 30 | 1000 | 1.0 | 1.3 | 3461 | 1000 | 5500 | 0.63 | Waigaoqiao power station |
| 7 | Energy-saving seal technology for rotary air preheater | 15 | 600 | 6.5 | 6.2 | 16315 | 10833 | 5500 | 4.94 | Kaifeng power station |
| 8 | Boiler combustion temperature monitor and performance optimization system | 30 | 300 | 2.5 | 2.1 | 5412 | 8200 | 5500 | 3.28 | Mudanjiang No.2 power station |
| 9 | Vacuum maintenance and energy-saving technology of power plants' condensers | 20 | 310 | 4.0 | 3.0 | 7920 | 12903 | 5500 | 4.65 | Huayin Zhuzhou power station |
| 10 | Scaling apparatus of spiral strips for condenser | 5 | 200 | 6.0 | 4.2 | 11088 | 30,000 | 7000 | 7.92 | Handan Cogen power station |
| 11 | intelligent optimization and online coking early warning system of utility boiler | 30 | 1000 | 1.8 | 1.3 | 3472 | 1800 | 5500 | 0.63 | Zouxian power station |
| 12 | Flue gas waste heat recovery and fan operation optimization technology of FGD | 15 | 1000 | 21.9 | 14.9 | 39349 | 21850 | 5500 | 7.15 | Waigaoqiao power station |
| 13 | Transformation of seal system of steam turbine | 9 | 300 | 5.0 | 3.4 | 8976 | 16667 | 5500 | 5.44 | Zhanjiang power station |
| 14 | Control technology of energy-saving and efficiency of electrostatic precipitation | 30 | 300 | 2.7 | 1.4 | 3696 | 9000 | 5500 | 2.24 | Anshun power station |
| 15 | High efficiency combined evaporative condenser | 20 | 660 | 33.9 | 15.9 | 41960 | 51409 | 5500 | 11.56 | Datang Luoyang power station |
| 16 | Integrated optimization system of flue gas and advanced heat recovery technology | 20 | 300 | 9.7 | 4.0 | 10534 | 13 | 5500 | 6.38 | Jinggangshan power station |
| 17 | Energy-saving technology of electric precipitation with quasi stable DC power | 30 | 600 | 14.4 | 0.6 | 1555 | 1 | 5000 | 0.52 | Jinglong power station |
| 18 | Modernized retrofit of flow passage of steam turbine | 30 | 300 | 38.4 | 4.0 | 10560 | 13 | 5500 | 6.40 | Shanghai Shidongkou power station |
| 19 | Coal Mill Energy Saving Retrofit | 10 | 360 | 1.5 | 1.3 | 3326 | 4 | 5500 | 1.68 | Huaneng Yueyang power station |
| 20 | Large thermal power unit liquid-coupled speed control electric feed pump frequency conversion technology | 30 | 330 | 7.8 | 3.2 | 8355 | 10 | 5500 | 4.60 | Ningdong Maliantai power station |

### 4.3 EC sub-technology category abatement contributions

Figure S9 shows the abatement contribution (Mt $CO_2$) of each EC sub-technology category across combinations of target abatement ratios and overlap coefficients. Under the zero-overlap assumption ($\delta_{ov}$ = 0%), the energy-saving effects of EC measures within the same category are fully additive. Boiler and milling technologies provide the largest contribution, accounting for approximately 60–80% of total EC abatement, followed by steam-turbine flow-path improvements (7–10%) and cooling and condensing systems (~4–11%). As the overlap coefficient increases to 50%, the achievable EC abatement declines noticeably—by up to approximately 25% at the highest target ratios—because the combined energy-saving effects of measures within the same category are increasingly reduced by technological overlap.

The boiler and milling category experiences the largest absolute reduction in abatement contribution, whereas auxiliary electricity and control systems and flue-gas heat recovery retain relatively stable contributions, reflecting lower assumed overlap among the measures represented in these categories. At higher target ratios, the optimized EC portfolios also become more diversified across technology categories as the model selects additional measures to compensate for the reduced combined effectiveness of overlapping technologies.

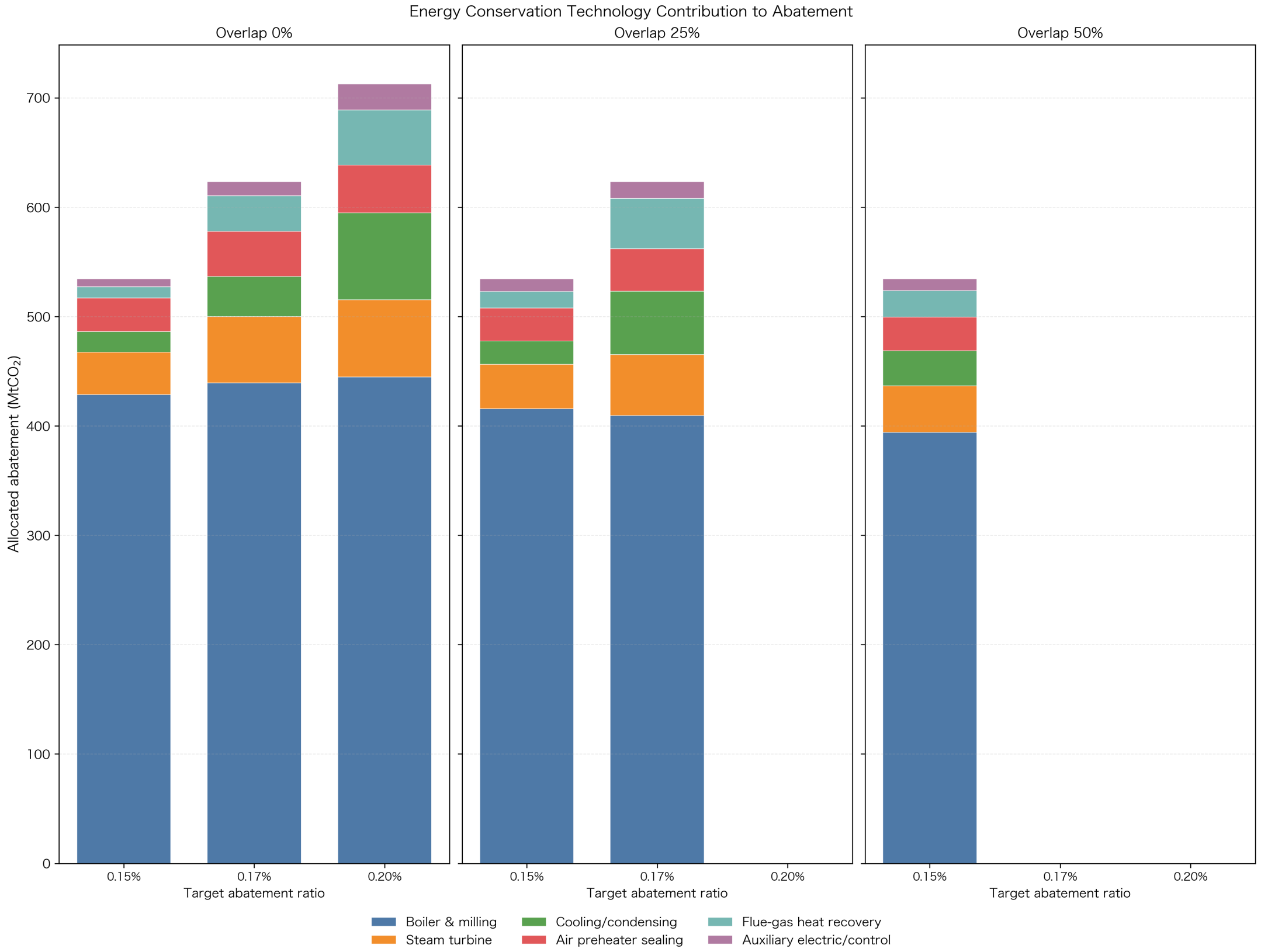

Fig. S9 Category-level composition of EC abatement contributions under varying overlap assumptions. Stacked bar charts show the contribution of each EC technology category to total EC abatement (Mt $CO_2$) across overlap coefficients ranging from 0% to 50%. Within each panel, bars from left to right represent progressively higher target abatement ratios. Increasing overlap reduces the combined abatement achievable from measures within the same category and consequently shifts the optimized EC portfolios toward a more diversified mix of technology categories.

Figure S10 presents the average net unit abatement cost of the optimized EC portfolio for each combination of target abatement ratio and overlap coefficient, after accounting for avoided coal-consumption costs. Combinations marked “No solution” occur when the required plant-level abatement target exceeds the maximum EC potential achievable under the specified overlap assumption. Such infeasibility becomes more frequent at high target ratios and high overlap coefficients. The highlighted cells indicate three representative EC configurations—15% target abatement with 50% overlap, 17.5% target abatement with 25% overlap, and 20% target abatement with no overlap—which are subsequently incorporated into the uncertainty analysis of the unified optimization framework (Fig. 4).

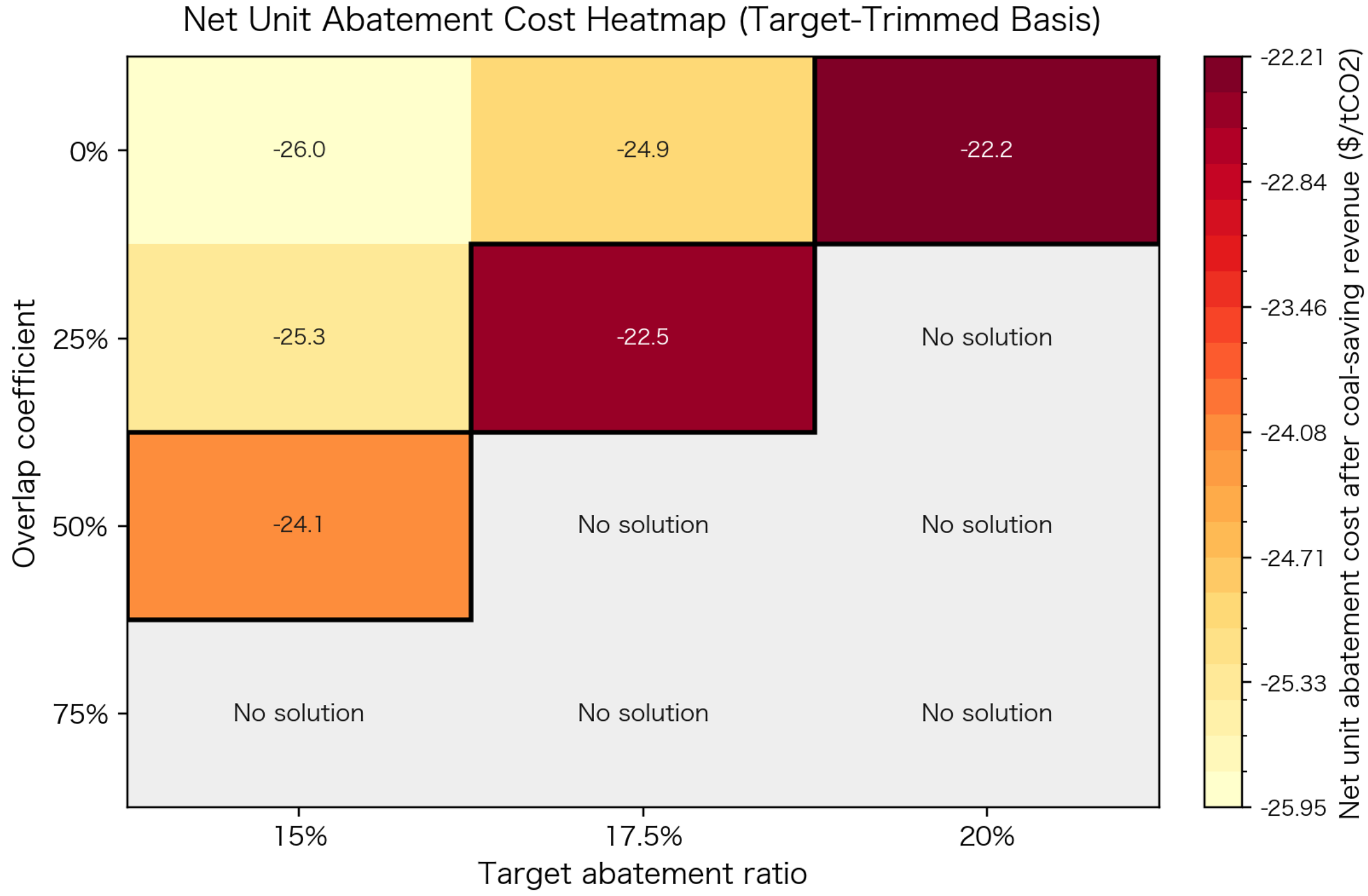

Fig. S10 **Net unit abatement cost of optimized EC portfolios across target abatement ratios and overlap coefficients.** The heatmap shows the average net unit abatement cost (US$ $tCO_2^{-1}$) of the optimized EC portfolio under combinations of plant-level target abatement ratios (x-axis) and within-category overlap coefficients (y-axis), after accounting for avoided coal-consumption costs. Cells labeled "No solution" indicate combinations for which the required plant-level abatement exceeds the maximum achievable EC potential under the specified overlap assumption. Highlighted cells denote the representative EC configurations selected for subsequent uncertainty analysis. Higher overlap coefficients generally reduce achievable EC abatement and increase the cost of meeting a given target, while progressively higher abatement targets require broader deployment of EC measures and eventually lead to infeasibility when the available EC potential is exhausted.

## 5. Additional models and data for CCS technology

### 5.1 Datasets for Carbon storage potential

Data on geological $CO_2$ storage resources were obtained from a publicly available national carbon storage database reported by fan et al. [19]. To ensure consistency with the spatial resolution of the optimization model, the original storage basins were spatially disaggregated into prefecture-level storage sites according to administrative boundaries (Fig. S11).

After excluding sites with limited storage capacity (<1 Mt $CO_2$), a total of **429** candidate storage sites were retained, including **deep saline aquifers, coal seams, depleted gas reservoirs, and depleted oil reservoirs**. The cumulative geological storage capacity exceeds **2,000 Gt** $CO_2$, providing sufficient storage potential for the long-term deployment of carbon capture and storage technologies in China's coal-fired power sector.

Deep saline aquifers dominate the national storage resource, accounting for approximately **99%** of the total storage capacity, whereas coal seams, depleted gas reservoirs, and depleted oil reservoirs contribute approximately **0.56%**, **0.20%**, and **0.24%**, respectively.

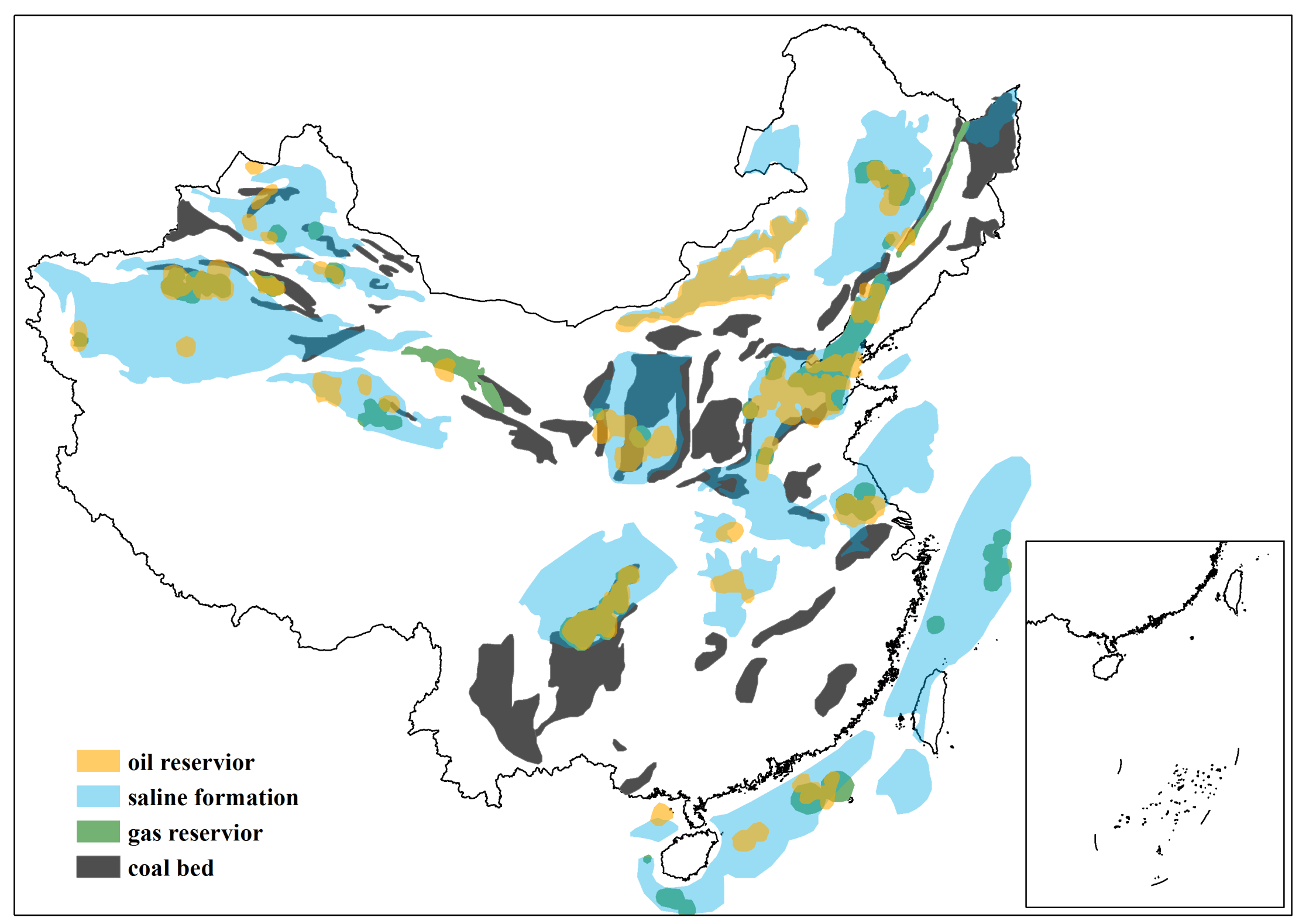

Fig.S11 $CO_2$ storage site

### 5.2 Cost adjustment for alternative $CO_2$ capture-rate scenarios

To evaluate the influence of higher $CO_2$ capture efficiencies on the optimized technology portfolios, two additional capture-rate scenarios (95% and 99.7%) were constructed based on the techno-economic assessment reported in literature[39]. All cost parameters were normalized relative to the baseline 90% capture scenario, which was assigned a scaling factor of 1.0.

The adjusted cost components include:

- **Capital expenditure (CAPEX).** Capital costs were scaled according to the reported Total Capital Requirement. Relative to the 90% baseline, the scaling factors are **1.005** for the 95% capture scenario and **1.014** for the 99.7% capture scenario, reflecting the additional investment required for larger absorbers, regenerators, and $CO_2$ compression systems.
- **Fixed operation and maintenance (O&M) costs.** Fixed O&M costs were adjusted using the reported annual fixed operating costs. Because these costs largely scale with installed equipment capacity, the resulting adjustment factors closely follow the CAPEX trend (**1.004** and **1.013**, respectively).
- **Variable operation and maintenance (O&M) costs.** Variable O&M costs were derived from the reported annual operating expenditures, resulting in scaling factors of **1.134** (95% capture) and **1.199** (99.7% capture). These larger increases primarily reflect higher solvent consumption, increased steam demand for solvent regeneration, additional process-water treatment, and other operating requirements associated with ultra-high capture efficiencies.

The adjusted cost coefficients were incorporated directly into the unified optimization framework, and the resulting marginal abatement cost curves were compared with the baseline 90% capture scenario.

## 5.3 Time-dependent cost reduction assumptions for carbon capture

To represent future technological learning, additional time-slice scenarios were developed by incorporating exogenous cost reductions for carbon capture technologies. The cost trajectories were adopted from the **China CCUS Technology Development Roadmap (2021)** [40], which projects continued reductions in capture and compression costs through technological innovation, industrial learning, and economies of scale.

Relative to the 2020 baseline cost, the investment cost of carbon capture is assumed to decrease by approximately **20% by 2040** and **40% by 2060**, corresponding to **80%** and **60%** of the baseline cost, respectively (Fig. S12). Linear interpolation was applied to estimate intermediate cost levels for the analyzed time slices.

These time-dependent cost assumptions were combined with alternative coal-plant retirement scenarios to generate a series of future fleet configurations, allowing the evolution of optimized technology portfolios and marginal abatement cost curves to be evaluated under progressively changing system conditions.

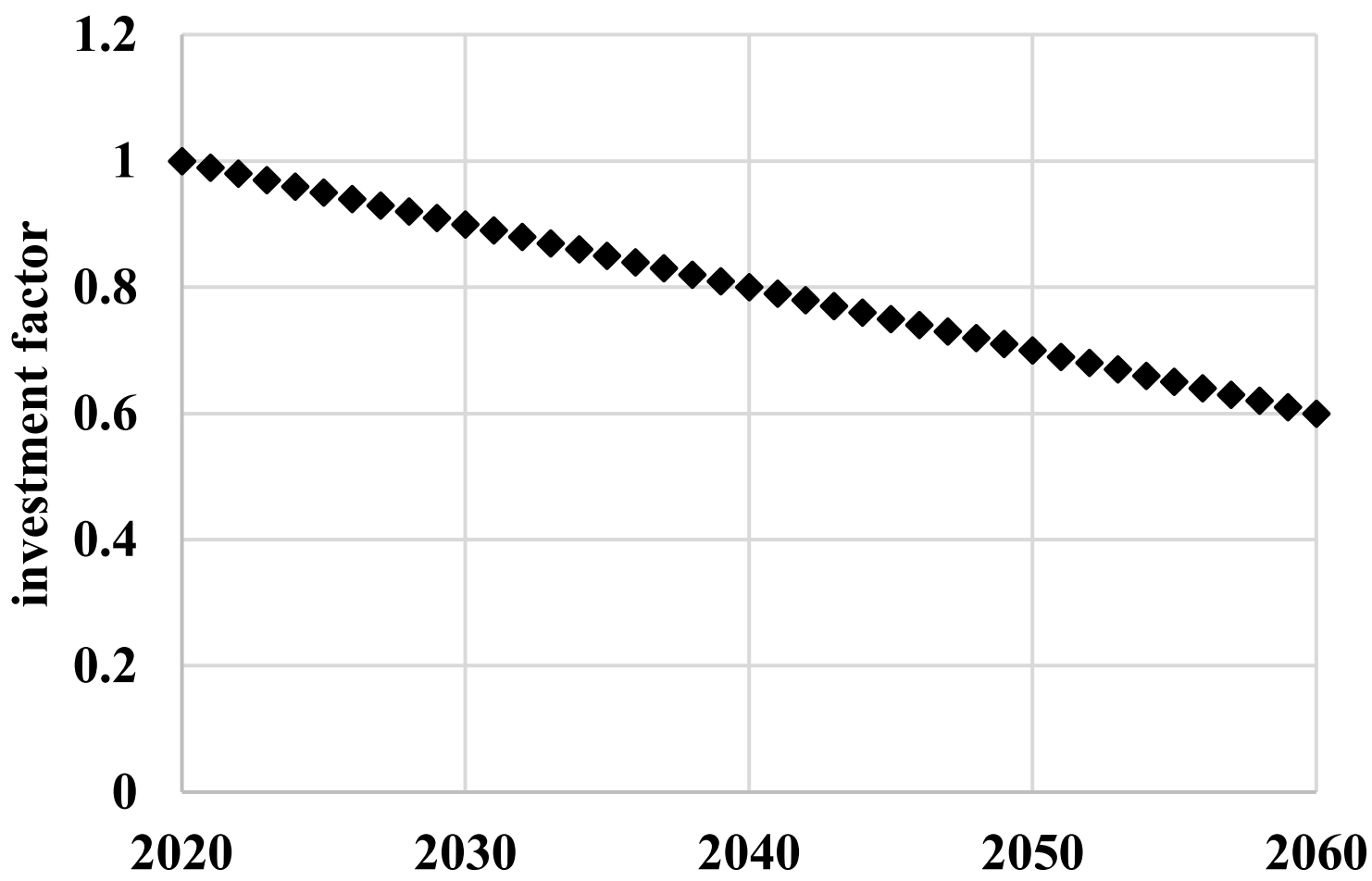

Fig. S12 Exogenous technology learning curves for CCUS capture and compression costs.

## 6. Calculation of EC–BC and BC–CC Interaction Cost Savings

To quantify the economic benefits generated by cross-technology interactions, we compared the optimized interaction-aware pathway with corresponding counterfactual pathways in which the upstream technology was excluded while all other modeling assumptions remained unchanged. Interaction-induced cost savings were calculated by multiplying the reduction in unit cost attributable to the interaction by the corresponding activity level (electricity generation or $CO_2$ captured).

(1) EC-BC interaction

The interaction between energy conservation (EC) and biomass co-firing (BC) lowers the unit cost of bioelectricity generation because energy conservation reduces the overall electricity generation cost before biomass is introduced. The resulting annual cost saving is calculated as

$$CS_{EC-BC} = \Delta LCOE_{BC} \times E_{BC} \quad (S33)$$

where

$CS_{EC-BC}$ is the total annual cost saving (USD/yr);

$\Delta LCOE_{BC}$ is the reduction in the unit unit bioelectricity generation cost (USD/MWh), calculated as

$$\Delta LCOE_{BC} = LCOE_{BC}^{without\ EC} - LCOE_{BC}^{with\ EC} \quad (S34)$$

Where $E_{BC}$ is the annual electricity generation from biomass co-firing (MWh/yr).

The values of $\Delta LCOE_{BC}$ are obtained from the interaction analysis presented in **Fig. 2f** of the main text, while $E_{BC}$ is determined endogenously by the joint optimization model for each abatement target.

(2) BC-CC interaction

The interaction between biomass co-firing (BC) and carbon capture (CC) reduces the unit abatement cost of carbon capture and storage, thereby generating additional economic benefits during deep decarbonization. The corresponding annual cost saving is calculated as

$$CS_{BC-CC} = \Delta UAC_{CC} \times A_{CC} \qquad (S35)$$

where

$CS_{BC-CC}$ is the total annual cost saving (USD/yr);

$A_{CC}$ is the annual amount of $CO_2$ captured (t $CO_2$/yr);

$\Delta UAC_{CC}$ is the reduction in the unit abatement cost (USD/t$CO_2$), calculated as

$$\Delta UAC_{CC} = UAC_{CC}^{without\ BC} - UAC_{CC}^{with\ BC} \qquad (S36)$$

The values of $\Delta UAC_{CC}$ are obtained from the interaction analysis shown in **Fig. 2e** of the main text, whereas $A_{CC}$ is determined by the optimized technology portfolio for each power plant.

Consequently, the calculated interaction cost savings represent the monetary benefits attributable exclusively to cross-technology interactions and do not include additional savings arising from changes in technology deployment or overall system configuration.

## 7. Decomposition of engineering and accounting contributions to interaction-induced changes in unit abatement cost

Because the unit abatement cost (UAC) is defined as the ratio between annualized mitigation cost and attributed $CO_2$ abatement, any upstream retrofit that changes either quantity will alter the reported UAC. Consequently, an observed change in UAC may arise from two fundamentally different mechanisms that cannot be distinguished from the aggregated UAC alone. The first reflects genuine engineering interactions, whereby upstream technologies modify fuel consumption, biomass demand, energy use, $CO_2$ transport requirements, or other physical resource flows that directly affect system costs. The second reflects an accounting effect, whereby upstream emission reductions alter the remaining $CO_2$ attributed to downstream technologies, thereby changing the denominator used to calculate UAC without necessarily representing any real change in engineering resource consumption. Distinguishing these two mechanisms is essential for correctly interpreting technology interactions and avoiding misleading conclusions regarding the economic performance of downstream retrofit measures.

To separate these two mechanisms, we decomposed the interaction-induced change in UAC into an engineering component and an accounting component using a two-factor Shapley decomposition. The Shapley approach provides an exact, path-independent attribution while satisfying the efficiency, symmetry, and null-player axioms

**1) Definition of Unit Abatement Cost**

Throughout this study, the unit abatement cost (UAC) is used as the primary indicator for comparing the cost-effectiveness of alternative retrofit technologies. For power plant *i*, the UAC is defined as

$$UAC_i = \frac{C_i}{A_i} \quad (S37)$$

where $C_i$ denotes the total annualized abatement cost (USD $yr^{-1}$) and $A_i$ denotes the annual $CO_2$ abatement (t $CO_2$ $yr^{-1}$). UAC is consequently expressed in units of USD t $CO_2^{-1}$. Both $C_i$ and $A_i$ are determined endogenously by the joint optimization model and vary with the prescribed fleet-wide emission reduction target.

**2) Shapley decomposition framework**

Suppose the optimization model generates two solutions corresponding to two emission-reduction targets, ($C_0$, $A_0$) and ($C_1$, $A_1$), and the change in unit abatement cost is:

$$\Delta UAC = \frac{C_1}{A_1} - \frac{C_0}{A_0} \quad (S38)$$

Because both mitigation cost and attributed $CO_2$ abatement change simultaneously, the observed ΔUAC combines two analytically distinct effects. To separate them, we employ a two-factor Shapley decomposition.

The Shapley method attributes the total change to each factor by averaging its marginal contribution over all possible orders in which the two variables change. Compared with simple sequential decomposition, the Shapley approach is path-independent and guarantees an exact additive decomposition without residual terms.

**3) Cost Component: Engineering component**

The engineering component quantifies the portion of ΔUAC that arises from genuine changes in mitigation cost while holding the attributed abatement at its Shapley-average level.

$$\Delta UAC_{\mathrm{eng}} = \frac{1}{2}(C_1 - C_0)\left(\frac{1}{A_0} + \frac{1}{A_1}\right) \qquad (S39)$$

This component represents the real engineering consequences of technology interactions, including changes in energy consumption, biomass utilization, fuel substitution, $CO_2$ transport and storage requirements, and other physical resource expenditures caused by upstream retrofit measures.

Positive values indicate that technology interactions increase the actual cost of delivering one additional tonne of $CO_2$ mitigation, whereas negative values indicate genuine cost savings resulting from improved system integration or resource sharing.

**4) Accounting component**

The accounting component quantifies the portion of ΔUAC that originates solely from changes in the amount of $CO_2$ attributed to the downstream technology while holding mitigation costs at their Shapley-average level.

$$\Delta UAC_{\mathrm{acc}} = \frac{1}{2}(C_0 + C_1)\left(\frac{1}{A_1} - \frac{1}{A_0}\right) \qquad (S40)$$

Unlike the engineering component, this term does **not** represent any physical change in resource consumption. Instead, it reflects an accounting effect caused by changes in the $CO_2$ baseline used to calculate UAC.

When an upstream technology removes part of the emissions before a downstream technology is deployed, the remaining $CO_2$ available for subsequent mitigation decreases. Even if the downstream technology incurs nearly identical engineering costs, the reduced denominator mechanically increases its reported unit abatement cost.

This accounting effect is expected to be particularly important for carbon capture, because the amount of $CO_2$ available for capture depends directly on the remaining flue-gas emissions after upstream mitigation measures have been implemented.

**5) Exact additivity**

By construction, the Shapley decomposition satisfies:

$$\Delta UAC = \Delta UAC_{\mathrm{eng}} + \Delta UAC_{\mathrm{acc}} \tag{S41}$$

The decomposition is exact for any finite difference between optimization solutions and therefore introduces neither residual terms nor interaction remainders. This property enables a transparent interpretation of whether observed changes in UAC originate primarily from genuine engineering interactions or from accounting effects associated with shifting emission baselines.

**6) Interpretation of decomposition results**

For visualization purposes, each plant-level observation is classified according to the joint signs of the engineering and accounting components. Positive values indicate that the corresponding mechanism increases the reported UAC, whereas negative values indicate that it decreases the reported UAC.

The resulting quadrant classification provides an intuitive interpretation of whether interaction-induced changes are dominated by engineering mechanisms, accounting effects, or their combined influence, thereby facilitating comparison across technologies and fleet-wide mitigation targets. The corresponding decomposition results for biomass co-firing and carbon capture are presented in Fig. S19.

## 8. Derivation of the optimal abatement response under carbon pricing

To investigate how carbon pricing influences the optimal decarbonization strategy of China's coal-fired power fleet, we derived the optimal fleet-wide abatement response as a function of an exogenous carbon price. Rather than solving a separate optimization problem for every possible carbon price, the analysis builds directly upon the discrete optimization results generated by the unified mixed-integer linear programming (MILP) framework. The resulting carbon-price response therefore represents the economically optimal technology portfolio that minimizes the net system cost under each carbon price while remaining fully consistent with the interaction-aware marginal abatement cost (MAC) curve presented in the main text.

### 8.1 Discrete representation of the MAC frontier

The optimized fleet-wide MAC curve is represented by a set of discrete optimization nodes,

$$S = \{(A_k, TAC_k)\}_{k=1}^{N} \tag{S42}$$

where

$A_k$ is the fleet-wide annual $CO_2$ abatement achieved at optimization node $k$ (t $CO_2$ /yr);

$TAC_k$ is the corresponding total annualized mitigation cost (USD /yr).

In this study, the optimization model was solved for **48** fleet-wide emission reduction targets, producing a high-resolution discrete representation of the system-wide MAC frontier.

### 8.2 Optimal abatement under a carbon price

For a given carbon price $p$ (USD/t $CO_2$), the economically optimal fleet configuration is obtained by minimizing the net system cost,

$$A^*(p) = argmin_k \, [TAC_k - p \cdot A_k] \quad (S43)$$

where the first term represents the annualized mitigation expenditure and the second term represents the economic value of avoided $CO_2$ emissions under the assumed carbon price.

The optimal solution therefore corresponds to the technology portfolio that minimizes total mitigation cost after accounting for carbon-price revenues. Because the underlying MAC frontier consists of discrete optimization solutions rather than a continuous analytical function, the optimal fleet-wide abatement response naturally takes the form of a **piecewise-constant step function**. Within each carbon-price interval, the same technology portfolio remains optimal, while the system shifts to a new portfolio once the carbon price exceeds a critical threshold.

### 8.3 Threshold carbon prices

The switching points of the step function are determined by the **threshold carbon price** between any two optimization nodes. This threshold is obtained by equating the net system costs of two candidate portfolios,

$$TAC_j - p \cdot A_j = TAC_k - p \cdot A_k \quad (S44)$$

which yields:

$$pj_k = \frac{TAC_k - TAC_j}{A_k - A_j} \quad (S45)$$

These threshold prices partition the continuous carbon-price domain into a series of intervals, within each of which the same optimization node remains optimal,

$$A^*(p) = A_{k^*}, \forall p \in \left[p_{low}, p_{high}\right] \quad (S46)$$

Consequently, the optimal abatement response is a **monotonically non-decreasing step function** of the carbon price. Each upward step represents the adoption of a new cost-effective technology portfolio once the carbon price exceeds the corresponding threshold. The locations and magnitudes of these steps therefore reflect the combined influence of technology interactions, resource constraints, and discrete retrofit opportunities represented in the unified optimization framework.

The resulting carbon-price response function is presented in **Fig. S22**, together with representative reference prices from China's national ETS, the EU ETS, and a benchmark carbon-tax scenario, enabling direct comparison between existing policy signals and the economically optimal fleet-wide mitigation response.

## 9. Other Data used in the model

Some of the data used in the model include:

- Table S4 Average Power Generation Hours by Power Rating for Thermal Power Plants in China by Province (2019)
- Table S5 Investment and O&M costs of coal power units in China
- Table S6 Coal Electricity Price Index by Province in China ((RMB/ton coal with a LHV of 5000 kcal/kg)
- Table S7 Standard coal consumption rates of coal-fired power plants by unit capacity, generation technology, and regional power grid (gce $kWh^{-1}$)

Table S4 Average Power Generation Hours by Power Rating for Thermal Power Plants in China by Province (2019)

| Power Rating MW | Average | >1000 | 600-1000 | 300-600 | 200-300 | 100-200 | 6- 100 |
|---|---|---|---|---|---|---|---|
| Grade difference multiplier | 1 | 1.08 | 1.05 | 0.95 | 0.92 | 0.76 | 1.02 |
| National | 4307 | 4652 | 4522 | 4092 | 3962 | 3273 | 4393 |
| Beijing | 3931 | 4245 | 4128 | 3734 | 3617 | 2988 | 4010 |
| Tianjin | 4028 | 4350 | 4229 | 3827 | 3706 | 3061 | 4109 |
| Hebei | 4851 | 5239 | 5094 | 4608 | 4463 | 3687 | 4948 |
| Shanxi | 4426 | 4780 | 4647 | 4205 | 4072 | 3364 | 4515 |
| Inner Mongolia | 5267 | 5688 | 5530 | 5004 | 4846 | 4003 | 5372 |
| Liaoning | 4070 | 4396 | 4274 | 3867 | 3744 | 3093 | 4151 |
| Jilin | 3767 | 4068 | 3955 | 3579 | 3466 | 2863 | 3842 |
| Heilongjiang | 3963 | 4280 | 4161 | 3765 | 3646 | 3012 | 4042 |
| Shanghai | 3253 | 3513 | 3416 | 3090 | 2993 | 2472 | 3318 |
| Jiangsu | 4329 | 4675 | 4545 | 4113 | 3983 | 3290 | 4416 |
| Zhejiang | 4075 | 4401 | 4279 | 3871 | 3749 | 3097 | 4157 |
| Anhui | 4838 | 5225 | 5080 | 4596 | 4451 | 3677 | 4935 |
| Fujian | 4299 | 4643 | 4514 | 4084 | 3955 | 3267 | 4385 |
| Jiangxi | 5153 | 5565 | 5411 | 4895 | 4741 | 3916 | 5256 |
| Shandong | 4443 | 4798 | 4665 | 4221 | 4088 | 3377 | 4532 |
| Henan | 3523 | 3805 | 3699 | 3347 | 3241 | 2677 | 3593 |
| Hubei | 4796 | 5180 | 5036 | 4556 | 4412 | 3645 | 4892 |
| Hunan | 3978 | 4296 | 4177 | 3779 | 3660 | 3023 | 4058 |
| Guangdong | 3841 | 4148 | 4033 | 3649 | 3534 | 2919 | 3918 |
| Guangxi | 4353 | 4701 | 4571 | 4135 | 4005 | 3308 | 4440 |
| Hainan | 4563 | 4928 | 4791 | 4335 | 4198 | 3468 | 4654 |
| Chongqing | 3584 | 3871 | 3763 | 3405 | 3297 | 2724 | 3656 |
| Sichuan | 3084 | 3331 | 3238 | 2930 | 2837 | 2344 | 3146 |
| Guizhou | 4237 | 4576 | 4449 | 4025 | 3898 | 3220 | 4322 |
| Yunnan | 2108 | 2277 | 2213 | 2003 | 1939 | 1602 | 2150 |
| Tibet | 297 | 321 | 312 | 282 | 273 | 226 | 303 |
| Shaanxi | 4383 | 4734 | 4602 | 4164 | 4032 | 3331 | 4471 |
| Gansu | 4236 | 4575 | 4448 | 4024 | 3897 | 3219 | 4321 |
| Qinghai | 2667 | 2880 | 2800 | 2534 | 2454 | 2027 | 2720 |
| Ningxia | 4603 | 4971 | 4833 | 4373 | 4235 | 3498 | 4695 |
| Xinjiang | 5069 | 5475 | 5322 | 4816 | 4663 | 3852 | 5170 |

Table S5 Investment and O&M costs of coal power units in China

| Unit Category | Power Generation Technology | Investment cost (RMB/kW) | Fixed O&M costs % | Variable O&M costs (RMB/kWh) |
|---|---|---|---|---|
| Subcritical units | <300MW | 3277 | 2.1 | 0.049 |
| | 300-600MW | 4492 | 2 | 0.068 |
| | ≥600MW | 3673 | 1.9 | 0.056 |
| Supercritical units | <300MW | 3975 | 2.1 | 0.059 |
| | 300-600MW | 3875 | 2 | 0.059 |
| | ≥600MW | 4050 | 1.9 | 0.059 |
| Ultra-supercritical units | 300-600MW | 3938 | 2 | 0.058 |
| | 600-1000MW | 3373 | 1.9 | 0.058 |
| | ≥1000MW | 3156 | 1.9 | 0.058 |

Table S6 Coal Electricity Price Index by Province in China ((RMB/ton coal with a LHV of 5000 kcal/kg)

| | 2015 | 2016 | 2017 | 2018 | 2019 | Average |
|---|---|---|---|---|---|---|
| Anhui | 458.0308 | 439.4375 | 605.6008 | 613.7358 | 584.365 | 601.2339 |
| Fujian | 423.5992 | 460.6208 | 588.665 | 616.3025 | 542.08 | 582.3492 |
| Gansu | 322.2675 | 318.5875 | 465.4692 | 489.9783 | 468.8683 | 474.7719 |
| Guangdong | 429.1775 | 470.4775 | 616.9775 | 633.0108 | 571.975 | 607.3211 |
| Guangxi | 438.08 | 468.0358 | 703.0942 | 731.035 | 688.055 | 707.3947 |
| Guizhou | 423.3958 | 434.1858 | 499.3625 | 500.8325 | 492.4483 | 497.5478 |
| Hainan | 521.3167 | 557.8342 | 580.5325 | 589.9667 | 505.685 | 558.7281 |
| Hebei | 334.7296 | 375.6629 | 487.8563 | 498.4238 | 473.1071 | 486.4624 |
| Henan | 423.4758 | 455.5617 | 575.9275 | 605.3975 | 527.8692 | 569.7314 |
| Heilongjiang | 371.3692 | 369.815 | 467.9883 | 509.8067 | 522.7025 | 500.1658 |
| Hubei | 369.4458 | 419.68 | 627.0783 | 647.4058 | 597.44 | 623.9747 |
| Hunan | 391.0008 | 451.5325 | 657.9733 | 682.0508 | 633.3367 | 657.7869 |
| Jilin | 398.4908 | 412.5275 | 505.7958 | 551.4025 | 540.5725 | 532.5903 |
| Jiangsu | 432.6708 | 434.4992 | 584.4242 | 596.8675 | 541.2975 | 574.1964 |
| Jiangxi | 422.0208 | 420.8367 | 690.4542 | 701.2958 | 656.7117 | 682.8206 |
| Liaoning | 217.2225 | 217.1633 | 535.5292 | 558.5492 | 540.2908 | 544.7897 |
| Inner Mongolia | 217.6113 | 231.1296 | 264.2433 | 263.7388 | 272.4408 | 266.8076 |
| Ningxia | 419.7275 | 406.6942 | 376.4867 | 379.1367 | 384.7433 | 380.1222 |
| Qinghai | 305.3725 | 317.4017 | 520.0075 | 491.4345 | 528.7408 | 513.3943 |
| Shandong | 498.3683 | 492.0283 | 605.32 | 560.7267 | 559.9833 | 575.3433 |
| Shanxi | 315.78 | 347.6475 | 373.8708 | 379.9067 | 352.1667 | 368.6481 |
| Shaanxi | 416.5092 | 404.6042 | 451.0375 | 443.2883 | 419.0667 | 437.7975 |
| Shanghai | 363.6758 | 351.1658 | 576.9458 | 599.9242 | 528.7008 | 568.5236 |
| Sichuan | 439.0625 | 435.725 | 627.4533 | 624.2575 | 589.2125 | 613.6411 |
| Tianjin | 373.4175 | 379.0308 | 542.8375 | 549.7667 | 491.235 | 527.9464 |
| Xinjiang | 253.24 | 241.6242 | 200.2975 | 248.9408 | 253.8892 | 234.3758 |
| Yunnan | 376.6942 | 360.625 | 494.6033 | 496.8975 | 479.7883 | 490.4297 |
| Zhejiang | 398.8025 | 436.9417 | 567.7633 | 602.16 | 582.7533 | 584.2256 |
| Chongqing | 455.5483 | 475.235 | 597.9458 | 604.2283 | 591.9667 | 598.0469 |

Table S7 Standard coal consumption rates of coal-fired power plants by unit capacity, generation technology, and regional power grid (gce $kWh^{-1}$)

| Generation technology | Unit capacity | North China | Southeast China | Central China | South China | Northwest China | Northeast China |
|---|---|---|---|---|---|---|---|
| Conventional pulverized coal (PC300L) | <300 MW | 317 | 319 | 335 | 342 | 349 | 349 |
| Subcritical (SUBC) | <300 MW | 312 | 314 | 330 | 337 | 344 | 344 |
| | 300 MW | 307 | 310 | 325 | 332 | 339 | 339 |
| | 600 MW | 299 | 301 | 315 | 322 | 328 | 328 |
| Supercritical (SC) | <300 MW | 304 | 306 | 320 | 327 | 334 | 334 |
| | 300 MW | 297 | 299 | 313 | 320 | 326 | 327 |
| | 600 MW | 285 | 287 | 302 | 308 | 315 | 315 |
| Ultra-supercritical (USC) | 300 MW | 284 | 286 | 301 | 307 | 313 | 313 |
| | 600 MW | 278 | 280 | 294 | 301 | 306 | 306 |
| | 1000 MW | 271 | 272 | 286 | 293 | 298 | 298 |

## 10. Supplementary results

### 10.1 Fleet-level LCOE distribution across increasing abatement targets

Figure S13 shows how the fleet-level levelized cost of electricity (LCOE) evolves as progressively more stringent emission reduction targets are imposed. The fleet weighted-average LCOE initially declines from \$45.2 $MWh^{-1}$ under the baseline to \$40.8 $MWh^{-1}$ at approximately 1.0 Gt $CO_2$ $yr^{-1}$ of abatement. This reduction is driven primarily by energy conservation (EC) and low-share biomass co-firing (BC), whose fuel savings more than offset their additional capital and operating expenditures. Consequently, during the early stage of decarbonization (0–1.5 Gt $CO_2$ $yr^{-1}$), retrofit measures not only reduce emissions but also improve the overall economic performance of the coal fleet.

As mitigation targets become more ambitious, the fleet-average LCOE increases steadily because carbon capture (CC) becomes the dominant mitigation technology. At the maximum feasible abatement level (4.7 Gt $CO_2$ $yr^{-1}$), the average LCOE reaches \$81.0 $MWh^{-1}$. Meanwhile, the spread of plant-level LCOEs widens substantially beyond approximately 2.5 Gt $CO_2$ $yr^{-1}$, reflecting increasing heterogeneity in carbon capture retrofit costs. Plants located close to geological storage sites and operating at higher capacity factors maintain relatively low generation costs, whereas plants facing longer transport distances or lower utilization experience substantially higher LCOEs.

Benchmarking against renewable electricity further illustrates the economic implications of deep retrofit strategies. During the early mitigation stage, the best-performing retrofitted coal plants achieve LCOEs comparable to utility-scale solar photovoltaics, whereas under deep decarbonization the fleet-average LCOE gradually exceeds the cost of offshore wind and eventually approaches the reported range of solar generation coupled with battery storage. These results indicate that although efficiency improvements and biomass co-firing can improve the short-term competitiveness of existing coal plants, deep decarbonization through large-scale carbon capture ultimately incurs generation costs comparable to—or higher than—firm low-carbon electricity supplied by renewable energy systems.

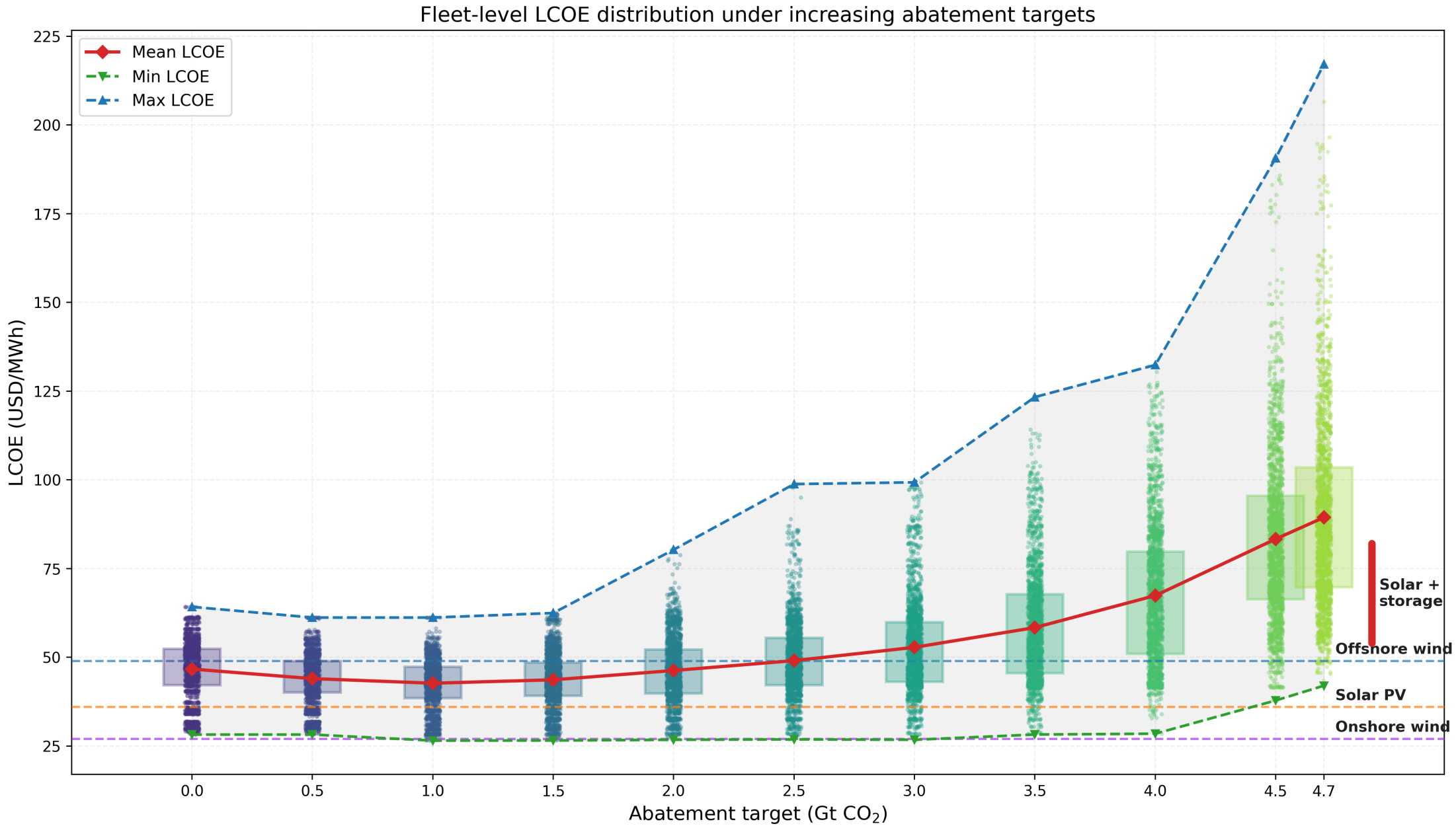

Fig. S13 Fleet-level LCOE evolution under progressively more stringent abatement targets. Distribution of plant-level levelized cost of electricity (LCOE; \$ $MWh^{-1}$) across representative fleet-wide abatement targets. Boxes denote the interquartile range, whiskers indicate the minimum and maximum values, and dots represent individual coal-fired power plants. The red line shows the fleet weighted-average LCOE. Horizontal dashed lines indicate benchmark LCOEs for onshore wind (\$27 $MWh^{-1}$), solar PV (\$36 $MWh^{-1}$), and offshore

wind ($49 $MWh^{-1}$) in China (IRENA 2024 [41]). The red vertical bar denotes the typical LCOE range of solar generation coupled with battery storage ($54–82 $MWh^{-1}$).

### 10.2 Evolution of plant-level technology portfolios across abatement targets

Figure S14 illustrates how plant-level technology portfolios evolve as increasingly stringent fleet-wide $CO_2$-abatement targets are imposed. At the baseline, all 1,885 coal-fired power plants remain in the no-retrofit configuration. At relatively low abatement targets, the optimized portfolios are dominated by energy conservation (EC), reflecting its generally negative or low abatement costs across much of the fleet. By approximately 1 Gt $CO_2$ $yr^{-1}$ of cumulative abatement, EC has been adopted across nearly the entire fleet.

As the mitigation target increases further, biomass co-firing (BC) becomes increasingly prevalent, and combined EC–BC configurations account for a growing share of the fleet between approximately 1 and 2.5 Gt $CO_2$ $yr^{-1}$ of abatement. Under more stringent targets, carbon capture (CC) expands substantially within the optimized portfolios, increasingly forming multi-technology configurations with EC and BC. At the maximum-abatement target, the combined EC–BC–CC configuration represents the largest share of the fleet.

Importantly, these changes in plant-level configurations are not prescribed through an exogenous technology ordering. Instead, each portfolio is determined endogenously by the unified optimization model through system-wide cost minimization under the corresponding abatement target. Figure S14 therefore illustrates how progressively more stringent mitigation requirements shift the cost-optimal fleet composition from predominantly low-cost EC measures toward increasingly integrated portfolios involving BC and CC.

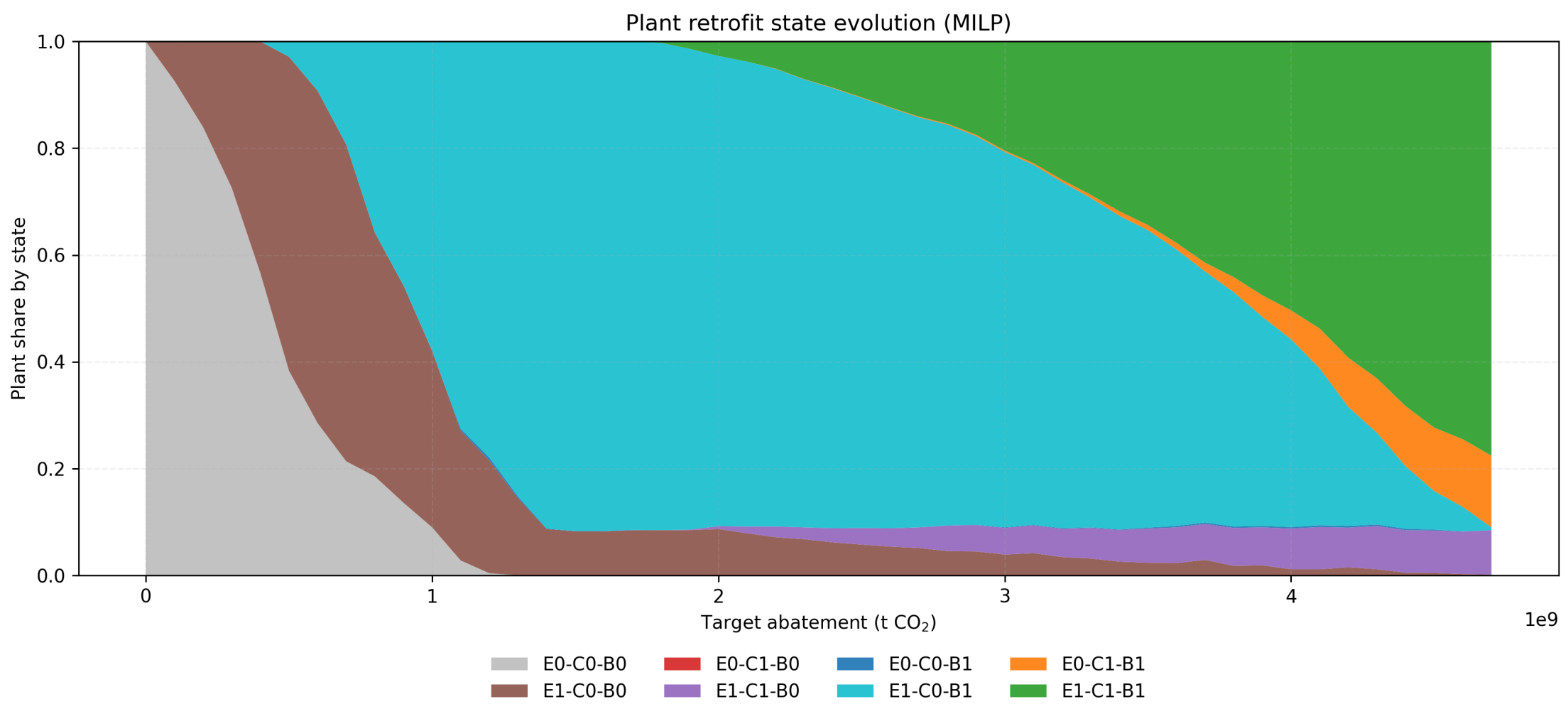

Fig. S14 Evolution of plant-level retrofit configurations across progressively more stringent abatement targets. Stacked area chart showing the proportion of coal-fired power plants assigned to each of the eight possible configurations defined by the adoption states of energy conservation (E), biomass co-firing (B), and carbon capture (C), where 0 and 1 indicate non-adoption and adoption, respectively. The changing shares illustrate the endogenous evolution of cost-optimal plant-level technology portfolios identified by the unified optimization framework as the fleet-wide $CO_2$-abatement target increases.

### 10.3 $CO_2$ storage and biomass transport logistics

Figure S15 quantifies how the logistical requirements associated with biomass supply and $CO_2$ transport evolve under progressively more stringent emission reduction targets. Total annual $CO_2$ storage increases almost linearly with carbon capture deployment, reaching approximately 2.9 Gt $CO_2$ $yr^{-1}$ at the maximum technically achievable abatement level.

As nearby storage sites become progressively saturated, the average weighted $CO_2$ transport distance increases from approximately 20 km under moderate mitigation to more than 120 km under deep decarbonization. Biomass logistics exhibit a similar pattern. Average biomass transport distance rises from approximately 20 km to over 100 km, reflecting the progressive expansion of biomass procurement beyond nearby agricultural and forestry residues as biomass demand increases.

These results demonstrate that deep decarbonization requires not only additional retrofit investments but also substantial expansion of transport infrastructure for both biomass and captured $CO_2$. The explicit optimization of biomass allocation and source–sink matching therefore becomes increasingly important as mitigation ambition increases, highlighting the value of representing transport and storage logistics directly within the optimization framework rather than treating them qualitatively.

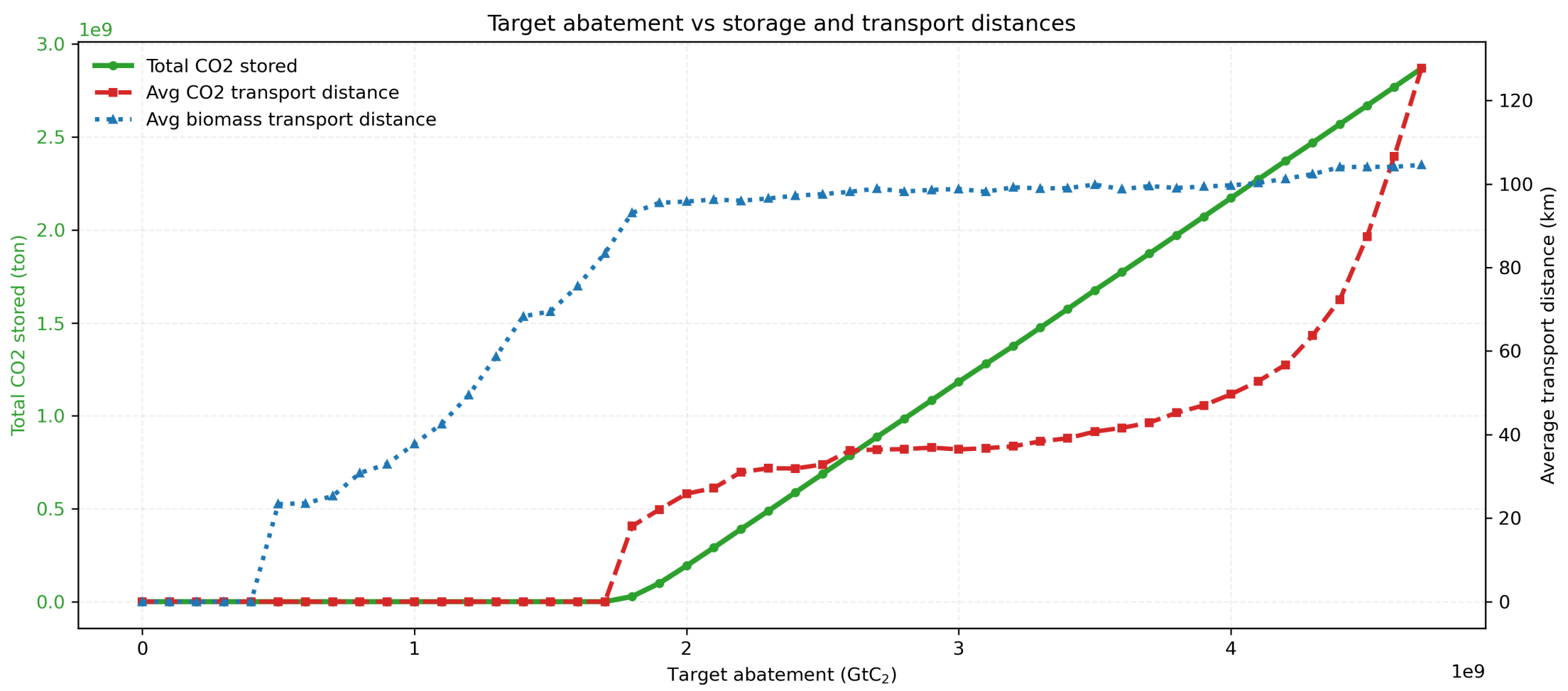

Fig. S15 Evolution of biomass and $CO_2$ transport logistics across fleet-wide abatement targets. Total $CO_2$ storage (left axis) and average weighted transport distances for captured $CO_2$ and biomass (right axis) under progressively more stringent mitigation targets. Increasing transport distances reflect the gradual utilization of more distant storage reservoirs and biomass resources as local low-cost resources become exhausted.

### 10.4 MAC curves under four subsidy scenarios

To quantify the impacts of policy incentives on the economic viability of various retrofitting technologies, this study constructs and compares fleet-wide marginal abatement cost (MAC) curves under four representative subsidy scenarios, comprising two levels of biomass power generation subsidies and two carbon capture and storage (CCS) subsidies, with all other model parameters held constant.

For biomass electricity subsidies, we adopt two tariff benchmarks: 0.15 RMB $kWh^{-1}$ and 0.25 RMB $kWh^{-1}$, where the 0.25 RMB $kWh^{-1}$ level serves as the baseline value referenced from China's historical biomass power support policies [42]. In terms of CCS subsidies, we set subsidy rates of 35 RMB t $CO_2^{-1}$ and 105 RMB t $CO_2^{-1}$. The baseline subsidy of 35 RMB t $CO_2^{-1}$ is determined based on the 20–50 RMB t $CO_2^{-1}$ incentive range for demonstration CCS projects implemented in Shenzhen, Shaanxi Province and Inner Mongolia.

As shown in Fig. S16, both biomass and carbon storage subsidies reduce the fleet-wide MAC relative to the no-subsidy baseline, but their impacts differ markedly across the abatement range. Biomass electricity subsidies primarily lower mitigation costs during the low- and medium-abatement stages (0–2 Gt $CO_2$ $yr^{-1}$), where biomass co-firing constitutes a major component of the optimal technology portfolio. Increasing the subsidy from 0.15 to 0.25 RMB $kWh^{-1}$ further shifts the MAC curve downward, although the benefit gradually diminishes under deep decarbonization because biomass availability rather than generation cost becomes the dominant constraint.

By contrast, carbon storage subsidies exert relatively little influence during the early transition stage but substantially reduce marginal abatement costs once carbon capture becomes widely deployed. Under the 105 RMB t $CO_2^{-1}$ subsidy scenario, the MAC decreases by approximately \$15 t $CO_2^{-1}$ over the 2.4–4.7 Gt $CO_2$ $yr^{-1}$ abatement range relative to the baseline. Meanwhile, the biomass subsidy scenarios converge toward the baseline beyond approximately 2 Gt $CO_2$ $yr^{-1}$, confirming that biomass co-firing alone cannot replace large-scale carbon capture under deep decarbonization targets.

Overall, these results suggest that different subsidy instruments support different stages of the transition. Biomass electricity subsidies are more effective for accelerating near-term deployment of low-cost mitigation options, whereas carbon storage subsidies become increasingly important for enabling deep decarbonization once carbon capture dominates the optimized technology portfolios.

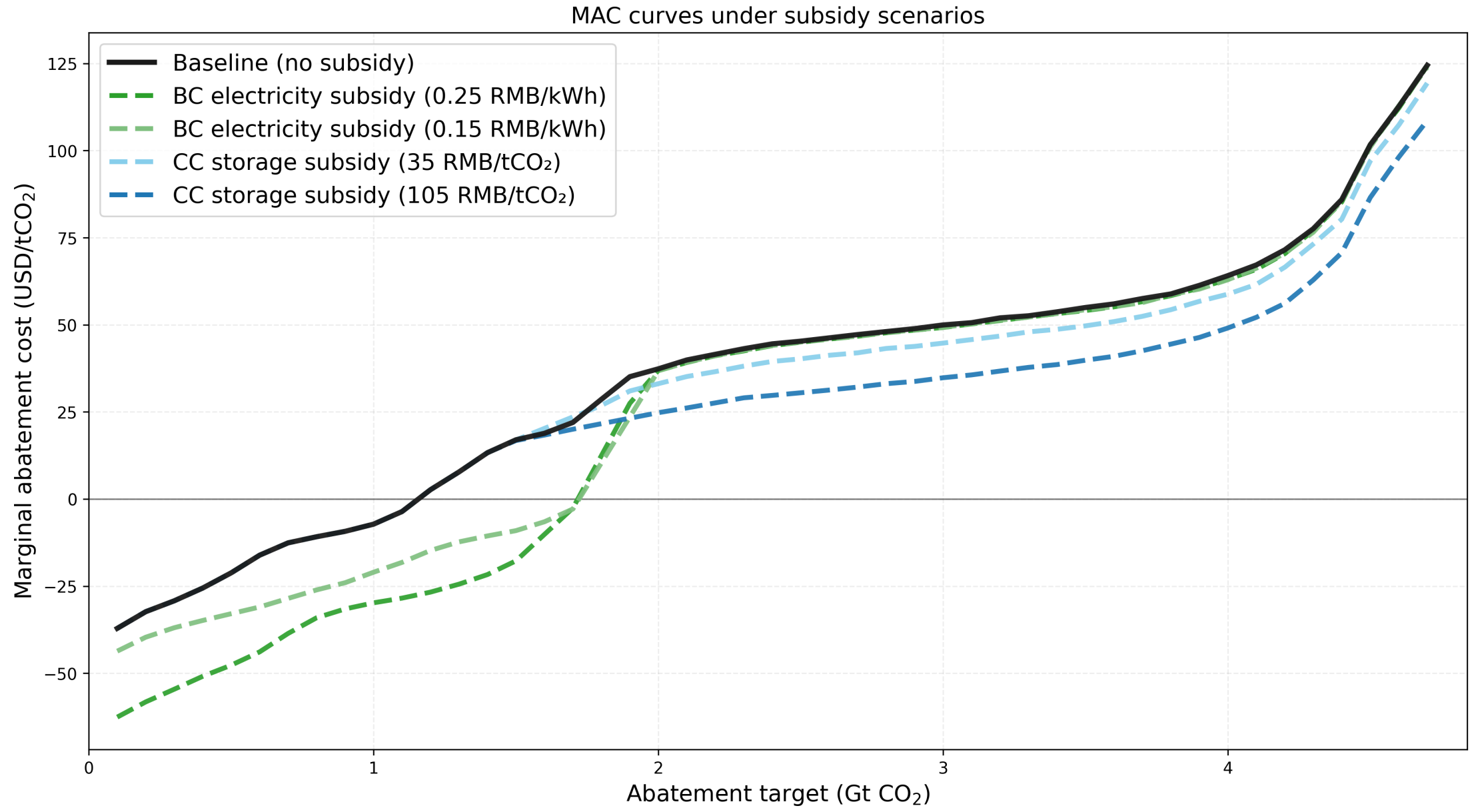

Fig. S16 Effects of alternative subsidy schemes on fleet-wide marginal abatement cost curves. Comparison of optimized MAC curves under four representative subsidy scenarios and the no-subsidy baseline. Green dashed lines denote biomass electricity subsidies of 0.15 and 0.25 RMB $kWh^{-1}$. Blue dotted lines denote carbon storage subsidies of 35 and 105 RMB t $CO_2^{-1}$. All curves were generated using the unified fleet-level optimization framework under identical technology and resource assumptions.

## 10.5 MAC curves under alternative target year and retirement assumptions

To explore how future fleet evolution may influence the optimized technology portfolios, we generated additional marginal abatement cost (MAC) curves for four representative time slices (2030, 2035, 2040, and 2045) under two alternative plant lifetime assumptions (30 and 40 years). These scenarios combine projected reductions in carbon capture investment costs with progressive retirement of aging coal-fired power plants.

As shown in Fig. S17, the maximum technically achievable abatement declines progressively as older generating units retire. Under the 40-year lifetime assumption, maximum fleet-wide abatement decreases only moderately, from 4.70 Gt $CO_2$ $yr^{-1}$ in the baseline to 4.07 Gt $CO_2$ $yr^{-1}$ by 2045. By contrast, adopting a 30-year lifetime substantially accelerates fleet contraction, reducing maximum abatement to only 1.54 Gt $CO_2$ $yr^{-1}$ in 2045, representing a 67% reduction relative to the current fleet.

Interestingly, although the maximum mitigation potential decreases considerably under earlier retirement, the terminal marginal abatement costs remain broadly comparable to the baseline. This is because the remaining fleet increasingly consists of newer and larger generating units, which generally exhibit more favorable retrofit economics for carbon capture. Nevertheless, the shrinking fleet substantially reduces the availability of profitable energy conservation and biomass retrofits, causing the negative-cost portion of the MAC curve to contract progressively and nearly disappear under the 30-year retirement scenario by 2045.

These results indicate that plant retirement policies fundamentally reshape both the attainable mitigation potential and the economic structure of coal fleet decarbonization. Earlier retirement reduces opportunities for low-cost retrofits and therefore increases reliance on alternative low-carbon electricity sources to compensate for the loss of retrofit potential.

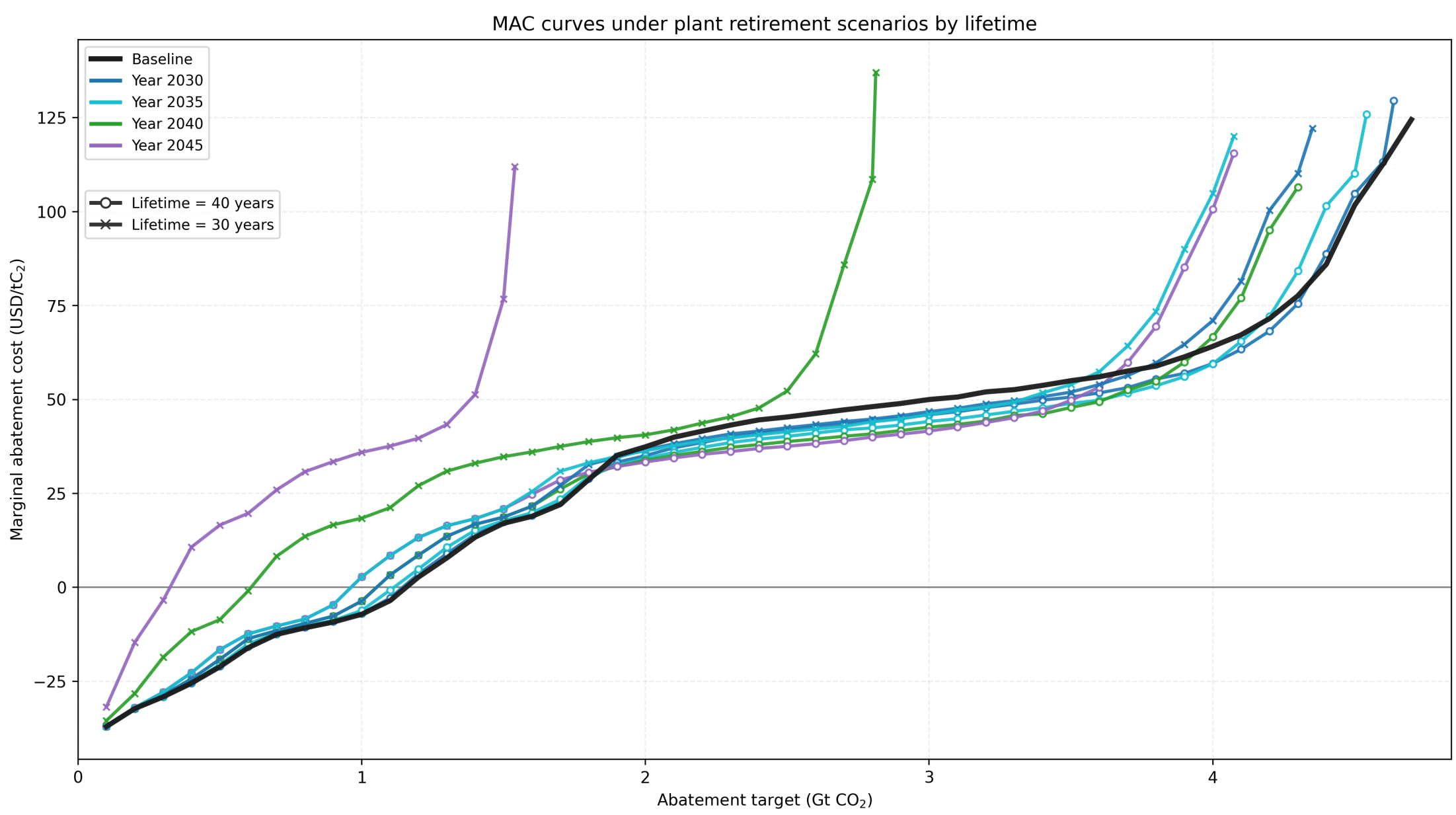

Fig. S17 Fleet-wide marginal abatement cost curves under alternative target years and plant retirement assumptions. MAC curves generated for four representative time slices (2030–2045) under 30-year and 40-year plant lifetime assumptions. The baseline represents the current fleet without mandatory retirement. Earlier retirement substantially reduces the maximum achievable emission reduction while altering the economic composition of the optimized technology portfolios.

### 10.6 Cost comparison under alternative technology-ordering constraints

To verify whether the recommended technology sequence is economically justified, we compared the unconstrained joint optimization with several alternative technology-ordering constraints while keeping all emission targets, parameters, and resource assumptions identical.

As shown in Fig. S18a, the unconstrained optimization establishes the global minimum-cost frontier across the entire mitigation range. Imposing the energy conservation → biomass co-firing → carbon capture sequence produces a cost frontier that is almost indistinguishable from the unconstrained optimum, recovering 99.86–99.997% of the global optimum across all abatement targets. The corresponding cost difference remains negligible throughout the mitigation range.

By contrast, alternative deployment sequences consistently increase system costs. Configurations that prioritize carbon capture before biomass co-firing become increasingly inefficient under high mitigation targets, whereas pathways that omit early energy conservation incur the largest economic penalties and substantially reduce the attainable emission reduction under comparable investment levels. As shown in Fig. S18b, these alternative sequences increase total system costs by $20–60 billion under deep decarbonization targets relative to the unconstrained optimum.

These comparisons provide an empirical falsification test of technology sequencing. Rather than being imposed a priori, the recommended energy conservation → biomass co-firing → carbon capture sequence emerges naturally from the joint optimization as the economically preferred transition pattern.

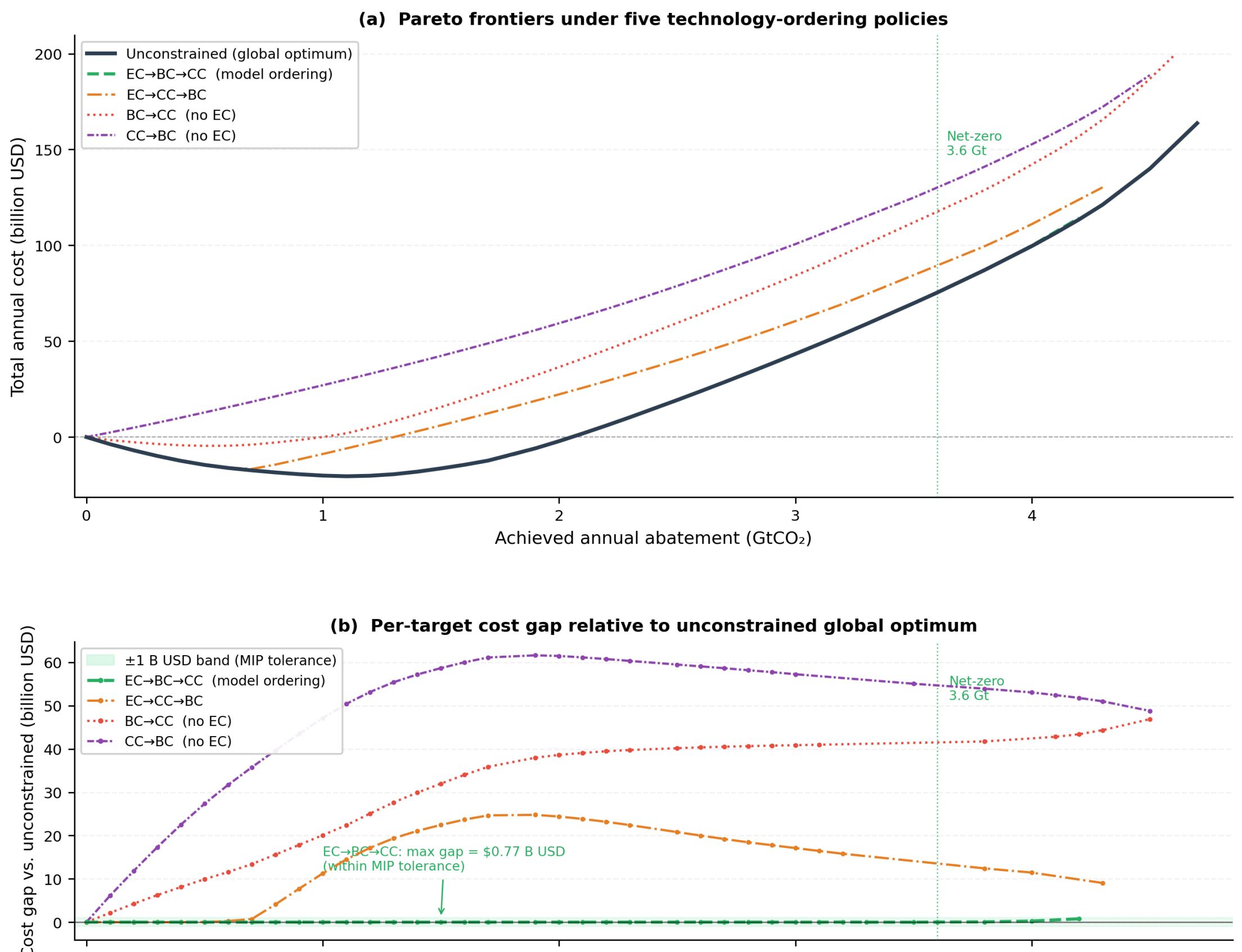

Fig. S18 Validation of alternative technology-ordering constraints. (a) Total annualized system cost under alternative technology-ordering constraints compared with the unconstrained global optimum. (b) Additional system cost relative to the unconstrained optimum. The recommended sequence (energy conservation → biomass co-firing → carbon capture) closely reproduces the unconstrained solution, whereas alternative deployment orders incur progressively larger economic penalties.

## 10.7 Decomposition of interaction-induced changes in unit abatement costs

To distinguish genuine engineering interactions from changes arising purely from accounting conventions, we decomposed the interaction-induced change in unit abatement cost (ΔUAC) into two additive components using a two-factor Shapley decomposition: a cost component, representing changes in actual system costs, and an abatement-baseline component, representing changes caused by shifts in the reference emissions used to calculate unit abatement costs.

As shown in Fig. S19a, biomass co-firing (BC) exhibits a predominantly positive cost component across the mitigation range, indicating that interaction with upstream technologies increases the actual resource cost of biomass deployment. This increase is partially offset by a negative abatement-baseline component, resulting in only a modest increase in the net ΔUAC. In other words, the interaction effect for BC is driven primarily by changes in physical system costs rather than by accounting effects.

Carbon capture (CC) exhibits the opposite pattern (Fig. S19b). The cost component becomes negative, reflecting genuine cost reductions associated with shared transport and storage infrastructure as well as lower operating costs. However, these savings are outweighed by a positive abatement-baseline component because upstream mitigation measures reduce the remaining $CO_2$ available for capture, thereby mechanically increasing the reported unit abatement cost. Consequently, the observed increase in CC unit abatement cost arises predominantly from the changing abatement baseline rather than from higher engineering costs.

The plant-level distributions shown in Fig. S19c,d further support these observations. Most biomass retrofit cases are characterized by positive cost contributions partially offset by negative baseline effects, whereas most carbon capture cases combine negative cost contributions with positive baseline effects. These contrasting decomposition patterns demonstrate that the interaction mechanisms differ fundamentally between biomass co-firing and carbon capture and confirm that changes in unit abatement costs cannot be interpreted solely as changes in engineering costs.

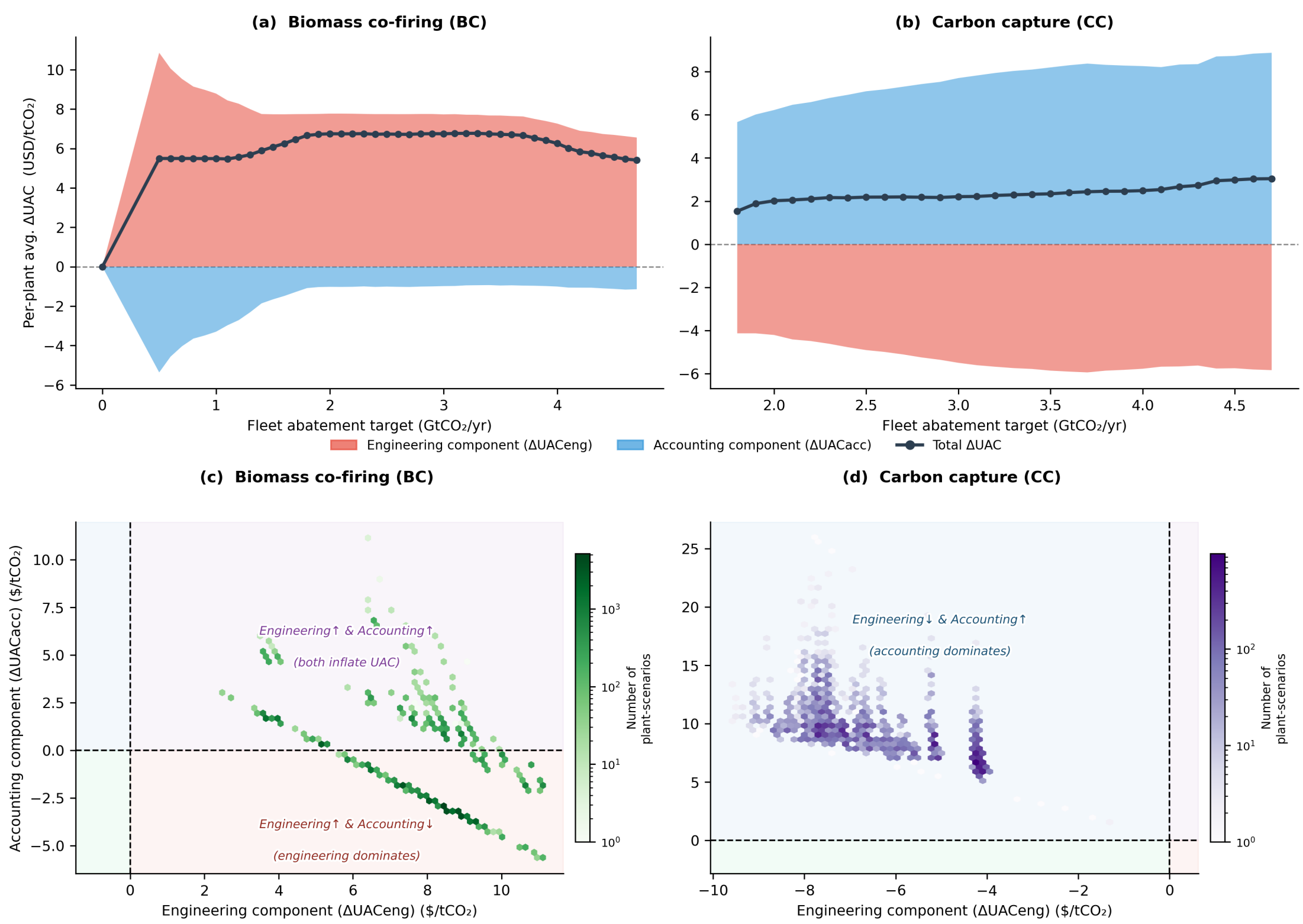

Fig. S19 Decomposition of interaction-induced changes in unit abatement costs. Interaction effects are decomposed into a genuine cost component and an abatement-baseline component using a two-factor Shapley decomposition. Panels (a) and (b) show the average contribution of each component across emission reduction targets for biomass co-firing (BC) and carbon capture (CC), respectively. Panels (c) and (d) present the corresponding plant-level distributions of the two components. The results demonstrate that interaction-induced changes in BC are dominated by genuine cost effects, whereas those for CC are primarily driven by changes in the abatement baseline.

## 10.8 Fleet fuel consumption breakdown

Figure S20 illustrates how the fleet-wide fuel mix evolves as progressively more stringent emission reduction targets are imposed. Under the baseline scenario, coal supplies almost the entire energy demand of the fleet, accounting for approximately 39.5 × $10^9$ GJ annually.

As mitigation targets increase, energy conservation progressively reduces total fuel demand while biomass co-firing increasingly substitutes for coal. By the maximum abatement target, cumulative energy savings from energy conservation reach 7.18 × $10^9$ GJ, corresponding to an 18.2% reduction in the original coal demand. At the same time, biomass consumption increases to 13.14 × $10^9$ GJ, accounting for approximately one-third of the total fuel input.

Consequently, coal consumption decreases from 39.54 × $10^9$ GJ under the baseline to 19.23 × $10^9$ GJ at the maximum abatement level, representing a reduction of 51.4%. Approximately one-third of this reduction is achieved through lower energy demand following efficiency improvements, while the remainder results from fuel substitution by biomass. The residual coal consumption is subsequently coupled with carbon capture to achieve the deep emission reductions illustrated by the fleet-wide marginal abatement cost curves in the main text.

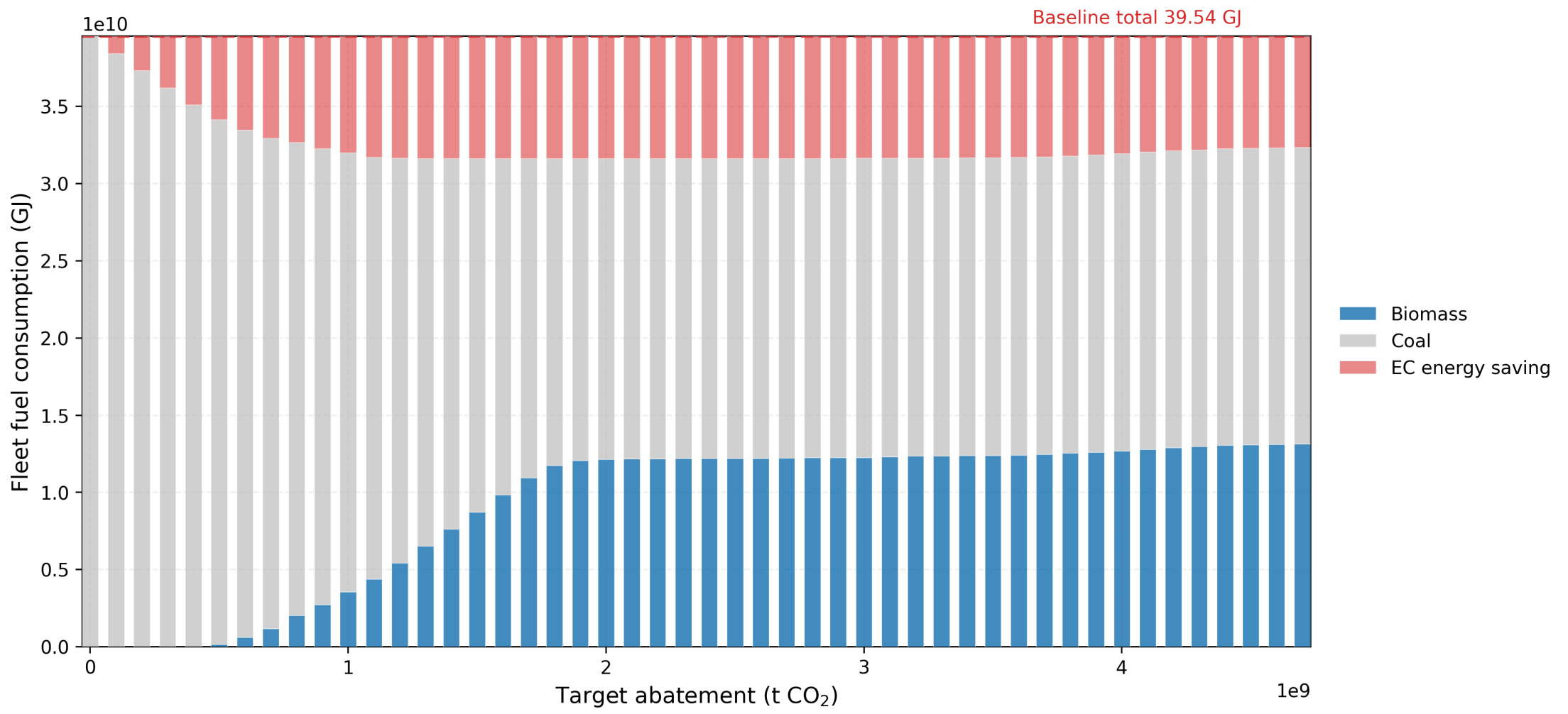

Fig. S20 Evolution of fleet-wide fuel consumption across emission reduction targets. Annual fleet fuel consumption is decomposed into coal consumption, biomass consumption, and energy savings from energy conservation across the 48 optimized emission reduction targets. The dashed horizontal line indicates the baseline total energy demand, which remains constant because energy conservation reduces fuel input rather than electricity generation.

### 10.9 Marginal cost deviation from baseline under parameter sensitivities

Figure S21 compares the pointwise deviation of the marginal abatement cost (MAC) from the baseline optimization under variations in key economic parameters.

Coal price produces the largest deviation throughout the mitigation range. Increasing the coal price by 25% decreases the MAC by up to approximately \$12 t $CO_2^{-1}$ over the intermediate abatement range (1.5–3.0 Gt $CO_2$ yr$^{-1}$), where biomass co-firing contributes substantially to the optimized portfolios. Conversely, lowering the coal price increases the MAC by approximately \$6–8 t $CO_2^{-1}$, reflecting reduced fuel-saving benefits from biomass co-firing and energy conservation.

Variations in carbon capture capital expenditure generate an approximately symmetric deviation of ±5–8 \$ t $CO_2^{-1}$, primarily affecting the high-abatement region where carbon capture dominates the optimized portfolios. Reductions in the discount rate trigger the mildest deviations, generally ranging from −5 to -3 \$ t $CO_2^{-1}$. The maximum influence emerges under the 12% discount rate scenario at an abatement target of roughly 1.6 Gt $CO_2$, when biomass gasification (GB) technologies begin large-scale deployment. Beyond this threshold, the gaps between scenarios gradually narrow to 5–7 \$ t $CO_2^{-1}$.

Across all parameter categories, deviations remain close to zero at low abatement levels, confirming that the economically attractive portion of the optimized mitigation portfolio dominated by energy conservation and biomass co-firing is relatively insensitive to these parameter uncertainties.

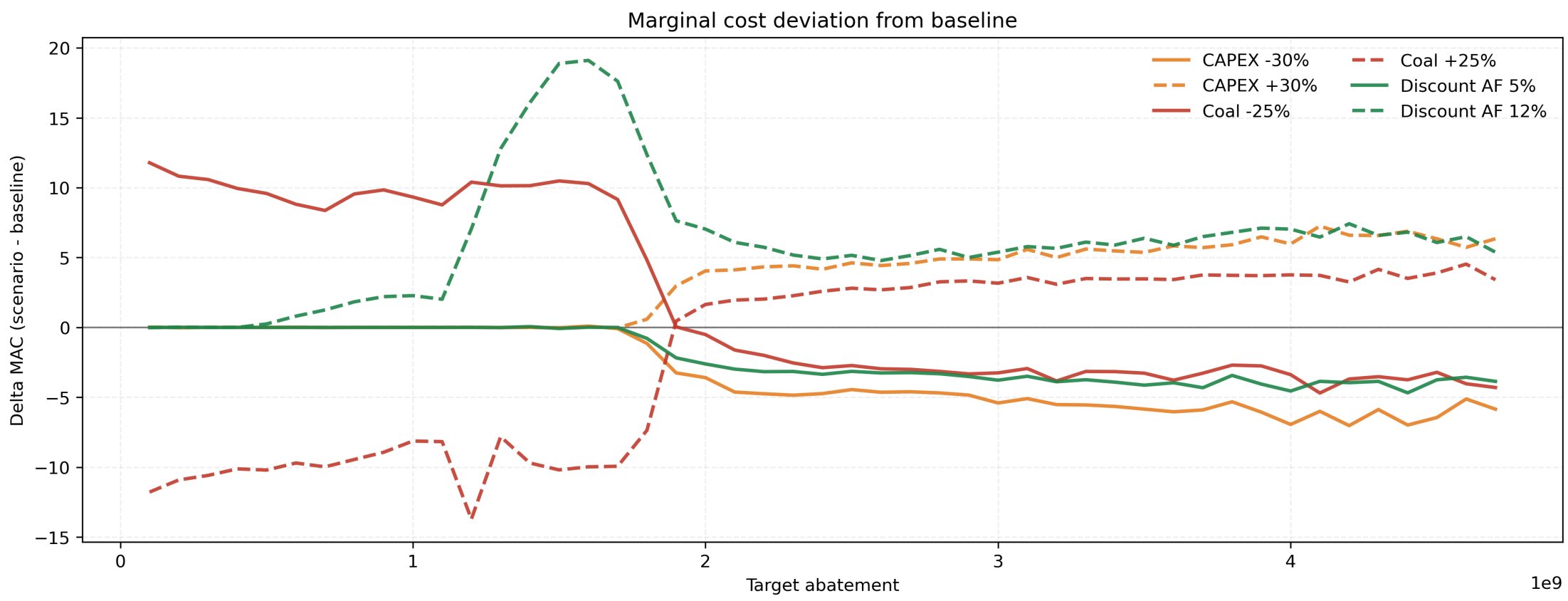

Fig. S21 Pointwise deviation of marginal abatement costs under parameter sensitivity scenarios. Deviations are calculated relative to the baseline optimization for alternative assumptions regarding coal prices (±25%), carbon capture capital expenditure (±30%), and discount rates (5% and 12%). Positive values indicate higher marginal abatement costs than the baseline.

### 10.10 Optimal abatement response under alternative carbon prices

To link the optimized marginal abatement cost curve with market-based climate policy, we derived the economically optimal fleet-wide emission reduction as a function of the carbon price. The optimal response is obtained by minimizing the net system cost under alternative carbon prices and therefore represents the cost-effective emission reduction level that would be voluntarily adopted by the coal-fired power fleet.

As shown in Fig. S22, the optimal abatement increases monotonically with carbon price but exhibits a stepwise rather than continuous response because technology portfolios change discretely as different retrofit combinations become economically competitive. Under very low or negative carbon prices, only a limited number of highly profitable energy conservation measures are adopted. As the carbon price increases to approximately \$18 t $CO_2^{-1}$, biomass co-firing gradually enters the optimized portfolios, increasing the economically optimal emission reduction to approximately 1.5 Gt $CO_2$ $yr^{-1}$.

A more pronounced transition occurs at a carbon price of approximately \$56 t $CO_2^{-1}$, beyond which carbon capture rapidly becomes economically competitive and the optimal abatement increases from less than 3.5 Gt $CO_2$ $yr^{-1}$ to approximately 3.6 Gt $CO_2$ $yr^{-1}$, corresponding to the fleet-wide carbon-neutrality threshold. Further increases in carbon price continue to expand carbon capture deployment until the maximum technically achievable mitigation potential (4.7 Gt $CO_2$ $yr^{-1}$) is reached at approximately \$125 t $CO_2^{-1}$.

Comparison with representative policy benchmarks further illustrates the practical implications of these results. Under the current China ETS carbon price (approximately \$9 t $CO_2^{-1}$), the economically optimal emission reduction reaches only 1.3 Gt $CO_2$ $yr^{-1}$, equivalent to approximately 28% of the fleet's maximum mitigation potential. In contrast, carbon prices comparable to those observed in the EU ETS would support substantially deeper decarbonization by making large-scale carbon capture economically attractive.

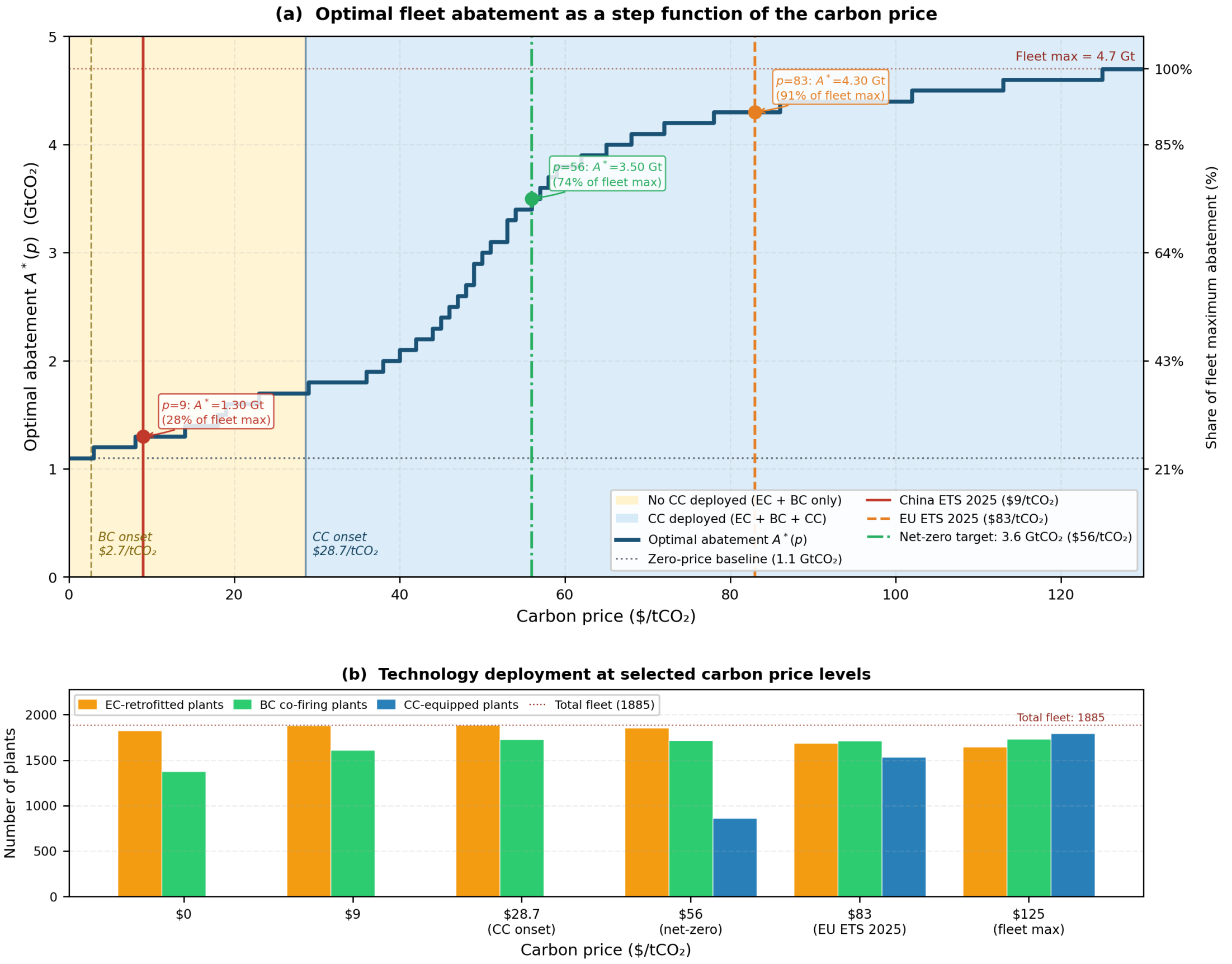

Fig. S22 Optimal fleet-wide abatement response under alternative carbon prices. The figure illustrates the economically optimal fleet-wide $CO_2$ abatement obtained by minimizing the net system cost under different carbon prices using the interaction-aware optimization framework. The left axis shows the marginal abatement cost (MAC) frontier, while the step function indicates the corresponding optimal abatement level under each carbon price. Representative reference prices, including the current China ETS, the EU ETS, and a benchmark carbon tax, are indicated for comparison. The discrete upward transitions reflect changes in the cost-optimal technology portfolio as progressively higher carbon prices make additional mitigation measures economically viable.

## 10.11 Provincial distribution of policy subsidy savings enabled by technology interactions

Output- and performance-based incentives, such as per-kWh support for low-carbon electricity from biomass co-firing and per-ton incentives for captured and geologically stored $CO_2$ [43,44], can improve the economic viability of mitigation technologies, but uniform subsidies may impose substantial fiscal burdens, particularly for high-cost plants. The pairwise interaction analysis indicates that technology co-deployment can reduce the level of policy support required: combining energy conservation (EC) with biomass co-firing lowers the unit cost of biomass-generated electricity, while combining biomass co-firing (BC) with carbon capture (CC) can reduce the unit abatement cost of CC through the contribution of captured biogenic $CO_2$. Figure S23 shows how the resulting potential subsidy savings are distributed across provinces under representative technology combinations.

For biomass co-firing, the presence of EC reduces the unit cost of biomass-generated electricity (Fig. 2f), thereby lowering the subsidy required to support biomass-based power generation, particularly for gasified biomass co-firing (GB). Across four representative EC–GB configurations combining 10% or 20% EC abatement with 10% or 20% biomass co-firing, subsidy savings are concentrated in provinces with large coal-fired generation capacity and substantial biomass utilization, particularly Inner Mongolia, Shandong, Jiangsu, Shanxi, and several eastern coastal provinces. Depending on the technology configuration, provincial annual savings range from approximately US$2 million to more than US$70 million.

For carbon capture, the presence of biomass co-firing reduces the UAC of CC through the contribution of captured biogenic $CO_2$ (Fig. 2e), thereby lowering the policy support required for $CO_2$ capture and geological storage. The largest savings occur for the configuration combining 20% biomass co-firing with 90% carbon capture and are concentrated in provinces with substantial deployment potential for both technologies, including Inner Mongolia, Xinjiang, Shanxi, and parts of northern China. Provincial annual savings range from approximately US$25 million to more than US$59 million.

Overall, these results indicate that cross-technology interactions can generate fiscal co-benefits in addition to changing mitigation costs. The substantial provincial heterogeneity in these savings suggests that accounting for local technology combinations and interaction effects could improve the cost-effectiveness and regional targeting of policy support for coal-power decarbonization.

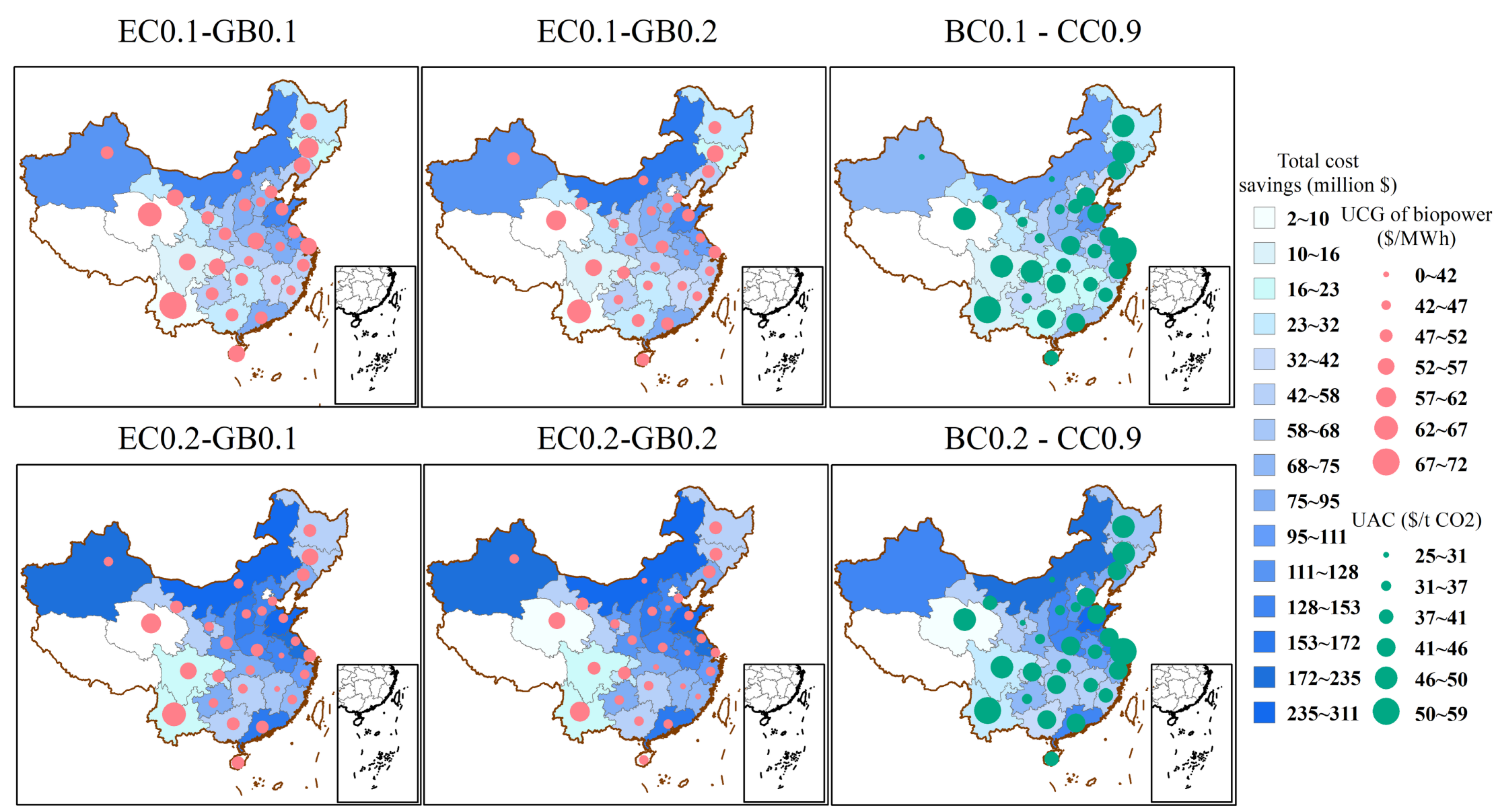

**Fig. S23** Provincial distribution of policy subsidy savings enabled by technology interactions. Spatial distribution of annual subsidy savings arising from changes in technology costs under representative paired technology configurations. The four left panels show reductions in biomass electricity subsidy requirements resulting from the lower unit generation cost of gasified biomass co-firing (GB) when combined with energy conservation (EC). The two right panels show reductions in carbon capture subsidy requirements resulting from the lower unit abatement cost (UAC) of carbon capture (CC) when combined with biomass co-firing (BC). Technology labels indicate the abatement level or deployment level of each technology in the paired configuration; for example, EC0.1–GB0.1 represents a configuration combining 10% $CO_2$ abatement EC with 10% $CO_2$ abatement GB. Total subsidy savings are calculated by multiplying the interaction-induced unit cost saving by the corresponding amount of biomass-generated electricity or CO2 abatement, as detailed in Note 6.

## 10.12 Comparative analysis for direct biomass co-firing and energy conservation retrofits

To further compare the economic characteristics of the two low-cost retrofit options, Fig. S24 presents a plant-level comparison between direct biomass co-firing (DB) and energy conservation (EC) under both 10% and 20% $CO_2$ reduction scenarios.

Panel (a) compares the unit abatement costs (UACs) of the two technologies against the one-to-one reference line (x = y). The results show that all power plants exhibit lower UACs for EC than for DB with 10% abatement, while only 0.35% of the installed capacity achieves a lower UAC under DB0.2 than under EC0.2. Furthermore, approximately 70% of generating capacity exhibits negative UACs for either EC or DB retrofits, indicating that these plants can recover retrofit investments through fuel savings or biomass fuel substitution. These economically attractive plants therefore represent the low-hanging fruits for near-term decarbonization.

Panels (b) and (c) further compare the initial capital expenditure (CAPEX) and simple payback period (PP) of these profitable retrofit opportunities. Although direct biomass co-firing generally requires lower upfront investment than energy conservation, its investment payback period is substantially longer for most plants. In contrast, energy conservation typically requires higher initial capital expenditure but achieves faster investment recovery because of greater fuel-cost savings. These results suggest different investment strategies under different financing conditions. For projects with limited access to capital, direct biomass co-firing may represent a more feasible entry option owing to its lower upfront investment. Where sufficient investment capital is available, however, energy conservation is generally preferable because it delivers faster capital recovery and superior long-term economic performance.

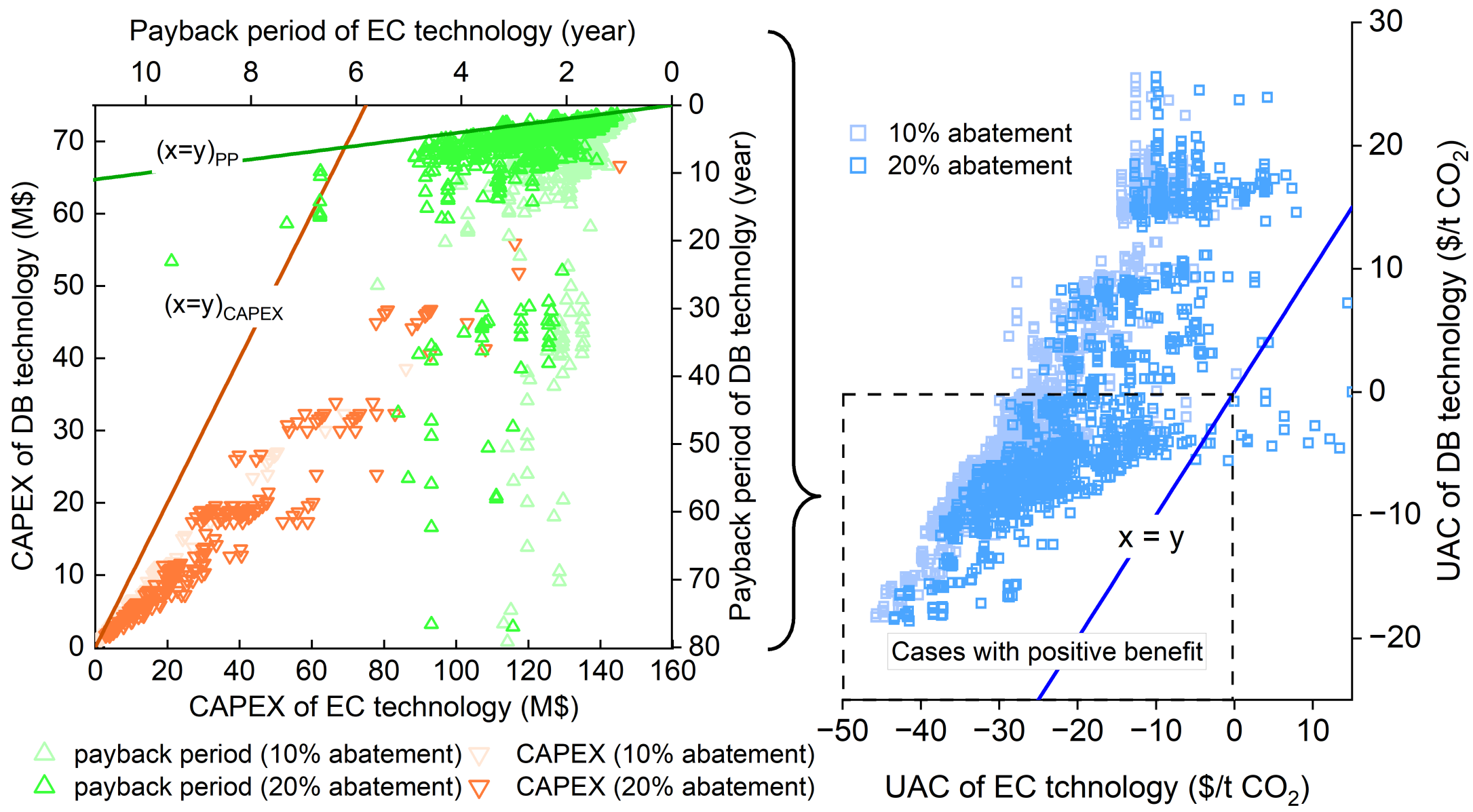

**Fig. S24** Comparative analysis of unit abatement cost (UAC), capital expenditure (CAPEX), and payback period (PP) for direct biomass co-firing (DB) and energy conservation (EC) retrofits under 10% and 20% $CO_2$ reduction scenarios. Right panel compares plant-level UACs between DB and EC using the one-to-one reference line. Left panels compare the corresponding initial investment costs and simple payback periods for retrofit projects with negative UACs (economically profitable retrofits).